\documentclass[12pt, a4paper]{article}

\usepackage[utf8]{inputenc}
\usepackage[T1]{fontenc} 
\usepackage[british]{babel}

\usepackage{amsmath, nicefrac}

\usepackage[textwidth=16cm, textheight=25cm]{geometry}
\usepackage{setspace}

\usepackage[title]{appendix}

\usepackage{pdflscape}

\usepackage{csquotes}
\usepackage[backend=biber, style = apa, natbib=true, maxcitenames=2, maxbibnames=99, url=false, doi=false, uniquename=false]{biblatex}
\usepackage{amssymb}
\usepackage{multirow}
\usepackage{makecell}
\usepackage{booktabs}
\usepackage{siunitx}
\newcolumntype{d}{S[
    input-open-uncertainty=,
    input-close-uncertainty=,
    parse-numbers = false,
    table-align-text-pre=false,
    table-align-text-post=false
 ]}

\usepackage{caption}
\usepackage{subcaption}
\usepackage{float}
\usepackage{adjustbox}
\usepackage{graphicx}

\usepackage{hyperref}
\hypersetup{
    citecolor=blue,
    colorlinks=true,
    linkcolor=blue,
    filecolor=magenta,      
    urlcolor=cyan,
    pdfpagemode=FullScreen,
    }

\title{Augmenting or Automating Labor? The Effect of AI Development on New Work, Employment, and Wages\footnote{I thank Michela Bia, Salvatore Di Salvo, Francis Kramarz, Christina Gathmann, Giuseppe Grasso, Terry Gregory, Mattia Laudi, Joel Machado, Fabien Petit, and Konstantinos Tatsiramos for constructive and helpful comments. I am also grateful to the participants of the EALE 2024 conference, the EEA 2024 conference, the IAB 15\textsuperscript{th} Interdisciplinary Ph.D. Workshop, 4\textsuperscript{th} Junior Research Day, 16\textsuperscript{th} IAAEU Workshop on Labour Economics, 72\textsuperscript{nd} Congress of the AFSE, the 4\textsuperscript{th} AI-Econ Lab conference, and the CORA 2024 conference for their valuable feedback. This research has been funded by the Luxembourg National Research Fund (AFR-15666136). All errors are my own. Research results and conclusions expressed are those of the author.}
}
\author{David Marguerit\footnote{Email: david.marguerit@liser.lu} \\
\small LISER
}
\providecommand{\keywords}[1]{\textbf{Keywords:} #1}
\providecommand{\JEL}[1]{\textbf{JEL classification:} #1}
\date{March 2025}

\begin{document}

\maketitle

\begin{abstract}
Artificial intelligence (AI) is reshaping the labor market by changing the task content of occupations. This study investigates the impact of AI development on the emergence of new work, employment, and wages in the United States from 2015 to 2022. I develop innovative methods to measure occupational and industry exposure to AI technologies that substitute labor (automation AI ) or enhance workers' output (augmentation AI), and to identify new work (i.e., new job titles). To address endogeneity, I use instrumental variable estimators, leveraging AI development in countries with limited economic ties to the United States. The findings indicate that automation AI negatively impacts new work, employment, and wages in low-skilled occupations, while augmentation AI fosters the emergence of new work and raises wages for high-skilled occupations. These results suggest that AI may contribute to rising wage inequality.
\end{abstract} \hspace{10pt}

 \par
\JEL{E24, J11, J21, J23, J24, J31, O33}

\clearpage

\begin{refsection}

\section{Introduction}

Artificial intelligence (AI) differs significantly from previous technological innovations due to its rapid advancements and its ability to perform a wide range of non-routine tasks. For instance, AI algorithms excel in image classification, visual reasoning, and language understanding (\cite{kiela_et_al_2021, stanford_ai_2024}), making it applicable across most occupations and industries. Consequently, these characteristics of AI introduce considerable uncertainty regarding the future of work \citep{korinek_2024}.

Investigating the impact of AI on the labor market requires data that capture both the extent of AI development in the economy and its specific uses. However, existing tools for measuring AI adoption remain scarce \citep{seamans_Raj_2018}. Moreover, even when data on AI adoption are available, they often fails to distinguish between automation AI and augmentation AI technologies \citep{acemoglu_al_2022, zolas_al_2020}. Automation AI refers to technologies that substitute for labor by automating tasks in occupations. Augmentation AI includes technologies that enhance workers' output by improving capabilities, quality, variety, or utility of outputs in occupations and industries \citep{autor_chin_salomons_seegmiller_2024}. This distinction is crucial, as these two applications may have opposing effects on the labor market \citep{acemoglu_restrepo_2018_aer, autor_chin_salomons_seegmiller_2024, acemoglu_2024, acemoglu_kong_restrepo_2024}.

This study investigates the effects of AI on the US labor market from 2015 to 2022, distinguishing between the impacts of automation AI and augmentation AI. Specifically, I assess occupational exposure to AI technologies suitable for task performance (automation AI) and complementing workers' output (augmentation AI). Then, I analyze how exposure to either of the AI technologies affects the emergence of new work, employment and wages.

Recent research underscores the importance of distinguishing between automating and augmenting technologies to understand the impact of technological change on the labor market, since they can have countervailing effects \citep{acemoglu_restrepo_2018_aer, autor_chin_salomons_seegmiller_2024, acemoglu_2024, acemoglu_kong_restrepo_2024}. The overall effect of automation is ambiguous: while it can displace labor, reducing labor demand and exerting downward pressure on wages, it may also increase productivity, stimulating labor demand and wage growth \citep{aghion_al_2022, autor_2015, acemoglu_restrepo_2018_aer}. In contrast, augmentation AI is expected to create new tasks in which labor has a comparative advantage, thereby boosting labor demand and wages.

The analysis relies on two novel measures to quantify exposure to AI and an innovative metric to capture the emergence of new work. The augmentation AI and automation AI indices measure occupational and industry exposure to AI development in the United States from 2015 to 2022. These measures are constructed by mapping AI-related questions posted on Stack Overflow—the leading Q\&A forum for developers—to occupational and industry descriptions. This mapping is performed using Semantic Textual Similarity, a natural language processing tool that quantifies documents similarity in terms of meaning \citep{Reimers_Gurevych_2019}. An AI-related question can be used to measure automation, augmentation, or both. While Stack Overflow data has been widely used in data science research to analyze developer discussions, it has not previously been employed in economic studies.

To construct the augmentation AI index at the occupational-industry level by year, I match AI-related questions with micro-titles listed in the 2016 Census Alphabetical Index (CAI). The CAI provides thousands of micro-titles associated with occupations and industries, offering detailed descriptions of the outputs produced by these roles (\cite{autor_chin_salomons_seegmiller_2024}). For example, the micro-title "Language translator", which belongs to the occupation "Interpreters and Translators", is matched to questions about AI technologies such as "language-translation" and "microsoft-translator". Similarly, the AI technology "vehicle-routing" is linked to the occupation "Couriers and Messengers", which includes micro-titles such as "Delivery driver, courier".

To construct the automation AI exposure index at the occupational level by year, I follow \textcite{felten_raj_seamans_2018, felten_raj_seamans_2021} by linking AI-related questions to the abilities workers require to perform tasks in specific occupations. The measure of occupational automation AI exposure is then built based on the importance of these abilities for task performance within each occupation, according to 2015 O*NET. For example, the AI technologies "voice-recognition" and "speech-recognition" are mapped to the ability "Speech Recognition", which is particularly important in the occupation of "Speech-Language Pathologists".\footnote{A Speech-Language Pathologist assesses and treats persons with speech, language, voice, and fluency disorders.}

The automation AI and augmentation AI measures are validated against external sources. I demonstrate that automation AI exposure is highly correlated with firms' adoption of AI technologies for task automation and the measure developed in \textcite{felten_raj_seamans_2021}. In the absence of an external benchmark for augmentation AI, I show that automation AI and augmentation AI capture distinct dimensions of AI technologies, despite differing only in the corpora used to measure input abilities and workers' outputs. Furthermore, I provide evidence that augmentation AI can be interpreted as a measure of enhancing technology.

Building on \textcite{lin_2011}, the third novel measure tracks the emergence of new work within occupations by comparing successive updates of the "Alternate titles" category in O*NET from 2015 to 2022. The "Alternate titles" category provides detailed job titles within occupations, offering more specific descriptions of professional roles. I identify new job titles by comparing updates of the "Alternate titles" category and measuring the textual similarity for each pair of job titles between two years. For example, the job titles "Sprinkler Design Engineer" and "Remote Pilot" were added in 2018 and 2021, respectively, signaling the emergence of new tasks or job roles.

Descriptive analysis reveals that AI differs from previous waves of technological change and that the emergence of new work is highly concentrated. High-skilled occupations, as well as those in sales, office, and administrative support, are more exposed to automation AI, whereas physical occupations are less affected. These patterns reflect differences in occupational ability specialization. Augmentation AI exposure is primarily concentrated in science, technology, engineering, and mathematics (STEM) occupations. When examining occupational content, I find that automation AI exposure is negatively correlated with routine tasks and positively associated with the level of task expertise. Augmentation AI exposure is associated with non-routine interpersonal tasks and task expertise. Finally, while new work emerges across all broad occupational categories, the highest concentration of new work creation is observed in computer and mathematical occupations.

The effects of automation AI and augmentation AI exposure on new work, employment, and wages are analyzed at the occupation-industry level by year from 2015 to 2022. Specifically, I use OLS estimators to regress automation AI and augmentation AI on labor market outcomes (share of new work, employment, and hourly wages). I control for a comprehensive set of covariates, including workforce characteristics and trade exposure. Additionally, I incorporate fixed effects to account for temporary industry-level shocks (e.g., robotization) and time-invariant occupation-industry characteristics.

Given the potential concerns of endogeneity, it is crucial to adopt a strategy that captures causal relationships. I employ an instrumental variables approach, utilizing five-year lagged AI development in countries with no significant economic ties to the United States. This approach addresses endogeneity threats for three reasons. First, by carefully selecting economies without direct ties to the US economy, I mitigate the risks of reverse causality and shocks that could simultaneously affect AI development and labor demand. Second, using a five-year lagged AI development alleviates the risk of anticipation in occupational sorting, as workers and firms are unlikely to predict AI developments that far in advance—especially given the significant variance and historical inaccuracies in AI development projections (\cite{armstrong_al_2014}). Finally, by employing five-year lagged AI development from countries with limited relationship to the US economy, I ensure that the instrumental variable does not directly affect the US labor market.

The first empirical analysis examines the impact of automation AI and augmentation AI exposure on the emergence of new work. The two forms of AI have distinct effects. Augmentation AI exposure significantly stimulates the creation of new work, whereas automation AI exposure has no measurable impact on this outcome. These results align with predictions that augmenting technologies foster new work by introducing novel processes, products, and services that demand specialized expertise and competencies \citep{acemoglu_restrepo_2018_aer, autor_chin_salomons_seegmiller_2024, acemoglu_kong_restrepo_2024}.

The second set of results investigates the role of automation AI and augmentation AI on employment size. Augmentation AI exposure positively affects employment, suggesting that firms are hiring to accommodate the emergence of new work. However, I do not observe any significant effect of automation AI exposure on employment. Three factors could explain this result. First, the displacement and productivity effects might be of similar magnitude, resulting in a net neutral effect. Second, despite recent advancements, AI technologies may not yet be advanced enough to fully displace labor in most occupations. Third, labor market rigidities might limit the displacement effect.

The third empirical analysis focuses on the effect of AI development on average hourly wages. I document an adverse effect of automation AI exposure on wages, suggesting that the displacement effect of automation outweighs any potential productivity gains. Conversely, I do not find any significant impact of augmentation AI exposure on wages.

Finally, I investigate the heterogeneous effects of AI by occupational skill group, reanalyzing the data by classifying occupations into low-, middle-, and high-skilled categories. Automation AI exposure has a negative impact on the emergence of new work, employment, and wages for low-skilled occupations, while augmentation AI exposure shows no significant effect in this group. In contrast, augmentation AI exposure fosters the emergence of new work and increases wages for high-skilled occupations, but it does not affect employment. These findings may be explained by labor shortages in high-skilled occupations and the slow workforce adjustment required to meet the growing demand for advanced skills. Middle-skilled occupations experience mixed effects, falling between low- and high-skilled occupations. The results support the hypothesis that automation AI technologies are more applicable to low-skilled occupations, while augmentation AI technologies are better suited for high-skilled occupations \citep{bloom_al_2024, acemoglu_restrepo_2018_jhc, autor_2024, acemoglu_2024}. Overall, these findings suggest that AI development may exacerbate wage inequality.

My paper contributes to three strands of economic literature. First, it builds on research developing AI exposure indices. \textcite{brynjolfsson_mitchell_rock_2018} propose a forward-looking index assessing the extent to which occupational tasks could be suitable for machine learning, a subfield of AI, in the near future. \textcite{felten_raj_seamans_2018, felten_raj_seamans_2021} leverage the AI Progress Measurement project by the Electronic Frontier Foundation to quantify exposure across 10 potential AI applications, while \textcite{engberg_2024} extends this work to a longitudinal framework covering 2010–2023. \textcite{webb_2020} utilizes patent data to estimate the extent to which occupational tasks can be performed by AI. \textcite{tolan_pesole_al_2021} construct an index linking AI research intensity to occupations based on expert assessments. Finally, \textcite{eloundou_al_2024} introduce an index measuring exposure to large language models.

While these approaches focus on the current and potential capabilities of AI algorithms, they do not directly capture the development of AI technologies. In contrast, I introduce new measures of AI exposure by analyzing the activities of developers, enabling to observe the types of AI technologies being developed and their intended applications. This approach approximates AI adoption more closely than previous indices based on AI capabilities. To achieve this, I propose a novel data source: posts on Stack Overflow.

Second, this work speaks to the large and growing literature studying the effect of technological changes on the labor market (\cite{autor_al_2003, Goos_Manning_Salomons_2009, Michaels_Ashwini_vanReenen_2014,autor_dorn_2013, Goos_Manning_2007, krueger_1993, machin_vanreenen_1998, Acemoglu_Restrepo_2020, Graetz_Michaels_2018, gregory_2022}). Focusing on AI and closer to my paper, \textcite{acemoglu_autor_hazel_restrepo_2022} find no significant relationship between AI exposure and employment and wage growth at the occupational level. In contrast, \textcite{babina_al_2024} show that firms investing in AI experience increased employment growth, and \textcite{bonfiglioli_al_2025} demonstrate that AI exposure negatively affects employment across US commuting zones. \textcite{mäkelä_stephany_2024} provide evidence of rising demand and wage premiums for AI-complementary skills, based on online job vacancy data. Using workers resumes and job postings, \textcite{hampole_al_2025} show that labor demand declines with exposure to AI technologies but increases with the dispersion of task exposure to AI. \textcite{gathmann_al_2025} show that the task shifts due to AI mainly occur within detailed occupations.

I contribute to this literature in three key ways. First, I distinguish between AI that substitutes for labor and AI that complements workers' output. Prior research has primarily examined the overall effects of AI without making this distinction, despite evidence that automating and augmenting AI may have opposing impacts \citep{acemoglu_Restrepo_2018, acemoglu_restrepo_2018_aer, acemoglu_2024, autor_chin_salomons_seegmiller_2024}. Second, I provide causal evidence by employing an instrumental variables strategy that leverages AI exposure in countries with limited economic ties to the United States. Third, I use longitudinal data, allowing for the inclusion of a rich set of fixed effects, strengthening the robustness of the analysis.

Finally, this paper contributes to the emerging literature on the sources of new work. \textcite{lin_2011} uses the Census Alphabetical Indexes of Industries and Occupations to show that new work emerges in areas dense with college graduates and industry variety between 1965 and 2000. \textcite{atalay_al_2020} analyze job ads from 1950 to 2000 to explore changes in job titles, reflecting real changes in occupational tasks. \textcite{kim_2022, kim_al_2024} study the impact of trade and local factors on new work emergence using O*NET and the Dictionary of Occupation Titles between 1980 and 2010. \textcite{autor_chin_salomons_seegmiller_2024} combine the Census Alphabetical Indexes with patent data to investigate innovation's impact on new work from 1930 to 2018, finding that task-complementing innovations are associated with new work emergence.

Unlike previous studies, this research is the first to examine the emergence of new work driven by AI. AI is expected to play a crucial role in new work creation, given its broad impact across occupations and industries. Additionally, this study extends prior research by identifying new work between 2015 and 2022, covering a more recent period than previous attempts. Finally, I introduce a novel approach based on Semantic Textual Similarity, a state-of-the-art method for measuring document similarity \citep{Reimers_Gurevych_2019}. In contrast, previous studies have relied on string distance measures and Word2Vec algorithms, which fail to account for contextual meaning.

The remainder of this study is structured as follows. Section 2 outlines the theoretical mechanisms through which AI affects the labor market. Section 3 describes the data and details the construction of AI exposure indices and new work. Section 4 presents descriptive statistics on AI exposure and the emergence of new work. In Section 5, I investigate the impact of AI exposure on new work, employment, and wages. Section 6 explores the heterogeneous effects of AI across occupational skill levels. Section 7 provides robustness checks, and Section 8 concludes.

\section{Theoretical mechanisms}

The task‐based framework provides a natural foundation for understanding the theoretical mechanisms through which AI affects the labor market. This framework has proven particularly useful in analyzing not only historical waves of technological changes but also the recent surge in AI technologies \citep{zeira_1998, autor_al_2003, acemoglu_autor_2011, acemoglu_kong_restrepo_2024, acemoglu_2024, bloom_al_2024, aghion_jones_jones_2018, acemoglu_Restrepo_2018}. 

Its core insight is that production can be decomposed into a bundle of tasks, each of which may be performed by labor or capital (machines). The allocation of tasks is determined by the relative costs and capabilities of these inputs—a dynamic that becomes particularly relevant as new technologies reshape these trade-offs.

As technological progress reduces the cost of capital relative to labor, the range of tasks performed by capital expands, shifting the allocation of tasks between labor and machines. For example, in modern manufacturing, robotic arms and automated inspection systems have replaced routine tasks, while workers increasingly focus on managing these systems or performing non-routine tasks \citep{acemoglu_autor_2011}.

Within this framework, AI technologies can be broadly categorized as either automating or augmenting. Automating AI technologies replaces tasks traditionally performed by labor, creating a displacement effect. As AI advances in image recognition, natural language processing, and decision-making, tasks such as data analysis and standardized customer service become increasingly automated. This shift is evident across industries, from warehouse logistics, where automated guided vehicles operate alongside workers, to financial services, where algorithmic trading systems have replaced roles previously held by analysts. Despite the potential decline in labor demand for these tasks, productivity gains from automation can lower production costs and enhance firm competitiveness, enabling expansion into new markets (productivity effect). This expansion may, in turn, increase labor demand, suggesting that the overall impact of automation AI on employment and wages depends on the pace of automation and the relative strength of the displacement and productivity effects \citep{acemoglu_restrepo_2018_aer, autor_2015, aghion_jones_jones_2018}.

Augmenting AI technologies, in contrast, are designed to complement rather than replace human labor. These technologies enhance workers' output, expanding the set of tasks that labor can perform by generating new work \citep{autor_chin_salomons_seegmiller_2024}. Because labor holds a comparative advantage in these new tasks, their emergence leads to increased labor demand and rising wages. Drawing on Census records, \textcite{lin_2011, autor_chin_salomons_seegmiller_2024} show that technological change gives rise to new work, as reflected in the emergence of new job titles. \textcite{acemoglu_restrepo_2018_aer} find that the introduction and expansion of new tasks and job titles accounted for nearly half of the employment growth from 1980 to 2010. In the context of AI, recent examples of new tasks and occupations include roles such as "prompt engineers" and "AI model trainers", which have emerged as firms increasingly integrate AI into their workflows \citep{acemoglu_Restrepo_2018}.\footnote{Beyond these direct effects, additional mechanisms have been identified, including capital accumulation and deepening automation (automation at the intensive margin) \citep{acemoglu_2024,acemoglu_Restrepo_2018, acemoglu_kong_restrepo_2024}. Historically, these mechanisms have had less impact on employment and wages than the displacement and productivity effects, and the emergence of new work. Therefore, Therefore, they are not examined in this study.}

The balance between automation and augmentation is critical in shaping AI's impact on employment and wage structures. A greater emphasis on automation may lead to labor displacement and downward wage pressures if the displacement effect outweighs the productivity effect. Conversely, augmenting AI technologies fosters the creation of new work, enabling employment growth and supporting wage increases.

\section{Data and measurement}

In this section, I present the data sources and methodology used to construct the measures of AI development for automation and augmentation and the measure of new work.\footnote{The development of the automation AI and augmentation AI indices involved several methodological choices, each of which warrants discussion. Numerous variants were tested during this process. However, modifying these parameters, whether by applying more or less restrictive definitions, does not alter the quality of the results presented in this study.} Detailed descriptions of the data sources and the methodology for constructing the indices are available in Appendix~\ref{app: Data sources and AI indices}.

Figure~\ref{fig: Schema AI indices} provides an overview of the construction of automation and augmentation AI exposures. I exploit questions about AI posted on Stack Overflow, a Q\&A website specializing in coding issues, to track AI development from 2015 to 2022. To measure AI that substitutes tasks in occupations, these questions are matched to the abilities required for task performance as specified in 2015 O*NET. For augmentation AI, I link AI-related questions from Stack Overflow to outputs produced at the micro-occupation and micro-industry levels, according to the 2016 Census Alphabetical Indexes of Industries and Occupations from the US Census Bureau.

The measure of new work is derived from updates to the "Alternate titles" rubric in O*NET from 2015 to 2022. After extensively cleaning the dataset to ensure that new work reflects new tasks and specializations rather than simple renaming or rewording, I compare alternate titles year by year to identify new work.

\subsection{Automation and augmentation AI exposures}

\subsubsection{AI Development: Stack Overflow}

To track the development of AI algorithms in the economy, I analyze the AI-related questions posted on Stack Overflow between 2010 and 2022. Stack Overflow, established in 2008, is the leading Q\&A platform for programming issues, with over 24 million questions asked and 20 million users as of early 2023. Approximately 80\% of these users report that coding is part of their job responsibilities (\cite{stackoverflow}).\footnote{See Appendix~\ref{app: Stack Overflow} for a detailed overview of Stack Overflow and a discussion of the advantages and limitations of using Stack Overflow to track AI development.}

Each member of Stack Overflow can freely post questions, which must be tagged with 3-5 tags related to technologies or tasks (e.g., scikit-learn, Python, text-to-speech, regex). These tags facilitate the classification of questions into categories. The community provides answers and comments to suggest solutions. Members also vote on the relevance of questions and answers.

I identify AI-related questions on Stack Overflow by utilizing the information provided by tags. Specifically, I employ a combination of keyword searches, tag co-occurrences, and ChatGPT queries to identify tags related to AI. A total of \num{1182} tags are classified as AI-related. Any question using one of these tags is considered AI-related. By identifying and analyzing AI-related questions, I aim to capture the types of AI algorithms developers develop for their organizations. I assume developers typically develop AI algorithms for their employers rather than for personal use.

Then, I determine the location of Stack Overflow members and focus on those residing in the United States and its primary trading partners.\footnote{The primary trading partners are identified by selecting the fifteen most significant countries for US trade in goods and services. These include Bermuda, Canada, China, France, Germany, Hong Kong, India, Ireland, Italy, Japan, Malaysia, Mexico, the Netherlands, South Korea, Switzerland, Thailand, the United Kingdom, and Vietnam.} The members' locations are identified using the information provided in their profiles. By concentrating on members living in the United States and its key trading partners, I aim to track the development of AI algorithms relevant to the US market. It is established that international trade and the activities of multinational corporations significantly contribute to technological transfers (\cite{bilir_morales_2020, keller_al_2013, buera_oberfield_2020}). 

This yields a total of \num{186498} AI-related questions posted from 2010 to 2022 by members of Stack Overflow living in the US and its primary trading partners (see Figure~\ref{fig: Yearly number of questions on SO} for the yearly number of AI-related questions posted on Stack Overflow).

\subsubsection{Occupational input: O*NET}

For occupational input content, I rely on the O*NET database, a comprehensive resource describing occupations across the US economy through various descriptors (\cite{Peterson_al_2001}). This database has been widely used in the literature to measure the content of occupations (see, for instance, \cite{acemoglu_autor_2011, Blinder_2009, brynjolfsson_mitchell_rock_2018, felten_raj_seamans_2021, firpo_fortin_lemieux_2011, Peri_Sparber_2009}). This study uses O*NET 20.0, released in August 2015, ensuring that the AI development measured in this research does not influence occupational descriptors.

Following \textcite{felten_raj_seamans_2018, felten_raj_seamans_2021}, I measure occupational content using 52 abilities (e.g., oral comprehension, fluency of ideas, finger dexterity). Abilities represent fundamental capacities required to perform a wide range of tasks (\cite{fleishman_1984, carroll_1993}). O*NET 20.0 provides detailed descriptions of 954 occupations in terms of these abilities. Each occupation-ability pair is assigned an importance and level score, indicating the significance and required proficiency of the ability for the occupation.

Abilities are chosen over other descriptors to measure AI exposure because AI algorithms are described broadly, making them more analogous to general abilities rather than specific tasks or activities.\footnote{There is no guidance in the literature regarding the descriptors of the occupations that should be used to measure AI exposure. While \textcite{webb_2020} relies on the description of more than \num{18000} tasks performed within occupations, \textcite{felten_raj_seamans_2018, felten_raj_seamans_2021} use 52 abilities required to perform occupations. In contrast, \textcite{brynjolfsson_mitchell_rock_2018, eloundou_al_2024} take advantage of 2069 detailed work activities, which are merged to tasks performed in occupations.} By aligning AI capabilities with occupational abilities, I assume that firms mobilize abilities derived from AI algorithms as inputs to perform tasks traditionally carried out by human labor. This assumption is consistent with the O*NET documentation, which states that each occupation requires a specific combination of abilities to effectively perform its tasks (\cite{Peterson_al_2001}).

\subsubsection{Automation AI Exposure}

The automation AI exposure index is designed to capture the extent to which AI technologies can automate tasks associated with specific occupational abilities. To construct this index, I combine information on AI algorithms developed in the US economy, derived from posts on Stack Overflow, with occupational abilities data from O*NET. 

The construction of the index starts by linking AI-related questions from Stack Overflow to occupational abilities from O*NET. To quantify this relationship, I use Sentence-BERT embeddings to create two transition matrices (\cite{Reimers_Gurevych_2019}). The first matrix links AI-related tags to occupational abilities based on their names, while the second does so using their descriptions. Sentence-BERT embeddings offer several advantages over traditional methods, such as Bag-of-Words and Term Frequency-Inverse Document Frequency, as they capture semantic meaning and contextual information. Each transition matrix is populated with cosine similarity scores, which measure how semantically similar AI-related tags are to occupational abilities.\footnote{Cosine similarity scores range from -1 (opposite meaning) to 1 (similar meaning). Negative scores are replaced by 0, as AI is unlikely to be developed for conceptually opposite tasks.} For example, the AI-related tag "Vision" is semantically closed to the abilities "Near Vision" and "Far Vision" (similarity scores of 0.70 and 0.66, respectively). Finally, I retain only similarity scores that are in the top 25\% in both transition matrices and compute an averaged transition matrix, taking the mean similarity score between the two matrices (name-based and description-based).\footnote{Altering this threshold or removing the constraint does not qualitatively affect the results presented in this study (see Appendix~\ref{app: full matrices ss}).} This approach follows standard practice in the literature, where filtering out weakly connected pairs improves linkage quality (\cite{autor_chin_salomons_seegmiller_2024, prytkova_al_2024, hampole_al_2025}).

Once the averaged transition matrix is constructed, I calculate AI exposure at the ability level over time by applying a yearly decay factor to smooth the scores of Stack Overflow questions. A 50\% decay factor is applied to account for the relevance of newer technologies, reducing the influence of older questions. For example, a question posted in 2020 with a score of 10 in 2022 would have smoothed scores of 5.7, 2.9, and 1.4 in 2020, 2021, and 2022, respectively. After smoothing, the scores are aggregated at the tag level and combined with the transition matrix to compute annual AI exposure at the ability level.

Then, occupational automation AI exposure is determined by weighting the AI exposure ability scores using O*NET importance and level ratings, which reflect the varying requirements for abilities across occupations. The importance and level scores are fixed at 2015 levels to ensure that all variations in automation AI exposure stem from changes in the questions asked on Stack Overflow over time. This process produces a yearly measure of automation AI exposure at the 8-digit occupational level.

Finally, I retain observations for 2015-2022 and convert the index into a 6-digit constant Standard Occupational Classification (cSOC) by using crosswalks from O*NET and the US Bureau of Labor Statistics (BLS). Additionally, I standardize the AI exposure index using the Z-score to facilitate interpreting the results. A negative Z-score indicates that AI exposure is below the average, while a positive Z-score indicates above-average exposure. Ultimately, this methodology yields automation AI exposure scores for 765 occupations from 2015 to 2022.

In appendix~\ref{app: Abilities and occupational AI exposure}, Figure~\ref{fig: AI exposure abilities} presents the exposure to AI automation at the ability level for 2022. The findings indicate that sensory and cognitive abilities are more exposed to AI than physical and psychomotor abilities. Specifically, abilities related to vision, language understanding, and comparison of elements are among the most exposed to AI. This pattern reflects the domains in which AI has experienced rapid advancements during the last decade (\cite{stanford_ai_2024}). 

Table~\ref{tbl: highest lowest automation AI exposure occupations} documents the occupations most and least exposed to AI automation. The most exposed occupations predominantly consist of high-skilled occupations (e.g., "Real Estate Brokers", "Judges, Magistrate Judges, and Magistrates", and "Loan Officers"). By contrast, low-skilled occupations are overrepresented among those with the lowest exposure (e.g., "Dancers", "Dishwashers", and "Tire Builders"). This distinction reflects the differing ability requirements of these occupations: high-skilled occupations rely more heavily on cognitive and sensory abilities, which are more susceptible to automation by AI. In contrast, low-skilled occupations require a greater emphasis on physical and psychomotor abilities, which are less exposed to automation AI.

\subsubsection{Occupational and industry output: Census Alphabetical Index}

The measure of augmentation AI is based on the types of services and goods produced by workers. Following \textcite{autor_chin_salomons_seegmiller_2024}, I use the 2016 CAI as the primary data source for measuring workers' output. The CAI provides a comprehensive list of micro-industry and micro-occupational titles reported by respondents of Census Bureau demographic surveys. \textcite{autor_chin_salomons_seegmiller_2024} demonstrate that those micro-titles effectively capture the services and goods rendered in micro-occupations and micro-industries rather than the specific tasks required to perform these services and goods. They show that matching micro-titles to innovations provides a measure of technology that augments occupations and industries. Within this framework, the augmentation AI measure reflects how AI enhances the capabilities, quality, variety, or utility of outputs in occupations and industries.

For this study, I use the 2016 CAI, which includes \num{27105} unique micro-occupational titles and \num{22101} unique micro-industry titles. Utilizing micro-titles from 2016 ensures that the measure of augmentation AI exposure is not affected by contemporaneous development of AI technologies. Each micro-industry title is assigned a North American Industry Classification System code (NAICS), and the Census Bureau assigns each micro-occupational title a Standard Occupational Classification (SOC) code. For example, the micro-titles "Insurance Counselor", "Legal Adviser", and "Tax Attorney" all belong to the occupation "Lawyers" (SOC: 23-1011), illustrating the services provided by these roles rather than the tasks they perform. Similarly, in industry, the micro-titles "Air Compressor Service", "Heating Equipment, Sale and Installation", and "Plumbing Shop" are part of the industry "Plumbing, Heating, and Air-Conditioning Contractors" (NAICS: 238220).

\subsubsection{Augmentation AI Exposure}

The augmentation AI exposure index measures the extent to which AI complements workers' output. This index combines posts on Stack Overflow with micro-titles sourced from the CAI for industries and occupations.

To construct this index, I follow the same approach as the automation AI exposure index. I use the Sentence-BERT model to generate transition matrices linking AI-related tags from Stack Overflow with micro-titles for occupations and industries. Following a smoothing and aggregation process like the one used for automation AI, I calculate augmentation AI exposure separately for industries and occupations and then merge them at the occupation-industry level by taking the average. Finally, I retain observations for 2015-2022. The result is a time series dataset providing augmentation AI exposure from 2015 to 2022, classified by the 6-digit cSOC and the 4-digit constant NAICS (cNAICS).\footnote{I used crosswalks from BLS to create the cNAICS.} This dataset covers 759 occupations in 220 industries.

In appendix~\ref{app: Abilities and occupational AI exposure}, Table~\ref{tbl: highest lowest augmentation AI exposure occupations} presents the occupations most and least exposed to augmentation AI. Occupations related to computing are highly represented among those with the most significant exposure, such as "Computer and Information Research Scientists", "Computer Programmers", and "Computer Systems Analysts". In contrast, the occupations with the lowest exposure predominantly consist of general medical practitioners, including "Orthodontists", "Oral and Maxillofacial Surgeons", and "Dental Hygienists".

\subsubsection{Validation of AI measures}

After constructing the two AI exposure indices, I compare them with external measures of AI adoption and exposure for validation. In Appendix~\ref{app: Validation AI exposure indices}, Figure~\ref{fig: automation AI VS indus automation} presents the relationship between the automation AI exposure index developed in this study and the share of firms adopting AI for task automation by industry from the 2019 Annual Business Survey \citep{ncses_2019}. The analysis reveals a strong positive correlation between the two measures, supporting the validity of the automation AI exposure index.

In the absence of firm-reported measures of AI adoption for augmenting labor, I assess whether automation AI exposure and augmentation AI exposure capture distinct aspects of AI development. Figure~\ref{fig: Automation AI VS augmentation AI} presents the relationship between these two measures, revealing a strong dispersion. This finding is significant, as it suggests that automation AI exposure and augmentation AI exposure reflect different dimensions of AI development.\footnote{For comparison, \textcite{autor_chin_salomons_seegmiller_2024} present a similar analysis of exposure to automation and augmentation innovations at the occupational level from 1980 to 2018, where they observe a much stronger positive correlation.} Furthermore, in Section~\ref{sect: results}, I demonstrate that automation AI and augmentation AI exposure have distinct effects on labor market outcomes and that augmentation AI exposure aligns with theoretical expectations for augmenting AI measures \citep{acemoglu_Restrepo_2018, acemoglu_2024}.

Finally, Table~\ref{tbl: comparison AI indices} reports the associations between the exposure measures developed in this study and those from \textcite{webb_2020}, \textcite{brynjolfsson_mitchell_rock_2018}, and \textcite{felten_raj_seamans_2021}. The results demonstrate that the indices developed here closely align with those from previous studies, particularly with the index from \textcite{felten_raj_seamans_2021}. This is expected since our approaches are close.

\subsection{New work}

The measure of new work is constructed by comparing updates of the "Alternate titles" rubric in O*NET between 2015 and 2022.\footnote{See \textcite{kim_2022, kim_al_2024} for a similar approach for measuring new work for 1980-2010.} The "Alternate titles" rubric was introduced with 20.1 O*NET in 2015. Alternate titles are job titles developed to enhance keyword searches in O*NET (\cite{gregory_lewis_2015}). These job titles provide a more detailed description of specific positions within occupations.

In O*NET 27.1, the latest version for 2022, there are \num{52772} entries (\num{42889} unique job titles) covering \num{1016} occupations. For example, the occupation "Software Developer" includes 132 job titles, such as "Application Developer", "Artificial Intelligence Specialist", and "Computer Software Engineer". These job titles are distinct and provide more granular details than the broader occupation categories. On average, there are 52 job titles per occupation, with most occupations having between 10 and 100 job titles.

The alternate titles rubric is regularly updated, with two updates realized in 2015 and four updates per year after that. Five sources are utilized to identify new job titles: incumbents and occupational experts, employer job postings, submitted job titles in the occupational code assignment process, analysis of search term data from customers, and requests from representative groups such as associations and professional organizations. When a new job title is identified, occupational analysts undergo a multi-step review process before adding it to the alternate titles rubric. This review ensures that new job titles which were not yet in the database, are sufficiently familiar to be included and adhere to style and formatting guidelines.

A major challenge in measuring new work is distinguishing job titles that reflect task creation and specialization from those resulting from mere renaming or rewording. To address this, I extensively clean the alternate titles to standardize their names and ensure the measurement captures only new work attributable to specialization and task creation. Specifically, I convert plural to singular forms and standardize gendered words to male forms, as male words are more commonly used in O*NET.\footnote{In 2020, O*NET updated its occupational classification to align with the 2018 Standard Occupational Classification provided by the US Bureau of Labor Statistics. This update introduced new occupations and involved splitting and merging some existing ones. To address these changes, I use the crosswalk provided by O*NET.}

I identify new work by comparing the cleaned versions of job titles within occupations across two years. This process involves exact matching of job titles between years \textit{t} and \textit{t-1}, supplemented by fuzzy matching based on the Semantic Textual Similarity of the job titles.

Fuzzy matching is necessary because the wording of some job titles might change over time, even if the task content remains the same. For instance, in the occupation "Nuclear Technicians" (SOC 19-4051), the alternate titles "Nuclear Technician Worker" and "Nuclear Technician" appear in 2016, while "Nuclear Operating Technician" and "Nuclear Reactor Technician" were already present in 2015. The titles appearing in 2016 are broader and do not reflect task content or specialization changes. Similarly, in 2019, the title "Medical Transport Driver" was added to the occupation "Ambulance Drivers and Attendants, Except Emergency Medical Technicians" (SOC 53-3011), even though "Transport Medic", "Medical Driver", and "Driver Medic" were already present. Fuzzy matching based on semantic similarity ensures that these new titles are considered variations of existing ones rather than being incorrectly identified as new work.

Fuzzy matching relies on Semantic Textual Similarity. To achieve this, I use the sentence-BERT model to create sentence embeddings for each alternate title (\cite{Reimers_Gurevych_2019}). Then, I compute the cosine similarity measures for each pair of alternate titles within occupations. Two alternate titles are considered identical if their similarity measure is 0.7 or higher. This threshold was determined by examining the matching results.\footnote{The following thresholds were tested: 0.95, 0.90, 0.80, 0.70, 0.60, and 0.50. The choice of threshold does not affect the results concerning the effect of AI exposure on creating new work.} A threshold set too high would not sufficiently account for rewording, causing some alternate titles to erroneously appear as new work despite no changes in task content. Conversely, a low threshold would fail to capture genuine instances of new work.\footnote{\textcite{kim_2022, kim_al_2024} use a similar approach but rely on the Continuous Bag of Words model instead of sentence embeddings. They fix a threshold of 0.85 and 0.75, respectively.} For example, the alternate title "Medical Transport Driver" has a similarity score of 0.72 with "Transport Medic", 0.84 with "Medical Driver", and 0.79 with "Driver Medic". Therefore, "Medical Transport Driver" is not considered new work when added in 2019.

Between 2015 and 2022, \num{2159} instances of new work were added within occupations, averaging \num{308} new work additions per year. This figure is slightly below the estimates reported by \textcite{autor_chin_salomons_seegmiller_2024}. Using updates of CAI from 1940 to 2018, \textcite{autor_chin_salomons_seegmiller_2024} identified \num{28315} instances of new work during the period, corresponding to an average of \num{363} new work additions per year.

Table \ref{tbl: new alternate titles} provides examples of new alternate titles added to O*NET between 2015 and 2022. Some new work can be directly attributed to technological advancements, such as "Autonomous Vehicle Design Engineer" introduced in 2018 and "Remote Pilot" in 2021. Other new works reflect emerging services unrelated to technological changes, such as "Culinary Artist" added in 2020 and "Cat Groomer" in 2022.

\subsection{Other data sources}

Assessing the effect of AI development on US labor market outcomes requires merging the measures of AI exposure on other data sources. Data on wages and employment are sourced from the Occupational Employment and Wage Statistics (OEWS) database (\cite{bls_2023}), maintained by the US Bureau of Labor Statistics. This database, built on establishment-based data series, provides more accurate estimates for wages and employment at the occupational level compared to household surveys (\cite{acemoglu_autor_hazel_restrepo_2022}).

The OEWS provides annual employment and wage estimates for industry-occupation cells at the 6-digit and 4-digit levels, respectively.\footnote{While most of the observations are provided at 4-digit industry, some of them are aggregated at a higher level (2 and 3-digit).} Wage estimates represent straight-time gross pay, excluding premium pay. I rely on the mean hourly wage, adjusted to 2022 dollars.\footnote{Hourly wages are computed by dividing total wages by total worked hours.} The data includes part-time and full-time employees who are paid a wage or salary. Data from establishments in farm industries and those in the Public Administration sector are excluded from the sample.\footnote{Excluded industries correspond to the following NAICS codes: 111, 112, 1131, 1132, 114, 1153, 814, and 92.} Both the mean hourly wages and employment size are converted to logarithmic values.

I utilize data from the Longitudinal Employer-Household Dynamics (LEHD) Program provided by the US Census Bureau to measure the yearly sociodemographic composition of the workforce across industries. Specifically, I use the Quarterly Workforce Indicators for 2015-2022 to compute the distribution of workers by gender, age groups (14-34, 35-54, and 55-99 years old), race (White, Black, Asian, and Other), ethnicity (Hispanic or Latino versus not Hispanic or Latino), and educational attainment (High school or lower, some College or Associate degree, Bachelor’s degree or advanced degree, and not available).

The yearly values of US imports from 2015 to 2022 are obtained from the UN Comtrade database. I select data on US imports from the rest of the world and match these imports to industries using the concordance table provided by \textcite{liao_al_2020}. Then, I compute the level of imports per capita (in log).

These indicators are converted into cSOC (6-digit) for occupations and cNAICS (4-digit) for industries.

Finally, I merge all the datasets together at the occupation-industry level by year. The final dataset comprises \num{202695} observations from 2015 to 2022. These observations cover 702 distinct occupations across 220 industries. Table~\ref{tbl: Descriptive statistics} provides descriptive statistics.

\section{Descriptive statistics}

This section presents descriptive statistics on AI exposure and the emergence of new work to identify where these phenomena are occurring. First, I examine which occupations are most exposed to automation AI and augmentation AI, as well as where new work emerges. Second, I analyze whether AI exposure is associated with different task content.

\subsection{Which occupations are more exposed to AI, and where does new work emerge?}

\begin{figure}[H]
	\centering
	\caption{AI exposure by broad occupation in 2022}
	\label{fig: AI exposure by broad occupations}
	\includegraphics[scale=0.6]{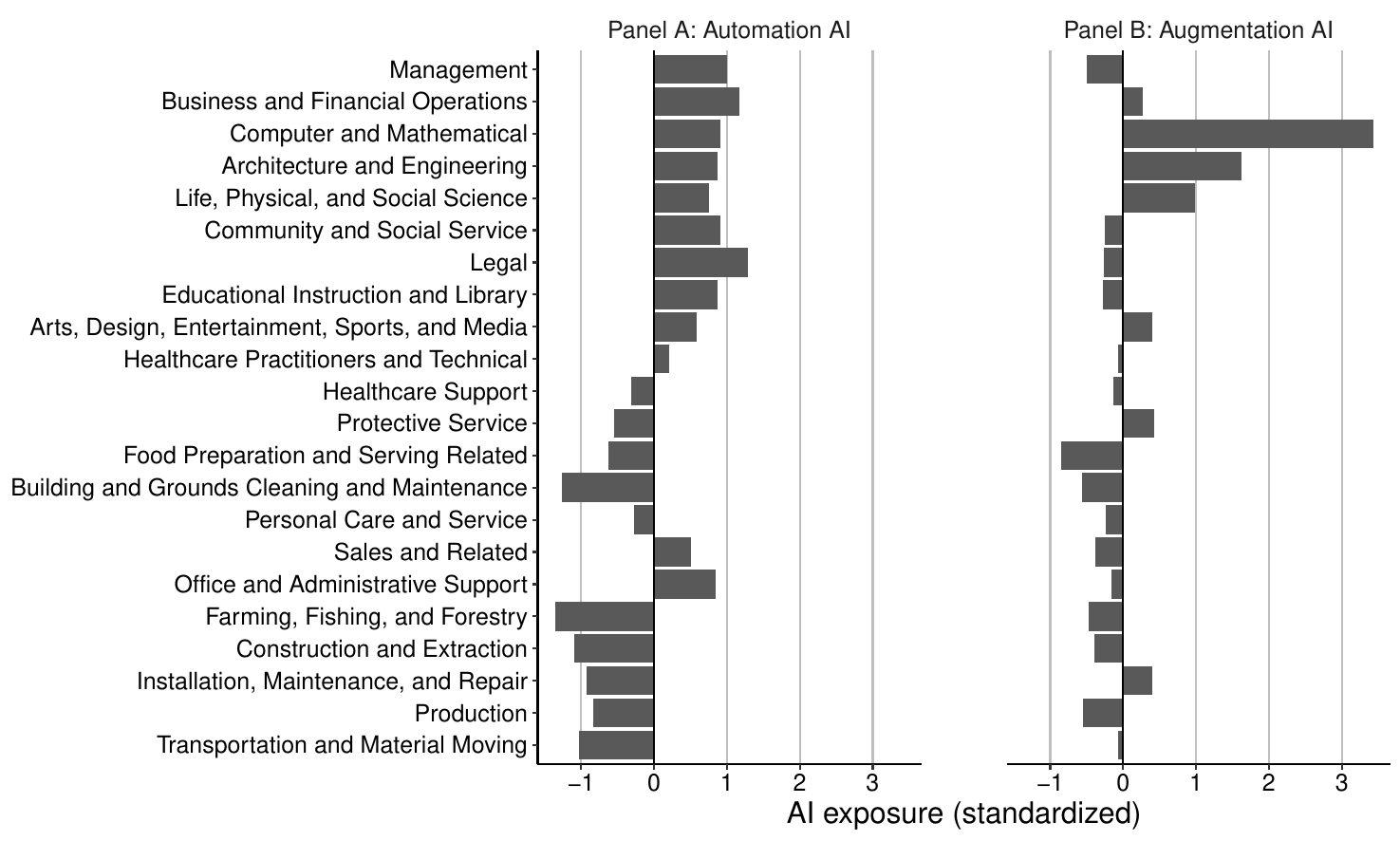}
	\caption*{\footnotesize{Note: The figure shows the exposure to automation AI (Panel A) and augmentation AI (Panel B) by broad occupation in 2022. Augmentation AI exposure is the weighted average exposure at the occupational level, using employment size in industries as weight. AI scores are the average of the exposure by broad occupation weighted by the employment size. The data includes part-time and full-time employees who are paid a wage or salary. Data from establishments in farm industries are excluded from the sample (NAICS: 111, 112, 1131, 1132, 114, 1153, and 814), as well as industries within sector 92 Public Administration.}}
\end{figure}

I start by exploring the heterogeneous exposure to automation AI and augmentation AI by broad occupation in Figure~\ref{fig: AI exposure by broad occupations}. Panel A presents the exposure to automation AI in 2022. Occupations in management, business, legal, and STEM exhibit the highest levels of AI automation exposure. In contrast, occupations that primarily involve physical and psychomotor activities, such as those in "Farming, Fishing, and Forestry", "Construction and Extraction", and "Transportation and Material Moving" show the lowest exposure. This pattern reflects differences in ability requirements, with management, business, legal, and STEM occupations relying more heavily on cognitive and sensory abilities, which are more susceptible to automation AI (see Appendix~\ref{app: Abilities and occupational AI exposure}).

Panel B focuses on augmentation AI exposure and shows that the development of complementing AI technologies is concentrated among STEM occupations. The figure reveals that "Computer and Mathematical" occupations are the most exposed, followed by "Architecture and Engineering" and "Life, Physical, and Social Science". In contrast, occupations in "Food Preparation and Serving Related", "Healthcare Support", and "Sales and Related" exhibit the lowest levels of exposure.

\begin{figure}[H]
	\centering
	\caption{Percentage of new work by broad occupation, 2015-2022}
	\label{fig:share new work by occupation}
	\includegraphics[scale=0.55]{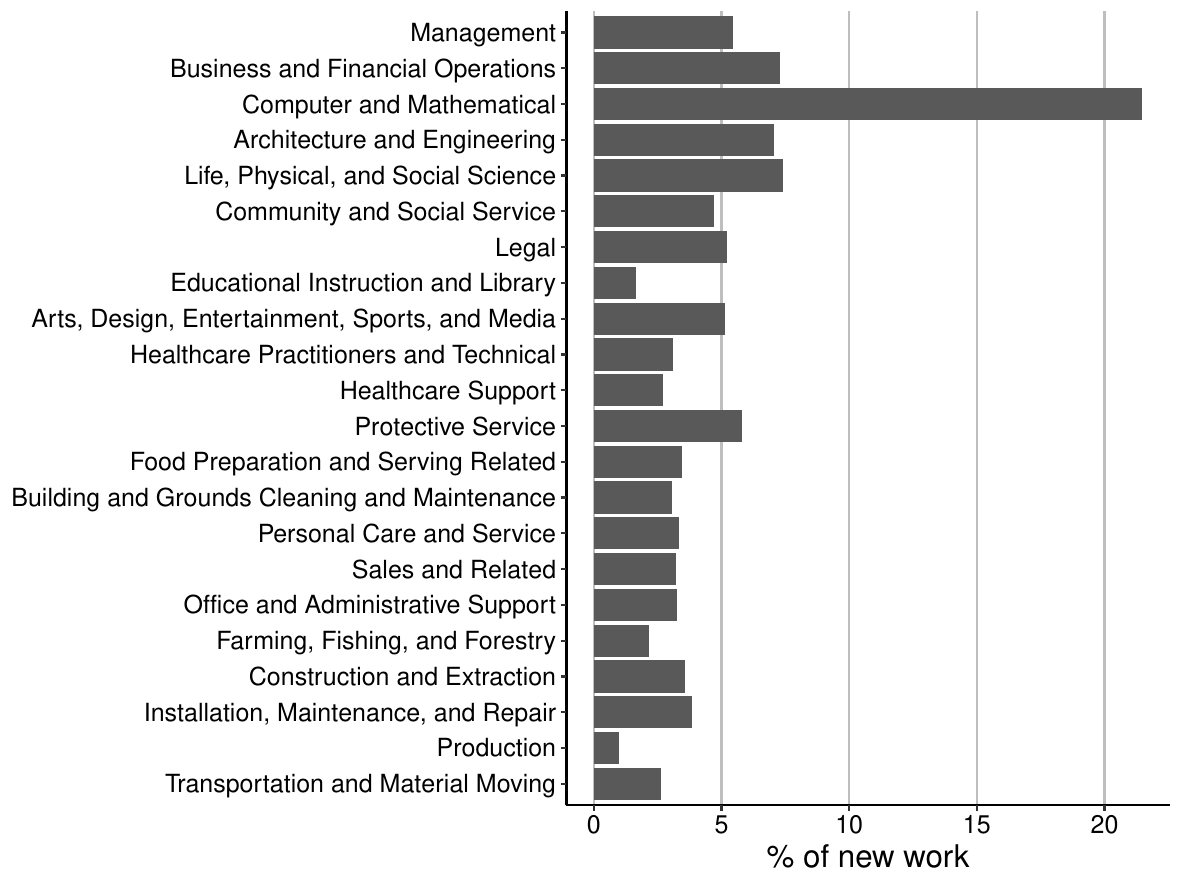}
	\caption*{\footnotesize{The figure displays the percentage of new work by broad occupation from 2015 to 2022. New work is identified by tracking the emergence of new job titles within occupations, based on the annual updates of the "Alternate rubric" in O*NET over this period.}}
\end{figure}

Figure \ref{fig:share new work by occupation} documents the percentage of new work by broad occupation for 2015-2022. It shows that new work emerges mainly in STEM, business, and managerial occupations. The occupation "Computer and Mathematical" has the highest percentage of new work: 24\% of the alternate titles in 2022 appeared during the last 7 years. Within this broad occupation, and not surprisingly, the percentage of new work for the occupations "Data Scientists" and "Computer and Information Research Scientists" reach 41\% and 21\%, respectively. This result reflects the recent progress in computer science and the creation of new tasks in this field. \textcite{deming_noray_2020} shows that STEM occupations have experienced the highest rates of skills change during the last decade, reflecting the rapid development of digital technologies related to software and data sciences.

This subsection highlights two key findings regarding AI exposure and the emergence of new work. First, high-skilled occupations tend to be more exposed to AI, consistent with findings from previous studies \citep{webb_2020, hampole_al_2025}. This contrasts with recent waves of technological change, which primarily affected middle-skilled workers \citep{autor_al_2003, Goos_Manning_Salomons_2009, Michaels_Ashwini_vanReenen_2014, autor_dorn_2013, Goos_Manning_2007, kogan_al_2023}. Second, augmentation AI exposure and the emergence of new work follow similar patterns: STEM occupations are more exposed to augmentation AI and experience higher levels of new work creation. \textcite{autor_chin_salomons_seegmiller_2024} further document a recent shift in new work creation toward high-skilled occupations.

\subsection{AI exposure and task content}

In this subsection, I examine whether AI exposure is associated with task content that have played a role in recent waves of technological change, such as robotics and computerization. Comparing AI exposure with these factors provides insights into how AI differs from previous technological advancements.

\begin{figure}[H]
	\centering
	\caption{AI exposure and task content}
	\label{fig: ai exposure and Routine tasks indices}
	\includegraphics[scale=0.55]{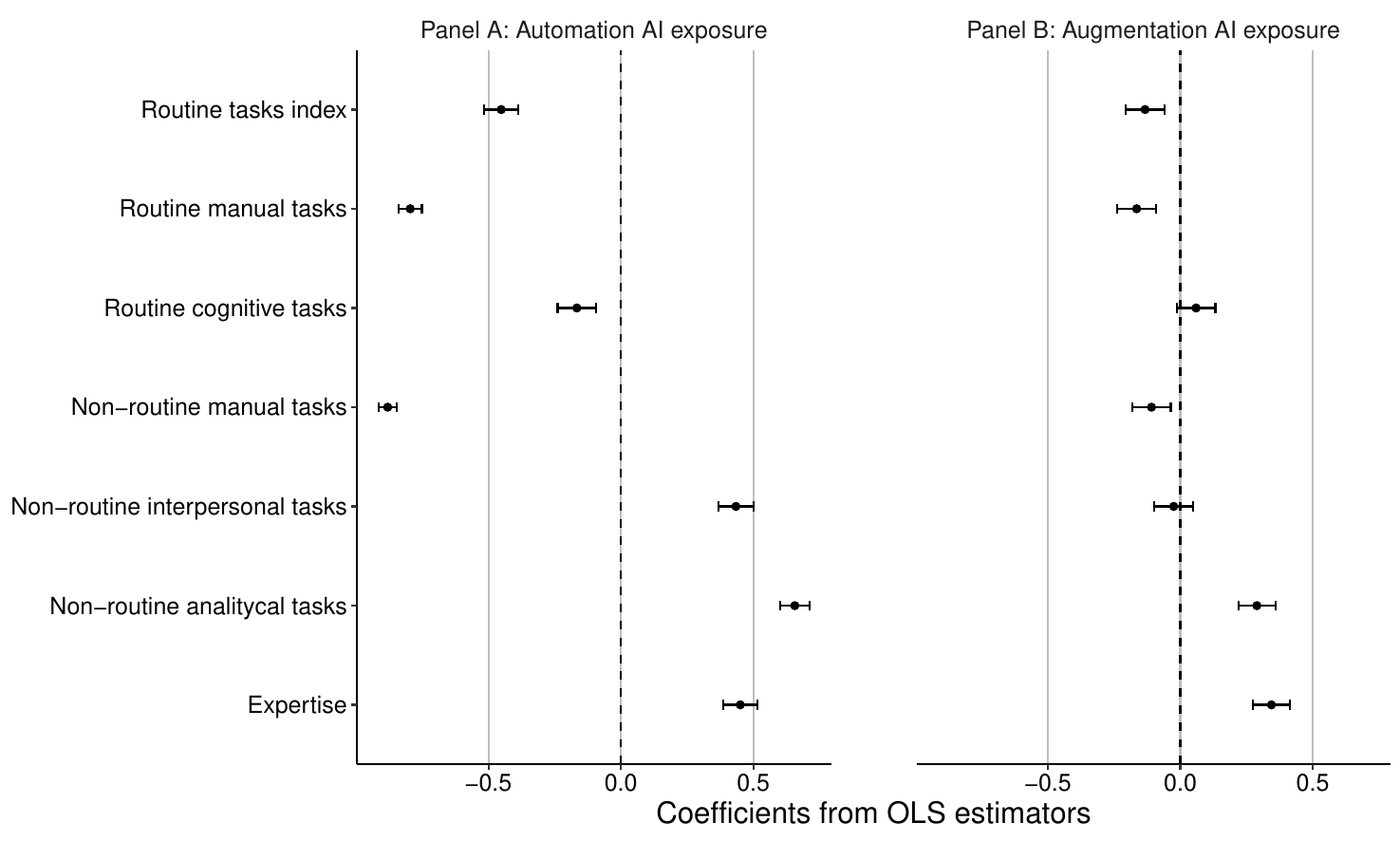}
	\caption*{\footnotesize{Note: The table presents coefficients from OLS estimations. Augmentation AI exposure is the weighted average exposure at the occupational level, using employment size in industries as weight. The routine and non-routine indices are constructed following \textcite{acemoglu_autor_2011}. The index of expertise is based on the Dale–Chall readability measure following \textcite{autor_thompson_2024}. Indices are converted into percentile ranks. The indices for automation AI exposure (Panel A) and augmentation AI exposure (Panel B) are for 2022. Panel A uses automation AI occupation for the dependent variable, while Panel B uses augmentation AI exposure. $\star\star\star$ Significant at the 1\% level; $\star\star$ significant at the 5\% level; $\star$ significant at the 10\% level.}}
\end{figure}

Figure~\ref{fig: ai exposure and Routine tasks indices} explores the relationships between the AI exposure measures developed in this study and indices measuring different types of task composition at the occupational level, all converted into percentile ranks. The coefficients are estimated with OLS regression.

Panel A presents the results with automation AI exposure as the dependent variable. The routine task index exhibits a negative relationship with automation AI exposure (point estimate = -0.44; SE = 0.03).\footnote{The task composition indices are constructed following \textcite{acemoglu_autor_2011}. Task expertise is measured using the Dale-Chall Readability Index, as described by \textcite{autor_thompson_2024}. A higher score indicates that occupations perform, on average, more expert-level tasks.} Decomposing the routine task index into subcategories reveals that routine tasks (both manual and cognitive) and non-routine manual tasks are negatively associated with automation AI exposure. In contrast, non-routine interpersonal tasks, non-routine analytical tasks, and the degree of task expertise exhibit a positive association.

These findings underscore the differences between AI and previous waves of technological change. Prior research has shown that computers and robots primarily automate routine tasks, as they are easier to codify \citep{autor_al_2003, acemoglu_autor_2011, autor_2015}. In contrast, non-routine tasks have historically been complemented by earlier waves of automation \citep{autor_2022}. However, consistent with the findings of this study, \textcite{gathmann_al_2025} provide evidence that AI reduces non-routine abstract tasks, signaling a shift from previous automation trends. Additionally, \textcite{autor_thompson_2024} demonstrates that automation processes are linked to task expertise, a pattern that appears to extend to AI-driven automation.

In Panel B, augmentation AI exposure is used as the dependent variable. Overall, the associations are smaller than for automation AI exposure and are sometimes not statistically significant, highlighting the differences between automation AI exposure and augmentation AI exposure. Augmentation AI exposure is negatively associated with the routine task index and routine manual tasks (point estimates: -0.14 and -0.10, respectively). Conversely, it positively correlates with non-routine analytical tasks (point estimate: 0.29). This result suggests that occupations relying on non-routine analytical tasks might benefit from AI, similar to the effects of robotization and computerization (\cite{autor_2022}). I find a positive association between the measure of augmentation AI exposure the degree of task expertise, which echoes evidence from \textcite{autor_thompson_2024}.

This subsection presents several key implications. The findings highlight that AI involves distinct tasks compared to robots and traditional computing technologies. Notably, AI exposure is strongly associated with non-routine analytical tasks, which were previously exclusive to workers.

\section{Labor market effect of AI exposure}

While the previous section has highlighted that AI exposure is heterogeneous across occupations and distinct from previous waves of technological change, the question of its effect on labor market outcomes remains crucial. This section addresses this question by studying the impact of AI exposure on three key labor market outcomes: the emergence of new work, employment levels, and wages. The analysis begins with a detailed description of the empirical strategy employed to assess these effects. Subsequently, I present the main findings, elucidating the relationship between AI exposure and these labor market outcomes.

\subsection{Empirical strategy}

I provide evidence of the effect of AI on new work by exploiting the yearly variations in AI exposure. This approach allows for full exploitation of the dataset's information. The following model is estimated using ordinary least squares (OLS) estimations:

\begin{equation}\label{eq: impact ai on new work}
	\begin{split} 
		\frac{\Sigma_{2015}^t NewWork_{ot}}{Work_{o2015}} = {} & \beta_1 autoAI_{ot} + \beta_2 augmAI_{oit} + \beta_3 log(Emp_{oit}) + \\ 
		& \beta_4 X_{it} + \alpha_{oi} + \gamma_{it} + {\varepsilon}_{oit}
	\end{split}
\end{equation}

In this equation, $o$ indexes occupation, $i$ industry, and $t$ year. The dependent variable is the cumulative share of new work within occupations. By considering the cumulative share, I assume that a job is tagged as new from the year it first appears in O*NET onward. This choice is justified by the fact that innovations contribute to the emergence of new work, and the adoption of these innovations by firms may take time (\cite{hall_khan_2003, bresnahan_yin_2017, goehring_al_2024, autor_chin_salomons_seegmiller_2024, brynjolfsson_mcAfee_2014}). Given the dataset's characteristics, this assumption means that a job is considered new for a maximum of six years. This duration is shorter than that used by \textcite{autor_chin_salomons_seegmiller_2024}, who define a job as new for up to ten years.

Automation AI exposure is measured by $autoAI_{ot}$, which varies at the occupation level (6-digit) by year. $augmAI_{oit}$ denotes the score for augmentation AI exposure at the occupation-industry level by year (6-digit for occupations and 4-digit for industries). Both AI indices are standardized to have a mean of zero and a standard deviation of one to facilitate the interpretation of the results.

I introduce a set of fixed effects to control for unobserved characteristics. The match between occupations (6-digit) and industries (4-digit) is controlled by $\alpha$. I capture temporary shocks at the industry level (3-digit) with $\gamma$.\footnote{While most of the observations are computed at 4-digit industry, a few of them are aggregated at 3-digit. For these aggregated observations, the fixed effects are introduced at 2-digit.} This fixed effect absorbs factors that have influenced labor market outcomes over recent decades, such as robotization (\cite{acemoglu_restrepo_2018_aer, Acemoglu_Restrepo_2020, Graetz_Michaels_2018, }). By incorporating fixed effects at the 3-digit industry-year level (n=41), I achieve more granular control over the impact of robotization compared to most studies analyzing robots using data from the International Federation of Robotics. This data typically measures robot adoption across approximately 20 broad industries.

I also control for the logarithm of employment size ($log(Emp)$). $X_{it}$ includes controls for trade per capita and several demographic characteristics of industries, which encompass age, gender, education, ethnicity, and race. The idiosyncratic error term is $\varepsilon$. The estimations are weighted by 2015 employment size, and standard errors are clustered at the occupation-industry level.

The coefficients of interest are $\beta_1$ and $\beta_2$. It is hypothesized that new work will emerge in occupations where AI augments output by affecting the quality and the variety of the goods and services (\cite{autor_chin_salomons_seegmiller_2024, acemoglu_kong_restrepo_2024, acemoglu_restrepo_2018_aer}). Conversely, automation AI exposure is not expected to positively influence new work creation.

Then, I analyze whether AI exposure affects employment and wages. To do so, I estimate the following OLS regression:

\begin{equation}\label{eq: impact ai on wages and employment}
	\begin{split} 
		log(y_{oit}) = {} & \beta_1 autoAI_{ot} + \beta_2 augmAI_{oit}  + \beta_3 log(A_{oit}) + \\ 
		& \beta_4 X_{it} +\alpha_{oi} + \gamma_{it} + {\varepsilon}_{oit}
	\end{split}
\end{equation}

In this context, $y$ represents either employment size or mean hourly wage. The coefficients of interest are $\beta_1$ and $\beta_2$. The sign of $\beta_1$ is \textit{a priori} unknown since it depends on which effect is the strongest between the displacement effect and the productivity effect (\cite{acemoglu_restrepo_2018_aer, acemoglu_Restrepo_2018}). Theoretically, the displacement effect tends to reduce labor demand and wages due to capital's comparative advantage, whereas the productivity effect increases the labor demand and wages. $\beta_2$ is expected to influence wages and employment positively. Technological changes that complement output have been shown to increase labor demand by creating new tasks and work where labor holds a comparative advantage (\cite{acemoglu_restrepo_2019, autor_chin_salomons_seegmiller_2024}).

Depending on whether I am studying the effect of AI exposure on employment or wages, $log(A_{oit})$ controls for the logarithm of wages or employment size, respectively.

Identifying the causal effect of AI exposures on labor market outcomes relies on the assumption that automation AI and augmentation AI exposure measures are exogenous to the error term. However, this assumption could be questionable for three main reasons. First, the estimates might be affected by reverse causality: higher wages could attract more AI investments rather than AI exposure affecting wages. Second, a shock could simultaneously influence both AI exposure and the outcome variables, affecting the quality of the estimates. Third, workers might anticipate the development of AI algorithms and adjust their occupational choices accordingly.

To mitigate potential identification threats, I employ an instrumental variables (IV) strategy to estimate the effect of AI exposures on labor market outcomes. I use five-year lagged AI exposures from countries with limited economic ties to the US as instrumental variables.

The AI exposure measures for the IV countries group (i.e., countries with no significant economic ties to the US) follow the same methodology used to construct the US indices, with two key differences. First, I use the 2010 CAI and O*NET 15.0 (released in July 2010) to measure workers' output and occupational inputs, respectively. This choice ensures that the descriptors remain unaffected by contemporaneous AI development. Second, instead of including members of Stack Overflow residing in the US and its primary trading partners, I exclude them and retain only members from countries with limited economic relationships with the US. I define these countries by analyzing the share of imports from the US as a percentage of their gross domestic product (GDP). Countries where US imports account for less than 5\% of their GDP are included in the IV countries group. Additionally, US primary trading partners are excluded from this group.\footnote{This approach yields 134 countries in the IV countries group. Appendix~\ref{app: Instrumental variable} shows the top 15 countries.} The most influential countries of the IV group are mainly high-income (e.g., Australia, Spain, Russian Federation, Poland, and Israel) and upper-middle income countries (e.g., Brazil, Pakistan, and Turkiye) (see Appendix~\ref{app: Instrumental variable} for the top 15 countries).

The identification strategy relies on three assumptions. First, countries in the IV group adopt AI technologies when they become available, similar to the US. In Appendix~\ref{app: Instrumental variable}, Figure~\ref{fig: Question on SO for US and IV} illustrates the yearly AI-related questions posted on Stack Overflow by country of residence from 2010 to 2022. A key finding from this figure is that a similar trend is observed across both groups of countries. While countries in the IV group tend to post fewer AI-related questions compared to the US and its primary trading partners, both trends show a sharp increase from 2010 to 2018. Since 2018, the number of yearly AI-related questions posted on Stack Overflow has declined in both groups.

Second, labor should not anticipate AI development five years in advance. Anticipation depends on the capacity to forecast accurately the improvement and usage of AI algorithms. However, forecasting AI is challenging due to rapid technological advancements and unpredictable breakthroughs. Studies show that predictions often display significant variance and have historically failed to accurately project AI milestones (\cite{armstrong_al_2014}). For example, unforeseen developments like the resurgence of deep learning have unexpectedly shifted the AI landscape around 2012 (\cite{lecun_al_2015}). Experts once predicted that AI defeating top human Go players was at least a decade away due to the game's complexity; yet, contrary to these forecasts, Google's AlphaGo defeated world champion Lee Sedol in 2016 (\cite{silver_al_2016}). Similarly, sudden advances in natural language processing, such as the introduction of transformer architectures like BERT, have caught many by surprise (\cite{devlin_al_2019}). 

Third, past AI development in the IV countries group should be a strong predictor of current AI development in the United States. Appendix~\ref{app: IV first-stage} presents the first-stage estimates, with Table~\ref{tbl: first_stage new work wages} reporting results for new work and wages and Table~\ref{tbl: first_stage employment} for employment. Both tables demonstrate a strong positive association between the automation AI exposure measure and its corresponding instrument. Similarly, the instrument for augmentation AI exhibits high predictive power. Furthermore, introducing both instruments simultaneously does not alter their respective relationships with AI exposure measures, reinforcing the validity of the approach. Finally, the F-statistic is large, confirming that AI exposure in countries with limited economic ties to the US serves as a robust instrument.

Finally, for the instrumental variables to be valid, they must satisfy the exclusion restriction. This condition requires that AI exposure in the IV countries group influences US labor market outcomes only through its relationship with US AI exposure. The strategy used to construct the instruments inherently satisfies this restriction for three key reasons. First, the IV countries group excludes the United States' largest trading partners, measured by US imports of goods and services.\footnote{On average, a country in the IV group accounts for 0.1\% of total US goods imports and 0.3\% of total US services imports.} This ensures that countries in the IV group do not significantly impact the US economy. Second, US exports account for less than 5\% of the GDP of the IV countries group. This condition ensures that these countries are not substantially influenced by the US economy. Finally, the instrumental variables use five-year lagged AI exposure, preventing them from being affected by global shocks that could also influence US AI exposure measures. These factors support the validity of the exclusion restriction and reinforce the robustness of the instrumental variable strategy.

\subsection{Results}\label{sect: results}

In this subsection, I present the main results concerning the effects of automation AI and augmentation AI exposure. First, I study whether AI exposure affects the emergence of new work. Then, I investigate whether the impact of AI exposure on new work translates into changes in employment size. Finally, I assess whether exposure to AI influences hourly wages.

\subsubsection{New work}

Does augmentation AI exposure spur the emergence of new work? Table~\ref{tbl: ai impact on new work} addresses this question by presenting estimates from equation~(\ref{eq: impact ai on new work}), which investigates the effect of AI exposure on the creation of new work. The dependent variable is the cumulative share of new work within occupations from 2015 to 2022.

Panel A presents the results using OLS estimators. In Column 1, I include automation AI exposure as the sole variable of interest, along with a set of controls (employment size, trade per capita, and demographic characteristics) and fixed effects for occupation-industry pairs. The coefficient on automation AI exposure is small and not statistically significant, as expected. Adding fixed effects to capture temporary shocks at the industry level does not qualitatively affect the coefficient (Column 2). In Column 3, I study the association with augmentation AI exposure. As expected, augmentation AI exposure is positively associated with the cumulative share of new work, with a point estimate of 0.011 (SE = 0.002). By enhancing workers' output, augmentation AI exposure requires more specialization and creates new work \citep{autor_chin_salomons_seegmiller_2024, acemoglu_kong_restrepo_2024, acemoglu_restrepo_2018_aer}. Further controlling for time-invariant occupation-industry characteristics strengthens this positive relationship, increasing the point estimate to 0.015 (SE = 0.002). In Columns 5 and 6, I include both automation AI and augmentation AI exposures simultaneously. The coefficients remain consistent with those obtained when each variable is included separately, reinforcing the robustness of the findings.

The IV approach further reinforces these conclusions (Panel B). The estimated coefficients for augmentation AI exposure increase slightly, ranging from 0.017 to 0.022, and remain statistically significant at the 1\% level (Columns 3–6). In contrast, automation AI exposure is not statistically significant, suggesting no effect on the emergence of new work. The F-Statistic yields a high value, confirming that the 2SLS estimations are not subject to weak instrument concerns. First-Stage estimations are reported in Table~\ref{tbl: first_stage new work wages}.

These findings support the expected positive effect of augmentation AI on the emergence of new work \citep{autor_chin_salomons_seegmiller_2024, acemoglu_kong_restrepo_2024, acemoglu_Restrepo_2018}. The development of AI technologies that complement outputs enhances the quality, variety, and utility of goods and services, thereby increasing the need for specialization and the performance of new tasks. The absence of an effect from automation AI exposure further validates the reliability of the measures developed in this study. Finally, these results align with the findings of \textcite{autor_chin_salomons_seegmiller_2024}, who focus on innovations rather than AI. Their research shows that new occupational tasks emerge in response to augmenting innovations, while automation innovation does not significantly contribute to the creation of new work.

\subsubsection{Employment}

Having demonstrated that augmentation AI exposure stimulates the creation of new work, I now aim to investigate whether this emergence translates into changes in employment levels. I also study whether automation AI generates displacement effect.

Table~\ref{tbl: ai impact on employment} reports the estimates of equation~(\ref{eq: impact ai on wages and employment}), using the logarithm of employment size as the dependent variable. Panel A presents results obtained using OLS estimators. When automation AI exposure is the sole explanatory variable of interest included (Columns 1 and 2), it shows no significant relationship with employment size, a result that remains robust across different specifications. In Columns 3 and 4, I test the relationship with augmentation AI exposure. The coefficient is statistically positive and increases with the inclusion of time-varying industry fixed effects (point estimate = 0.040; SE = 0.016; Column 4). Finally, including both AI exposure measures simultaneously does not alter the results.

Panel B presents the results using 2SLS estimators, which confirms the previous findings, though the magnitude of the coefficients reduces. The coefficients for automation AI exposure decrease and remain statistically insignificant (Columns 1–2 and 5–6). The positive effect of augmentation AI exposure on employment is also confirmed when the full set of fixed effects is introduced, though its coefficient decreases and is significant only at the 10\% level. In the most refined specification (Column 6), a one standard deviation increase in augmentation AI exposure leads to a 3.1\% increase in employment size. The estimations present a high F-Statistic, ensuring the validity of the instrumental variables. First-stage estimations are reported in Table~\ref{tbl: first_stage employment}.

The evidence suggests that the emergence of new work leads to increased employment. AI technologies that complement output create labor-intensive tasks, prompting firms to hire workers to perform this new work, resulting in higher employment levels.

These findings align with the results of \textcite{babina_al_2024}, which show that firms investing in AI experience higher employment growth. They also echo \textcite{acemoglu_autor_hazel_restrepo_2022}, who find no discernible relationship between AI exposure and employment, using the AI exposure measure from \textcite{felten_raj_seamans_2021}, which closely resembles the automation AI measure used in this study. In contrast, \textcite{bonfiglioli_al_2025} estimate robust negative effects of AI exposure on employment across commuting zones. However, their study does not distinguish between AI that augments output and AI that automates tasks. Additionally, it does not account for industry time-varying shocks, which appear to play a critical role in shaping employment outcomes.

Not specifically focusing on AI, but still related to this study, \textcite{autor_chin_salomons_seegmiller_2024} and \textcite{kogan_al_2023} find a positive and statistically significant effect of labor-augmenting technologies and innovations on employment. However, unlike this study, \textcite{autor_chin_salomons_seegmiller_2024} report a negative effect of automation innovation on employment. This discrepancy may stem from their long-term perspective, as they analyze a century of innovation, allowing more time for innovations to mature, diffuse throughout the economy, and affect employment trends.

\subsubsection{Wages}

Does AI development impact wages? To investigate this question, I estimate equation~(\ref{eq: impact ai on wages and employment}), using mean hourly wages as the dependent variable. Results are presented in Table~\ref{tbl: ai impact on wages}.

Panel A presents the estimates using OLS estimators. Automation AI exposure exhibits a negative and statistically significant relationship with hourly wages, a result that remains robust across different specifications (Columns 1 and 2). In Columns 3 and 4, automation AI exposure is replaced by augmentation AI exposure. The coefficient is negative but statistically insignificant. Including both automation AI and augmentation AI exposures simultaneously does not alter the results for automation AI and augmentation AI exposure (Column 6).

The IV approach reinforces the findings for automation AI exposure and augmentation AI exposure. The negative effect of automation AI exposure on hourly wages remains statistically significant at the 1\% level across all specifications (Columns 1–2 and 5–6). A one standard deviation increase in automation AI exposure reduces by 7.7\% the hourly wages (Column 6). The effect of augmentation AI exposure is negative when I control only for the time-invariant characteristics of occupation-industry (Column 3). However, it becomes statistically insignificant after including industry-year fixed effects (Column 4). However, unlike in the OLS estimations, this result remains robust when both automation AI and augmentation AI exposures are included simultaneously (Column 6). The F-Statistic is high, suggesting the absence of weak instruments. First-stage estimations are shown in Table~\ref{tbl: first_stage new work wages}.

These results echo the findings in \textcite{hui_al_2024}, which find that freelancers offering tasks in an online platform and competing with AI experienced a decrease in their earnings after the release of the large language model ChatGPT. \textcite{acemoglu_autor_hazel_restrepo_2022} also find a negative association with AI exposure when they use the measure of AI exposure from \textcite{felten_raj_seamans_2021}, which is close to the measure of automation AI exposure developed in this study. Finally, \textcite{bonfiglioli_al_2025} provide evidence of a negative effect of AI exposure on wages.

In summary, this section confirms some predictions of the task-based framework (\cite{autor_al_2003, acemoglu_autor_2011, acemoglu_restrepo_2018_aer, autor_chin_salomons_seegmiller_2024, acemoglu_Restrepo_2018, acemoglu_2024}). First, when AI complements workers' output, it creates new work and tasks. Second, by creating new work and tasks for which labor has a comparative advantage, augmentation AI exposure increases the demand for labor. However, the emergence of new work and increased labor demand do not result in a significant wage increase. This absence of effect on wages might come from the type of occupations that are complemented and the availability of labor supply. If augmenting AI creates new work and tasks in occupations for which there is a labor surplus, new work can be performed without creating a shortage in the labor market and leaving wages unaffected. This question is explored in the next section.

Automation AI exposure reduces hourly wages but does not affect employment size. Three elements could explain this result. First, the displacement and productivity effects might be of similar magnitude, resulting in a net neutral effect. Second, AI is adopted to perform specific tasks, reducing the marginal return to labor without entirely displacing workers. The current capabilities of AI algorithms may not yet be advanced enough to entirely replace the content of occupations, which explains the negligible impact on employment size. Finally, labor market rigidities might limit the displacement effect.

\section{Effects of AI by skill group}
 
Building on the task-based framework, recent studies suggest that the impact of technological change may depend on occupational skill requirements \citep{bloom_al_2024, acemoglu_restrepo_2018_jhc, autor_2024, acemoglu_2024}. Automation AI technologies are hypothesized to be more applicable to low-skilled occupations, whereas augmentation AI technologies may be better suited for high-skilled occupations. In this section, I test this hypothesis and provide evidence on the heterogeneous effects of AI exposure based on occupational skill requirements.

Table~\ref{tbl: ai impact education} presents the effects of augmentation AI and automation AI exposure on three labor market outcomes—the share of new work, employment size, and hourly wages—differentiated by occupational skill requirements.\footnote{Skill groups are derived from the Job Zones provided by O*NET. Low-skilled occupations correspond to those requiring "Little or No Preparation Needed" or "Some Preparation Needed", while middle-skilled occupations fall under "Medium Preparation Needed". High-skilled occupations require "Considerable Preparation Needed" or "Extensive Preparation Needed".} The results are estimated using 2SLS estimators, leveraging AI exposure in countries with no significant economic ties to the US as an instrumental variable. OLS estimates are provided in Appendix~\ref{app: ai impact education}.

Panel A examines low-skilled occupations and shows that automation AI exposure negatively impacts labor market outcomes. Automation AI exposure has a detrimental effect on the share of new work (Column 1), employment (Column 2), and wages (Column 3), suggesting that the displacement effect is stronger than the productivity effect. In contrast, augmentation AI exposure yields imprecise coefficients, suggesting no significant effect for low-skilled occupations.

Panel B focuses on middle-skilled occupations and reveals mixed effects of automation AI and augmentation AI exposure. Automation AI exposure negatively impacts the share of new work (point estimate = -0.014; SE = 0.007, Column 3) and hourly wages (point estimate = -0.034; SE = 0.008, Column 3), but it has no significant effect on employment (Column 2). This suggests that automation AI technologies compete with labor but do not fully displace it. Augmentation AI exposure positively influences both the emergence of new work (Column 1) and hourly wages (Column 3), while its effect on employment remains statistically insignificant (Column 2). These results may be explained by the inelasticity of labor supply in middle-skilled occupations, where demand for skilled workers increases wages without necessarily expanding employment.

Panel C investigates high-skilled occupations mainly affected by augmentation AI exposure. Unexpectedly, automation AI exposure has a positive coefficient for the share of new work (Column 1), though it is only significant at the 10\% level. However, it does not significantly affect employment size (Column 2) or hourly wages (Column 3). The findings for augmentation AI exposure confirm that AI technologies complementing workers' output foster the emergence of new work (point estimate = 0.024; SE = 0.003, Column 1). However, this increase in new work does not translate into employment growth (Column 2), likely due to inelastic labor supply in high-skilled occupations. In contrast, augmentation AI exposure positively affects hourly wages, with a point estimate of 0.007 (SE = 0.002) (Column 3).

These results echo the findings in \textcite{gathmann_al_2025}. Using a patent-based measure of AI, they find evidence that low-skilled workers suffer some wage losses, while high-skilled incumbent workers experience wage gains.

These findings highlight three key economic implications. First, they confirm the expectation that automation AI technologies are more applicable to low-skilled occupations, while augmentation AI technologies are better suited for high-skilled occupations \citep{bloom_al_2024, acemoglu_restrepo_2018_jhc, autor_2024, acemoglu_2024}. Middle-skilled occupations fall into an intermediate category, experiencing mixed effects. Second, among middle- and high-skilled occupations, augmentation AI exposure fosters the emergence of new work, leading to higher wages but not an increase in employment size. This discrepancy may be explained by inelastic labor supply in these occupations. Due to the scarcity of workers with the necessary skills, firms compete by raising wages to attract talent rather than expanding employment. Third, these results suggest that AI may contribute to rising wage inequality.

\section{Robustness checks}

One might be concerned that the findings documented in this study are sensitive to different methodological choices in constructing the indices or in the empirical approach. To assess the robustness of these results, I alter key methodological decisions and perform several robustness checks.

First, one might be concerned about the selection of countries included in the IV group, particularly that including countries with lower levels of development than the United States could bias the results. To address this concern, I re-estimated the analysis using only countries classified as high-income by the World Bank in 2010 (n = 43) within the IV group. The findings, presented in Appendix~\ref{app: IV_high_income_countries}, are qualitatively consistent with those obtained using the full IV group, suggesting that the results are robust to excluding lower-income countries.

Second, I demonstrate that using all similarity scores in the transition matrices instead of keeping only the top 25\% does not qualitatively affect the results. In Appendix~\ref{app: full matrices ss}, I provide results where all similarity scores are included in the transition matrices. The magnitude of the coefficients increases, especially for mean hourly wages; however, the overall quality of the estimates remains similar.

Third, one might be concerned that using ChatGPT to provide descriptions of AI-related tags could affect the results. To assess this, I recompute the AI exposure indices using the original technical descriptions provided by Stack Overflow at the time each tag was created. This adjustment reduces the number of tags by 21\%. I then rerun the analysis and present the findings in Appendix~\ref{app: first_excerpt}. The quality of the estimates remains consistent for new work and wages, although the coefficients tend to increase slightly. Regarding employment size, the effect of augmentation AI exposure becomes statistically insignificant. In contrast, automation AI exposure shows a positive effect significant at the 10\% level (Table~\ref{tbl: app first excerpt employment}, Panel B, Column 6).

Finally, I rerun the analysis using current employment size to weight the regressions, rather than fixing it at its 2015 value (Appendix~\ref{app: current weights}). This alternative weighting scheme does not qualitatively alter the results, though the coefficients tend to be higher. Augmentation AI exposure continues to spur the emergence of new work and increase employment size. Automation AI exposure maintains its negative effect on mean hourly wages.

This section presents a series of robustness checks, demonstrating that the findings remain consistent across different methodological choices. Modifying the countries included in the IV, the construction of transition matrices, the tags' descriptions, or the weighting scheme has only marginal effects on the results.

\section{Conclusion}

In conclusion, this study provides robust evidence on the nuanced effects of AI exposure on the labor market, highlighting the importance of distinguishing between automation and augmentation AI. Utilizing novel measures derived from Stack Overflow data, the analysis reveals that augmentation AI fosters the emergence of new work and increases overall employment size. Conversely, automation AI exposure negatively affects wages.

The heterogeneity analysis shows that the impact of AI exposure varies according to occupational skill requirements. Specifically, low-skilled occupations are negatively affected by AI technologies that automate tasks. These technologies reduce the emergence of new work, employment, and hourly wages. Conversely, AI technologies that augment workers' output increase the emergence of new work and hourly wages for high-skilled occupations. Middle-skilled occupations are in an intermediary, where they are negatively affected by automation AI but benefit from augmentation AI.

These results have several policy implications. Policymakers must consider tailored strategies to mitigate the adverse effects of AI automation, particularly for low—and middle-skilled workers, while promoting the beneficial aspects of augmentation AI that can spur job creation and economic growth. Additionally, this research highlights the need to continuously monitor AI advancements and their diverse impacts across occupations, ensuring that labor market policies remain adaptive and responsive to technological progress.

This paper contributes significantly to the existing literature by introducing innovative methods to measure AI exposure and disentangling the effects of different types of AI on new work, employment, and wages. Future research should build on these findings to explore the long-term impacts of AI on various economic sectors and to develop comprehensive policy frameworks that address the dynamic nature of technological change. The ongoing evolution of AI technologies necessitates a proactive approach to understanding and managing their implications for the workforce, ensuring that the benefits of AI are broadly shared and potential disruptions are effectively mitigated.

\clearpage

\printbibliography

\clearpage

\begin{figure}[H]
    \centering
    \caption{Measuring automation AI and augmentation AI}
    \label{fig: Schema AI indices}
    \includegraphics[trim={0 10cm 0 0},scale=0.6]{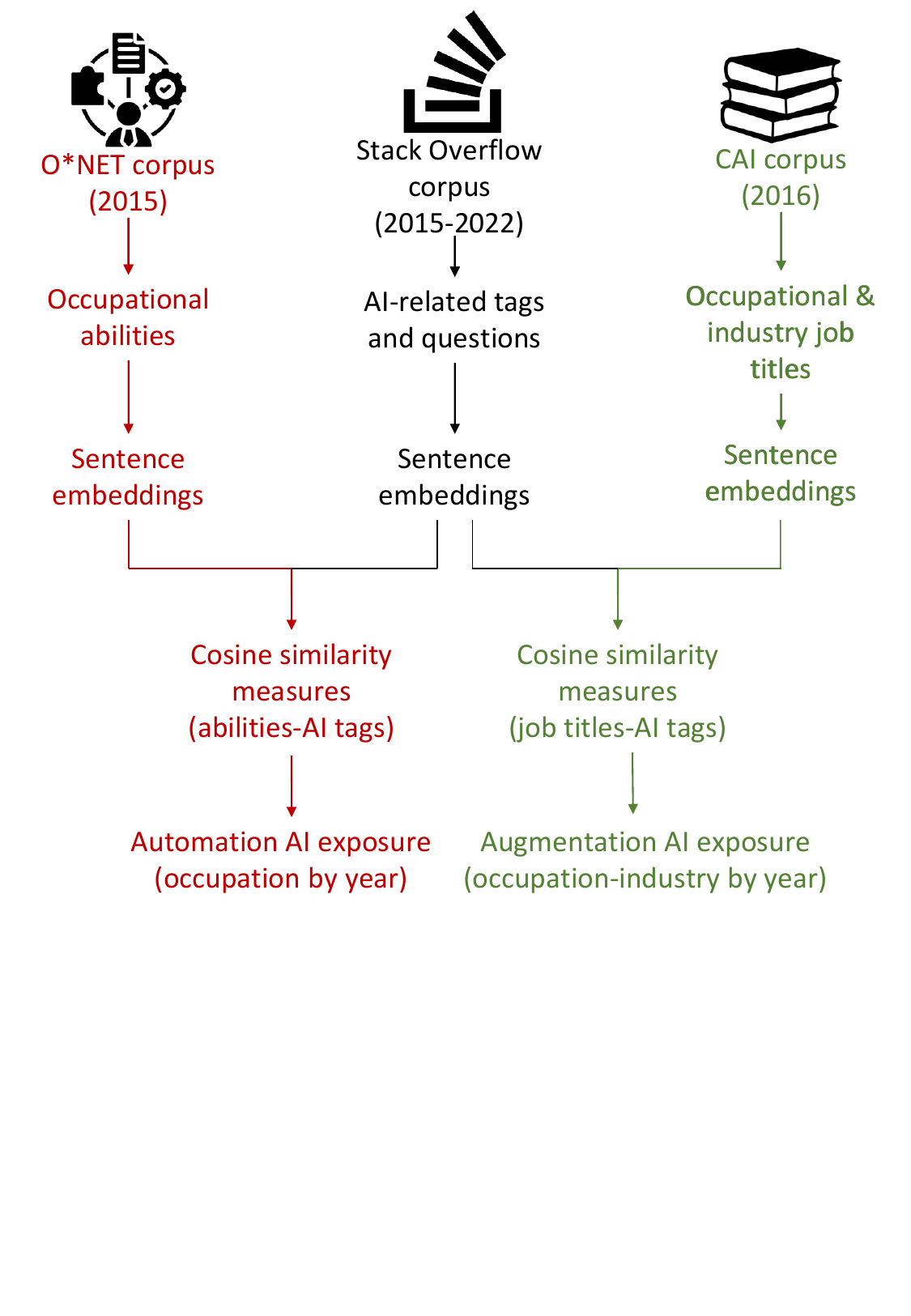}
\end{figure}

\begin{table}[H]
\caption{Example of new alternate titles per year, 2016-2022}
\label{tbl: new alternate titles}
\resizebox{\textwidth}{!}{
\centering
\begin{tabular}{lcc}
Year & Alternate titles & Occupations \\
\toprule
\multirow{2}{*}{2016} & Family Reunification Specialist & Social and Human Service Assistants (21-1093) \\
                      & Scrum Master & Computer Occupations, All Other (15-1299) \\
                      \cline{2-3}
\multirow{2}{*}{2017} & Executive Cyber Leader & Chief Executives (11-1011) \\
                      & Online Health and Fitness Coach & Health Education Specialists (21-1091) \\
                      \cline{2-3}
\multirow{2}{*}{2018} & Sprinkler Design Technician & Civil Engineering Technologists and Technicians (17-3022) \\
                      & Autonomous Vehicle Design Engineer & Engineers, All Other (17-2199) \\
                      \cline{2-3}
\multirow{2}{*}{2019} & Safety Research Professional & Occupational Health and Safety Technicians (19-5012) \\
                      & Route Diver & Commercial Divers (49-9092) \\
                      \cline{2-3}
\multirow{2}{*}{2020} & Blockchain Penetration Tester & Computer Occupations, All Other (15-1299) \\
                      & Culinary Artist & Cooks, Private Household (35-2013) \\
                      \cline{2-3}
\multirow{2}{*}{2021} & Solar Site Surveyor & Surveyors (17-1022) \\
                      & Remote Pilot & Life, Physical, and Social Science Technicians, All Other (19-4099) \\
                      \cline{2-3}
\multirow{2}{*}{2022} & Recruiter Sourcing & Human Resources Specialists (13-1071) \\
                      & Cat Groomer & Animal Trainers (39-2011) \\    
\bottomrule
\end{tabular}
}
\caption*{\footnotesize{Note: Examples of new alternate titles added in O*NET by year from 2016 to 2022. New work is identified by tracking the emergence of new job titles within occupations, based on the annual updates of the "Alternate rubric" in O*NET over this period. Occupations correspond to the Standard Occupational Classification 2010, and codes are given in parentheses.}}
\end{table}

\begin{table}[H]
	\centering
	\caption{Effect of AI exposure on the cumulative share of new work}
	\label{tbl: ai impact on new work}
	\resizebox{\textwidth}{!}{%
		\begin{tabular}[t]{lcccccc}
			\toprule
			& (1) & (2) & (3) & (4) & (5) & (6)\\
			\midrule
			&  &  &  &  &  & \\
			& \multicolumn{6}{c}{\textit{Panel A: OLS estimators}} \\
			\cline{2-7}
			&  &  &  &  &  & \\
			Automation AI   & \num{-0.003}  & \num{-0.001}  &                 &                 & \num{-0.003}   & \num{-0.001}   \\
			& (\num{0.003}) & (\num{0.003}) &                 &                 & (\num{0.003})  & (\num{0.003})  \\
			Augmentation AI &                &                & \num{0.011}*** & \num{0.015}*** & \num{0.011}*** & \num{0.015}*** \\
			&                &                & (\num{0.002})  & (\num{0.002})  & (\num{0.002})  & (\num{0.002})  \\
			$R^2 \ adj.$ & \num{0.644}   & \num{0.651}   & \num{0.648}    & \num{0.658}    & \num{0.648}    & \num{0.658}\\
			$R^2 \ within \ adj.$ & \num{0.024}   & \num{0.007}   & \num{0.036}    & \num{0.027}    & \num{0.036}    & \num{0.027}\\
			
			&  &  &  &  &  & \\
			
			& \multicolumn{6}{c}{\textit{Panel B: 2SLS estimators}} \\
			\cline{2-7}
			&  &  &  &  &  & \\
			Automation AI   & \num{-0.004}  & \num{-0.004}  &                 &                 & \num{-0.005}   & \num{-0.003}   \\
			& (\num{0.003}) & (\num{0.003}) &                 &                 & (\num{0.003})  & (\num{0.003})  \\
			Augmentation AI &                &                & \num{0.017}*** & \num{0.022}*** & \num{0.017}*** & \num{0.022}*** \\
			&                &                & (\num{0.002})  & (\num{0.002})  & (\num{0.002})  & (\num{0.002})  \\
			
			\addlinespace
			F-Stat (auto) & \num{14950} & \num{1694} & & & \num{15100} & \num{1583} \\
			F-Stat (augm) &  &  & \num{1606} & \num{247} & \num{1816} & \num{244} \\
			
			&  &  &  &  &  & \\
			
			\textit{Covariates included} & \checkmark & \checkmark & \checkmark & \checkmark & \checkmark & \checkmark\\
			\textit{Fixed effects:}  &  &  &  &  &  & \\
			cNAICS*cSOC & \checkmark & \checkmark & \checkmark & \checkmark & \checkmark & \checkmark \\
			cNAICS*year (3-digit) & & \checkmark & & \checkmark & & \checkmark \\
			
			\midrule
			Observations & \num{202695} & \num{202695} & \num{202695} & \num{202695} & \num{202695} & \num{202695}\\
			Unique cSOC (6-digit) & \num{702} & \num{702} & \num{702} & \num{702} & \num{702} & \num{702}\\
			Unique cNAICS (4-digit) & \num{220} & \num{220} & \num{220} & \num{220} & \num{220} & \num{220}\\
			
			\midrule
			Mean outcome & \num{0.02} & \num{0.02} & \num{0.02} & \num{0.02} & \num{0.02} & \num{0.02}\\
			SD outcome & \num{0.04} & \num{0.04} & \num{0.04} & \num{0.04} & \num{0.04} & \num{0.04}\\
			\bottomrule
		\end{tabular}
	}
	\caption*{\footnotesize{Note: The table presents the outputs for the regressions (\ref{eq: impact ai on new work}) when the dependent variable is the cumulative share of new work. Panel A shows the output when OLS estimators are used, whereas Panel B is when 2SLS estimators is applied. First-stage estimations are presented in Table~\ref{tbl: first_stage new work wages}. Augmentation AI exposure and automation AI exposure are standardized to have a mean of 0 and a standard deviation equal to 1. Covariates included are: employment size (log), trade per capita and demographic characteristics (age, gender, education, ethnicity, and race) for industry (4-digit). The estimations are weighted by 2015 employment size. Standard errors are reported in brackets and are clustered at the occupation-industry cell. $\star\star\star$ Significant at the 1\% level; $\star\star$ significant at the 5\% level; $\star$ significant at the 10\% level.}}
\end{table}

\begin{table}[H]
	\centering
	\caption{Effect of AI exposure on employment (log)}
	\label{tbl: ai impact on employment}
	\resizebox{\textwidth}{!}{%
	\begin{tabular}[t]{lcccccc}
		\toprule
		& (1) & (2) & (3) & (4) & (5) & (6)\\
		\midrule
		&  &  &  &  &  & \\
		& \multicolumn{6}{c}{\textit{Panel A: OLS estimators}} \\
		\cline{2-7}
		&  &  &  &  &  & \\
		Automation AI   & \num{0.031}   & \num{0.020}   &                &                & \num{0.031}   & \num{0.021}   \\
		& (\num{0.030}) & (\num{0.025}) &                &                & (\num{0.030}) & (\num{0.025}) \\
		Augmentation AI &                &                & \num{0.030}** & \num{0.040}** & \num{0.030}** & \num{0.041}** \\
		&                &                & (\num{0.015}) & (\num{0.016}) & (\num{0.015}) & (\num{0.016}) \\
		$R^2 \ adj.$ & \num{0.994}   & \num{0.995}   & \num{0.994}   & \num{0.995}   & \num{0.994}   & \num{0.995}\\
		$R^2 \ within \ adj.$ & \num{0.052}   & \num{0.026}   & \num{0.053}   & \num{0.028}   & \num{0.054}   & \num{0.029}\\
		
		&  &  &  &  &  & \\
		
		& \multicolumn{6}{c}{\textit{Panel B: 2SLS estimators}} \\
		\cline{2-7}
		&  &  &  &  &  & \\
		Automation AI   & \num{0.024}   & \num{0.015}   &                &                & \num{0.023}   & \num{0.016}   \\
		& (\num{0.031}) & (\num{0.025}) &                &                & (\num{0.031}) & (\num{0.025}) \\
		Augmentation AI &                &                & \num{0.024}   & \num{0.031}*  & \num{0.024}   & \num{0.031}*  \\
		&                &                & (\num{0.018}) & (\num{0.019}) & (\num{0.018}) & (\num{0.019}) \\

		\addlinespace
		F-Stat (auto) & \num{18210} & \num{1573} & & & \num{17940} & \num{1583} \\
		F-Stat (augm) &  &  & \num{1904} & \num{248} & \num{2145} & \num{246} \\
		
		&  &  &  &  &  & \\
		
		\textit{Covariates included} & \checkmark & \checkmark & \checkmark & \checkmark & \checkmark & \checkmark\\
		\textit{Fixed effects:}  &  &  &  &  &  & \\
		cNAICS*cSOC & \checkmark & \checkmark & \checkmark & \checkmark & \checkmark & \checkmark \\
		cNAICS*year (3-digit) & & \checkmark & & \checkmark & & \checkmark \\
		
		\midrule
		Observations & \num{202695} & \num{202695} & \num{202695} & \num{202695} & \num{202695} & \num{202695}\\
		Unique cSOC (6-digit) & \num{702} & \num{702} & \num{702} & \num{702} & \num{702} & \num{702}\\
		Unique cNAICS (4-digit) & \num{220} & \num{220} & \num{220} & \num{220} & \num{220} & \num{220}\\
		
		\midrule
		Mean outcome & \num{11.0} & \num{11.0} & \num{11.0} & \num{11.0} & \num{11.0} & \num{11.0}\\
		SD outcome & \num{2.3} & \num{2.3} & \num{2.3} & \num{2.3} & \num{2.3} & \num{2.3}\\
		\bottomrule
	\end{tabular}
	}
	\caption*{\footnotesize{Note: The table presents the outputs for the regressions (\ref{eq: impact ai on wages and employment}) when the dependent variable is the employment size in log. Panel A shows the output when OLS estimators are used, whereas Panel B is when 2SLS estimators is applied. First-stage estimations are presented in Table~\ref{tbl: first_stage employment}. Augmentation AI exposure and automation AI exposure are standardized to have a mean of 0 and a standard deviation equal to 1. Covariates included are: mean hourly wages (log), trade per capita and demographic characteristics (age, gender, education, ethnicity, and race) for industry (4-digit). The estimations are weighted by 2015 employment size. Standard errors are reported in brackets and are clustered at the occupation-industry cell. $\star\star\star$ Significant at the 1\% level; $\star\star$ significant at the 5\% level; $\star$ significant at the 10\% level.}}
\end{table}

\begin{table}[H]
	\centering
	\caption{Effect of AI exposure on mean hourly wages (log)}
	\label{tbl: ai impact on wages}
	\resizebox{\textwidth}{!}{%
	\begin{tabular}[t]{lcccccc}
		\toprule
		& (1) & (2) & (3) & (4) & (5) & (6)\\
		\midrule
		&  &  &  &  &  & \\
		& \multicolumn{6}{c}{\textit{Panel A: OLS estimators}} \\
		\cline{2-7}
		&  &  &  &  &  & \\
		Automation AI   & \num{-0.083}*** & \num{-0.081}*** &                &                & \num{-0.083}*** & \num{-0.081}*** \\
		& (\num{0.006})   & (\num{0.005})   &                &                & (\num{0.006})   & (\num{0.005})   \\
		Augmentation AI &                  &                  & \num{-0.003}  & \num{0.001}   & \num{-0.003}    & \num{0.000}     \\
		&                  &                  & (\num{0.003}) & (\num{0.003}) & (\num{0.002})   & (\num{0.003})   \\
		$R^2 \ adj.$& \num{0.994}     & \num{0.994}     & \num{0.993}   & \num{0.994}   & \num{0.994}     & \num{0.994}\\
		$R^2 \ within \ adj.$ & \num{0.151}     & \num{0.077}     & \num{0.093}   & \num{0.019}   & \num{0.151}     & \num{0.077}\\
		
		&  &  &  &  &  & \\
		
		& \multicolumn{6}{c}{\textit{Panel B: 2SLS estimators}} \\
		\cline{2-7}
		&  &  &  &  &  & \\
		Automation AI   & \num{-0.085}*** & \num{-0.080}*** &                  &                & \num{-0.085}*** & \num{-0.080}*** \\
		& (\num{0.006})   & (\num{0.006})   &                  &                & (\num{0.006})   & (\num{0.006})   \\
		Augmentation AI &                  &                  & \num{-0.010}*** & \num{-0.002}  & \num{-0.007}*** & \num{0.000}     \\
		&                  &                  & (\num{0.003})   & (\num{0.003}) & (\num{0.003})   & (\num{0.003})   \\

		\addlinespace
		F-Stat (auto) & \num{14950} & \num{1694} & & & \num{15100} & \num{1583} \\
		F-Stat (augm) &  &  & \num{1606} & \num{247} & \num{1816} & \num{244} \\
		
		&  &  &  &  &  & \\
		
		\textit{Covariates included} & \checkmark & \checkmark & \checkmark & \checkmark & \checkmark & \checkmark\\
		\textit{Fixed effects:}  &  &  &  &  &  & \\
		cNAICS*cSOC & \checkmark & \checkmark & \checkmark & \checkmark & \checkmark & \checkmark \\
		cNAICS*year (3-digit) & & \checkmark & & \checkmark & & \checkmark \\
		
		\midrule
		Observations & \num{202695} & \num{202695} & \num{202695} & \num{202695} & \num{202695} & \num{202695}\\
		Unique cSOC (6-digit) & \num{702} & \num{702} & \num{702} & \num{702} & \num{702} & \num{702}\\
		Unique cNAICS (4-digit) & \num{220} & \num{220} & \num{220} & \num{220} & \num{220} & \num{220}\\
		
		\midrule
	    Mean outcome & \num{3.2} & \num{3.2} & \num{3.2} & \num{3.2} & \num{3.2} & \num{3.2}\\
		SD outcome & \num{0.5} & \num{0.5} & \num{0.5} & \num{0.5} & \num{0.5} & \num{0.5}\\
		\bottomrule
	\end{tabular}
	}
	\caption*{\footnotesize{Note: The table presents the outputs for the regressions (\ref{eq: impact ai on wages and employment}) when the dependent variable is the mean hourly wages in log. Panel A shows the output when OLS estimators are used, whereas Panel B is when 2SLS estimators is applied. First-stage estimations are presented in Table~\ref{tbl: first_stage new work wages}. Augmentation AI exposure and automation AI exposure are standardized to have a mean of 0 and a standard deviation equal to 1. Covariates included are: employment size (log), trade per capita and demographic characteristics (age, gender, education, ethnicity, and race) for industry (4-digit). The estimations are weighted by 2015 employment size. Standard errors are reported in brackets and are clustered at the occupation-industry cell. $\star\star\star$ Significant at the 1\% level; $\star\star$ significant at the 5\% level; $\star$ significant at the 10\% level.}}
\end{table}

\begin{table}[H]
	\centering
	\caption{Effect of AI exposure by skill group - 2SLS estimators}
	\label{tbl: ai impact education}
	\resizebox{\textwidth}{!}{%
		\begin{tabular}[t]{lccc}
			\toprule
			
			\textit{Dependent variable:} & \makecell{Share New Work} & \makecell{Employment (log)} & \makecell{Hourly Wages (log)} \\
			& (1) & (2) & (3) \\

			\midrule
			\addlinespace
			& \multicolumn{3}{c}{\textit{Panel A: Low-skilled occupations}} \\
			\cline{2-4}
			\addlinespace
			
			Automation AI   & \num{-0.011}** & \num{-0.078}* & \num{-0.053}*** \\
			& (\num{0.004})  & (\num{0.041}) & (\num{0.007})   \\
			Augmentation AI & \num{-0.013}   & \num{0.088}   & \num{-0.010}    \\
			& (\num{0.009})  & (\num{0.056}) & (\num{0.012})   \\
			\addlinespace
			F-Stat (auto) & \num{719} & \num{666} & \num{719} \\
			F-Stat (augm) & \num{57} & \num{55} & \num{57} \\
			
			\addlinespace
			& \multicolumn{3}{c}{\textit{Panel B: Middle-skilled occupations}} \\
			\cline{2-4}
			\addlinespace
			
			Automation AI   & \num{-0.014}*  & \num{-0.023}  & \num{-0.038}*** \\
			& (\num{0.007})  & (\num{0.033}) & (\num{0.008})   \\
			Augmentation AI & \num{0.024}*** & \num{0.021}   & \num{0.019}***  \\
			& (\num{0.006})  & (\num{0.016}) & (\num{0.006})   \\
			\addlinespace
			F-Stat (auto) & \num{1443} & \num{1594} & \num{1443} \\
			F-Stat (augm) & \num{353} & \num{351} & \num{353} \\
			
			\addlinespace
			& \multicolumn{3}{c}{\textit{Panel C: High-skilled occupations}} \\
			\cline{2-4}
			\addlinespace
			
			Automation AI   & \num{0.023}*   & \num{0.115}   & \num{-0.039}   \\
			& (\num{0.014})  & (\num{0.106}) & (\num{0.028})  \\
			Augmentation AI & \num{0.024}*** & \num{-0.039}  & \num{0.007}*** \\
			& (\num{0.003})  & (\num{0.026}) & (\num{0.002})  \\
			\addlinespace
			F-Stat (auto) & \num{288} & \num{265} & \num{288} \\
			F-Stat (augm) & \num{237} & \num{203} & \num{237} \\
			
			\addlinespace
			\midrule
			\textit{Covariates included} & \checkmark & \checkmark & \checkmark \\
			\textit{Fixed effects:}  &  &  & \\
			cNAICS*cSOC & \checkmark & \checkmark & \checkmark \\
			cNAICS*year (3-digit) & \checkmark & \checkmark & \checkmark \\
			
			\bottomrule
		\end{tabular}
	}
	\caption*{\scriptsize{Note: The table presents the outputs for regressions (\ref{eq: impact ai on new work}) (Column 1) and (\ref{eq: impact ai on wages and employment}) (Column 2 and 3). In Columns 1, the dependent variable is the cumulative share of new work; in Column 2, it is the logarithm of employment size; and in Column 3, it is the logarithm of mean hourly wages. Panel A retains occupations requiring a low level of skill according to O*NET job zones ("Little or No Preparation Needed" or "Some Preparation Needed"). Panel B includes only occupations with medium skill requirements ("Medium Preparation Needed"). Panel C consists of occupations requiring a high skill level ("Considerable Preparation Needed" or "Extensive Preparation Needed"). N = \num{73722} in Panel A; N = \num{53391} in Panel B; N = \num{66277} in Panel C. Covariates included are: employment size (log, in Panel A and C), mean hourly wages (log, in Panel B)trade per capita and demographic characteristics (age, gender, ethnicity, and race) for industry (4-digit). The results are estimated with 2SLS estimators. The estimations are weighted by 2015 employment size. Standard errors are reported in brackets and are clustered at the occupation-industry level. $\star\star\star$ Significant at the 1\% level; $\star\star$ significant at the 5\% level; $\star$ significant at the 10\% level.}}
\end{table}

\end{refsection}
\clearpage

\begin{refsection}

\appendix

\topskip0pt
\vspace*{\fill}
\centerline{\huge{APPENDIX}}
\vspace*{\fill}

\clearpage

\section{Data sources and AI indices}\label{app: Data sources and AI indices}

\setcounter{table}{0}
\renewcommand{\thetable}{A\arabic{table}}

\setcounter{figure}{0}
\renewcommand{\thefigure}{A\arabic{figure}}

\subsection{Stack Overflow}\label{app: Stack Overflow}

Established in 2008, Stack Overflow is a Q\&A platform dedicated to resolving programming issues, software algorithms, and developer tools. Access to Stack Overflow is free, allowing anyone to create an account and participate by asking, answering, or commenting on posts. As of early 2023, the platform hosts 24 million questions, approximately 35 million answers, and 20 million users.

Stack Overflow is the most prominent website for debugging code and is the primary resource for programmers seeking assistance. In 2022, it attracted 250 million monthly visitors, more than three times the traffic of its closest competitors, such as w3schools.com (70.4 million monthly visits) and geeksforgeeks.com (64.7 million monthly visits).\footnote{For more details, see Similarweb, an online company that measures online audiences: \url{https://www.similarweb.com/website/stackoverflow.com/competitors}.}

A crucial feature of Stack Overflow for this research is the significant presence of developers among its users. The high proportion of professional developers ensures that the questions on Stack Overflow are related to algorithms deployed in the labor market rather than those related to leisure activities. This aspect is critical for this study, which aims to measure AI development in the labor market. According to Stack Overflow's annual survey, 73\% of its users are developers by profession, and 80\% write code as part of their work (\cite{stackoverflow}).

Stack Overflow offers several advantages over alternative data sources commonly used in the literature, such as patents and surveys completed by workers or AI experts (see, for example, \cite{brynjolfsson_mitchell_rock_2018, tolan_pesole_al_2021, felten_raj_seamans_2021, webb_2020}). Patents may provide an incomplete measure of AI exposure for two main reasons. First, it is well-documented that AI systems are often protected as trade secrets, as copyright and patent laws present challenges in safeguarding these innovations (\cite{Foss-Solbrekk_2021, Hattenbach_Snyder_2018, Shuijing_Tao_2019}). Second, strong incentives exist to release AI algorithms as open-source, allowing developers to quickly test and prototype AI solutions and gain insights without extensive in-house development. The release of the large language model Llama2 by Meta as open-source and the popularity of platforms like HuggingFace exemplify these incentives. Finally, surveys filled out by workers or experts typically suffer from small sample sizes and often focus more on the potential of AI rather than its actual development and implementation in the economy.

However, Stack Overflow also has some limitations. The first limitation pertains to AI algorithms developed internally by firms, which Stack Overflow might not capture. Nevertheless, this limitation is likely minimal and should not significantly affect the results of this study. An AI algorithm typically comprises various sub-algorithms and software, some likely to be referenced on Stack Overflow. For example, a developer building an internal chatbot for a company might utilize Python libraries like TensorFlow, NLTK, or ChatterBot. These packages are referenced on Stack Overflow, and developers may seek help during the chatbot's development. A second potential limitation is the firms' use of commercial AI solutions. However, this limitation seems minor, as creators of commercial AI solutions often seek to be referenced on Stack Overflow for two reasons: to gain popularity and to provide support to clients, allowing them to debug their code independently. For instance, there are 650 tags related to commercial Google solutions on Stack Overflow, including AI solutions such as Google Speech-to-Text, Google Translate, and Google Natural Language.

Stack Overflow data has already been used to explore topics related to this study in computer science and technology research. For instance, questions on Stack Overflow have been analyzed to understand discussion topics among developers (\cite{yang_al_2016, barua_al_2014, rosen_al_2016}). \textcite{moutidis_community_2021} use Stack Overflow to document technologies used by developers, and \textcite{montandon_al_2021} examine job advertisements posted on Stack Overflow to study skill demand from IT companies.

Despite these examples, the economics literature has underutilized Stack Overflow data. \textcite{gallea2023} use Stack Overflow questions to study the impact of AI on work dynamics, and \textcite{delriochanona2023} investigate the potential threats posed by large language models to digital public goods. Closely related to my study, the OECD.AI Policy Observatory has developed a set of indicators to measure AI knowledge flows using questions and answers on Stack Overflow (\cite{OECD.AI2023}).

In this project, I focus on questions asked between 2010 and 2022. Although it is possible to include more recent data, the release of ChatGPT in November 2022 significantly diminished the website's attractiveness (\cite{delriochanona2023}), rendering Stack Overflow less relevant for more recent years.

\subsection{Identifying members location and keeping those living in the US and primary trading partners}\label{app: Identifying users location}

The location of Stack Overflow members is crucial for this study, as it serves as a proxy for where AI algorithms are developed. Identifying member locations involves several steps, which are detailed below.

The geolocation of members is determined using the information provided in their Stack Overflow profiles. Upon registering, members can fill in personal details, including their location, a personal description, and a website link. Although this information is provided voluntarily, it is often in the developers' interest to complete these fields, as Stack Overflow has become a significant platform for hiring developers.

The process of identifying members' locations involves querying Google Maps for geographical information, which is derived from three primary sources. The first source is the place of residence specified in the members' profiles. This data is entered as free text and can range from particular locations, such as exact addresses, to broader areas, like regions or countries. The second source of geographical information is extracted from the members' descriptions, where they might mention their location of residence or work. For instance, some members explicitly mention in their descriptions that they work for a specific company located in the US. To extract this data, I use ChatGPT to identify any geographical details or information that could indicate the members' locations from their descriptions.\footnote{I use ChatGPT 3.5 Turbo with the following prompt: \textit{I will provide you with personal descriptions of members registered on StackOverflow. Please analyze the text and extract any explicit information that indicates where they live, such as country, city, region, or any geographical references mentioned. If you do not find any such information in the text, please return 'none'.}} The third source involves the domain of the personal website provided by the members. After retrieving the geographical information, it is passed to Google Maps to identify the country of residence. Using this method, 35.2\% of users are successfully geolocated, accounting for 38.0\% of the AI-related questions asked globally.

Once the members' locations are identified, I retain AI-related questions posted by members residing in the US and its primary trading partners. This methodological choice ensures that the AI development measures are relevant to the US market. It is well-established that international trade and the activities of multinational corporations significantly facilitate technological transfers (\cite{bilir_morales_2020, keller_al_2013, buera_oberfield_2020}). For instance, an AI algorithm developed by a firm in the EU is likely to be utilized by its US counterpart.

I select the fifteen most important countries for US imports of goods and services to identify the primary trading partners. This results in eighteen countries: Bermuda, Canada, China, France, Germany, Hong Kong, India, Ireland, Italy, Japan, Malaysia, Mexico, the Netherlands, South Korea, Switzerland, Thailand, the United Kingdom, and Vietnam. These countries account for between 0.1\% (Bermuda) and 19.7\% (China) of US imports of goods and between 0.1\% (Vietnam) and 13.3\% (The United Kingdom) of US imports of services.

\subsection{Identifying AI-related tags and questions}\label{app: Identifying AI-related questions}

To identify AI-related questions, I leverage the information provided by the tags attached to each question on Stack Overflow. I specifically identify tags related to AI and consider any questions associated with these tags as AI-related.

Tags are keywords that categorize questions, typically referring to a technology, programming language, or task developers aim to accomplish. For example, common tags on Stack Overflow include Python, GitHub, web scraping, indexing, and TensorFlow. To post a question, users must select between 3 to 5 tags. Stack Overflow contains over \num{63000} tags, each usually accompanied by a brief technical description.

The process of identifying AI-related tags involves three steps. In the first step, I search for \num{164} AI-related keywords within the tags and their technical descriptions. The list of AI keywords is derived from those identified in \textcite{alekseeva_al_2021}, supplemented by additional keywords from the computer science and technology literature and AI-specialized websites. Table~\ref{tbl: AI keywords} provides the complete list of keywords. This initial search yields \num{934} tags.

While this approach is straightforward, it has limitations. The accuracy of tag identification depends on the comprehensiveness of the keyword list, and some relevant tags may be overlooked if the list is not exhaustive. This issue is particularly pertinent in AI, which evolves rapidly with the frequent introduction of new algorithms and methods.

To address this limitation, I enhance the initial identification with a second step involving tag co-occurrence analysis. This process involves adding tags that frequently appear alongside those identified in the first step. This ensures that all AI-related tags are captured. For instance, tags such as "seaborn", "csv", "imbalanced-data", and "scikit-multilearn" are added through this procedure. This step adds \num{21837} tags, increasing the total number of potential AI-related tags to \num{22771}. By the end of this step, I have a comprehensive list of potential AI-related tags.

However, this second step also introduces a significant number of false positives—tags incorrectly classified as related to AI. For example, tags like "seaborn" and "csv", identified through the co-occurrence method, are not inherently related to AI and should not be included.

In the final step, I filter out false positives from the list of potential AI-related tags with the assistance of ChatGPT 3.5 Turbo. For example, the tag "Python" was identified during the co-occurrence step. Although Python is extensively used in AI algorithm development, it is not exclusively related to AI and thus should be excluded from the final selection. To refine the list, I prompt ChatGPT 3.5 Turbo with the following instructions:
\begin{displayquote}
	\textit{I will provide a tag and its description from the website www.stackoverflow.com. Please return 1 if the tag is related to artificial intelligence and 0 otherwise.}
\end{displayquote}
I retain only those tags that ChatGPT 3.5 Turbo classifies as AI-related. This final step results in \num{1182} AI-related tags, encompassing \num{687174} questions worldwide and \num{88600} for the United States and its primary trading partners from 2010 to 2022.\footnote{In a previous version of this paper, I manually classified the tags and arrived at a lower number of AI-related tags and questions. However, re-running the analysis with this earlier selection does not affect the results of the study.}

\begin{figure}[H]
	\centering
	\caption{Yearly number of AI-related questions asked on Stack overflow worldwide}
	\label{fig: Yearly number of questions on SO}
	\includegraphics[scale=0.5]{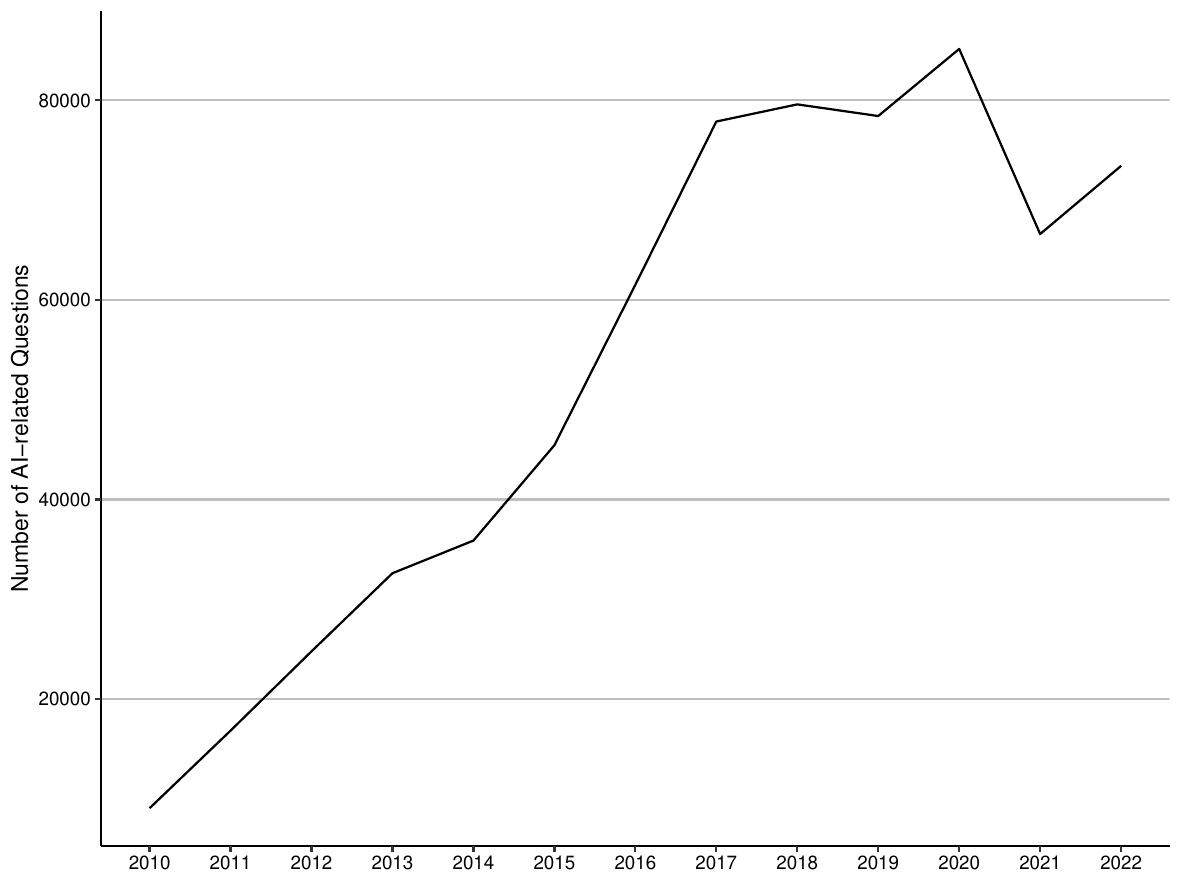}
\end{figure}

Figure~\ref{fig: Yearly number of questions on SO} shows the yearly number of AI-related questions asked on Stack Overflow. The number of questions increases sharply between 2010 and 2017. During this period, AI has seen considerable progress with the introduction of Generative Adversarial Networks (2014), Residual Networks (2015), Recurrent Neural Networks (2015), Long Short-Term Memory (2015), and Transformer Structure (2017). The number of yearly questions stabilizes around \num{80000} from 2017 to 2020 and drops in 2021. In 2022, the number of newly AI-related questions reaches \num{73430}.

\begin{figure}[H]
	\centering
	\caption{Distribution of questions per AI-related tag}
	\label{fig: Density number of questions per tag}
	\includegraphics[scale=0.5]{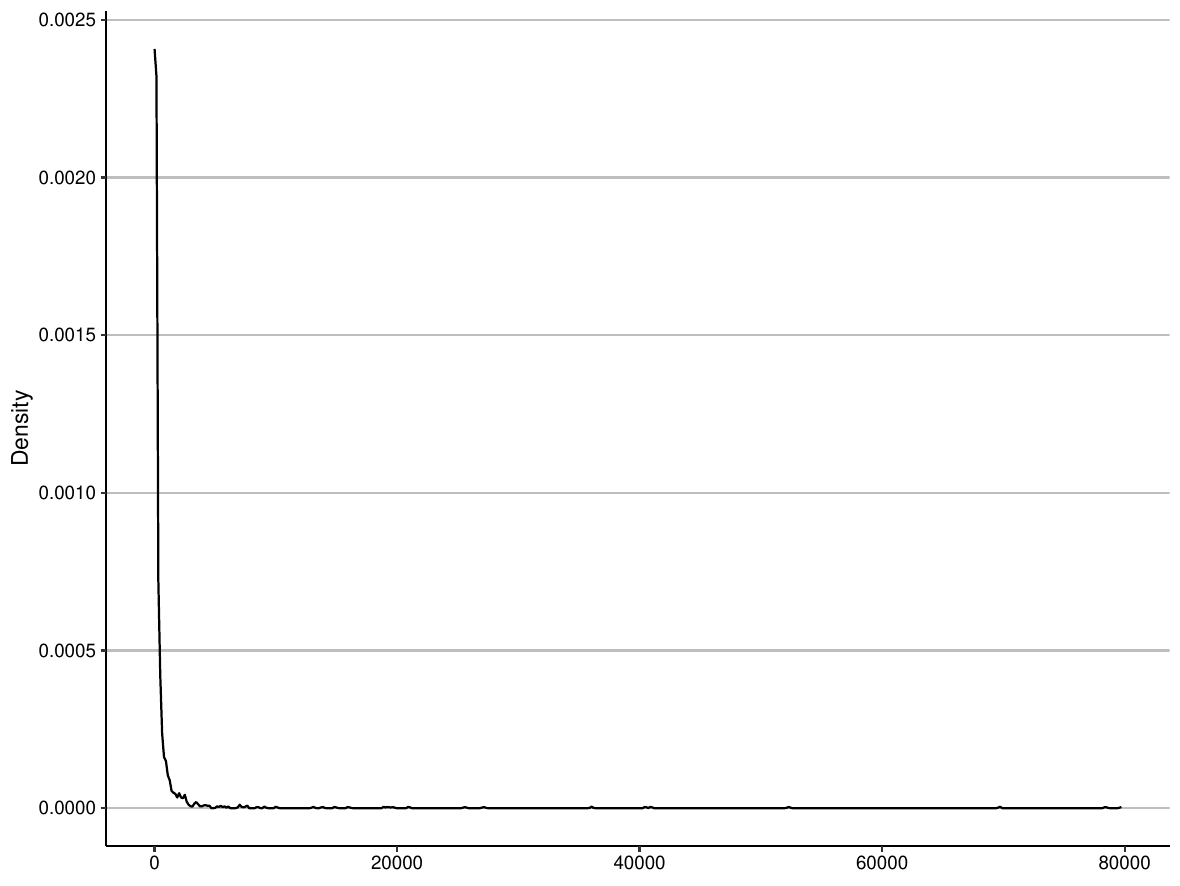}
\end{figure}

Figure~\ref{fig: Density number of questions per tag} shows the density of questions for AI-related tags. The distribution is right-skewed, indicating that some tags are highly used. The top 5 AI-related tags are Tensorflow (\num{79729} questions), Apache Spark (\num{78435}), OpenCV (\num{69694}), Machine Learning (\num{52299}), and Keras (\num{40940}).

\begin{table}[H]
	\centering
	\caption{AI keywords}
	\label{tbl: AI keywords}
	\resizebox{\textwidth}{!}{\begin{tabular}{|l|l|l|l|}
			\hline
			Activity-Recognition          & Generative AI                          & Mahout                              & Sdscm                                         \\ \hline
			Ai Chatbot                    & Gesture-Recognition                    & Marf                                & Semi-Supervised Learning                      \\ \hline
			Ai Kibit                      & Google Cloud Machine Learning Platform & Microsoft Cognitive Toolkit         & Semantic Driven Subtractive Clustering Method \\ \hline
			Antlr                         & Google-Cloud-Ml                        & Microsoft-Cognitive                 & Sentence-Transformers                         \\ \hline
			Apache-Spark-Ml               & Google-Colaboratory                    & Microsoft-Translator                & Sentiment Analysis                            \\ \hline
			Apache-Spark-Mllib            & Google-Speech-Api                      & Midjourney                          & Sentiment Classification                      \\ \hline
			Apertium                      & Gpt-2                                  & Mlpack                              & Sentiment-Analysis                            \\ \hline
			Artificial-Intelligence       & Gpt-3                                  & Mlpy                                & Sklearn-Pandas                                \\ \hline
			ASR                           & Gpt-4                                  & Modular Audio Recognition Framework & Spacy-Transformers                            \\ \hline
			Automatic Speech Recognition  & Gradient Boosting                      & Moses                               & Speech Recognition                            \\ \hline
			Automl                        & H2O                                    & Multilabel-Classification           & Speech-Recognition                            \\ \hline
			Azure-Cognitive-Services      & Handwriting-Recognition                & Mxnet                               & Speech-Synthesis                              \\ \hline
			Azure-Language-Understanding  & Huggingface-Transformers               & Named-Entity-Recognition            & Speech-To-Text                                \\ \hline
			Bert                          & Ibm Watson                             & Natural Language Processing         & Stable Diffusion                              \\ \hline
			Bert-Language-Model           & Image Processing                       & Natural Language Toolkit            & Stanford-Nlp                                  \\ \hline
			Caffe Deep Learning Framework & Image Recognition                      & Nd4J                                & Supervised Learning                           \\ \hline
			Chatbot                       & Image-Processing                       & Nearest Neighbor Algorithm          & Support Vector Machines                       \\ \hline
			Chatgpt                       & Image-Segmentation                     & Neural-Network                      & SVM                                           \\ \hline
			Chatgpt-Api                   & Information-Extraction                 & NLP                                 & Tensor                                        \\ \hline
			Classification                & Information-Retrieval                  & NLTK                                & Tensorflow                                    \\ \hline
			Computational Linguistics     & Ipsoft Amelia                          & Object Recognition                  & Tensorflow2.0                                 \\ \hline
			Computer Vision               & Iris-Recognition                       & Object Tracking                     & Text Mining                                   \\ \hline
			Computer-Vision               & Ithink                                 & Object-Detection                    & Text To Speech                                \\ \hline
			Conv-Neural-Network           & Keras                                  & Object-Detection-Api                & Text-Classification                           \\ \hline
			Copilot                       & Language-Detection                     & Opencv                              & Text-Extraction                               \\ \hline
			Dall-E                        & Languages Modeler                      & Opencv3.0                           & Text-To-Speech                                \\ \hline
			Decision Trees                & Large-Language-Model                   & Opennlp                             & Tf.Keras                                      \\ \hline
			Deep Learning                 & Latent Dirichlet Allocation            & Opinion Mining                      & Tokenization                                  \\ \hline
			Deep-Learning                 & Latent Semantic Analysis               & Pattern-Recognition                 & Topic-Modeling                                \\ \hline
			Deeplearning4J                & Lexalytics                             & Progen                              & Torch                                         \\ \hline
			Dialogflow-Es                 & Lexical Acquisition                    & Pybrain                             & Transformer                                   \\ \hline
			Distinguo                     & Lexical Semantics                      & Python-Imaging-Library              & TTS                                           \\ \hline
			Edge-Detection                & Libsvm                                 & Pytorch                             & Unsupervised Learning                         \\ \hline
			Emgucv                        & Llama                                  & Random-Forest                       & Virtual Agents                                \\ \hline
			Face-Api                      & LSTM                                   & R-Caret                             & Visual-Recognition                            \\ \hline
			Face-Detection                & Machine Learning                       & Recommender Systems                 & Voice-Recognition                             \\ \hline
			Face-Recognition              & Machine Translation                    & Recurrent-Neural-Network            & Vowpal                                        \\ \hline
			Facial-Identification         & Machine Vision                         & Reinforcement-Learning              & Wabbit                                        \\ \hline
			Feature-Extraction            & Machine-Learning                       & Roberta-Language-Model              & Word2Vec                                      \\ \hline
			Feature-Selection             & Machine-Translation                    & Scikit-Image                        & Word-Embedding                                \\ \hline
			Form-Recognizer               & Madlib                                 & Scikit-Learn                        & Xgboost                                       \\ \hline
	\end{tabular}}
	\caption*{\footnotesize{Note: This list has been built on the keywords from \textcite{alekseeva_al_2021} and keywords found in the computer science and technology literature and AI-specialized websites.}}
\end{table}

\subsection{Descriptions of the AI-related tags}\label{app: Descriptions of AI-related tags}

The link between discussions on Stack Overflow and the measures of occupational input content and output produced is partially based on the descriptions of AI-related tags. However, the descriptions available on Stack Overflow are often technical and provide limited detail. For example, the description for the tag "antlr3", created in 2010, is simply "Version 3 of ANTLR", which does not explain the type of tasks users aim to perform with this technology. A more detailed description is necessary to match the tag to measures of occupational input content and output produced. Moreover, 20\% of the tags lack any description when they are created, further complicating the matching process.\footnote{In the Appendix~\ref{app: first_excerpt}, I rerun the analysis using the technical descriptions from Stack Overflow. The results are not affected by this choice.}

I use ChatGPT 4o-mini to generate a description for each AI-related tag to mitigate this limitation. These descriptions are designed to explain the purpose of the technology referred to by the tag. I accomplished this task using the following set of prompts:
\begin{displayquote}
	\textit{System: You are an AI expert assistant in the year \{year\}. I will provide an AI-related tag and its description from www.stackoverflow.com. Based on this, you must clearly identify and describe the specific tasks someone would likely want to accomplish using this tag. Ensure your response is concise, fitting within a 60-token limit.}
\end{displayquote}

\begin{displayquote}
	\textit{User: speech-recognition: Speech recognition is a capability that enables a program to process human speech into a written format.}
\end{displayquote}

\begin{displayquote}
	\textit{Assistant: Speech recognition involves understanding a person's words or language and then converting that content into text.}
\end{displayquote}

\begin{displayquote}
	\textit{User:  \{tag\}: \{excerpt\}}
\end{displayquote}

In this set of prompts, I begin by outlining the context and specifying the type of information that ChatGPT must provide. In this context, I include the year when the tag was created on Stack Overflow, denoted as {year}. Providing the year ensures that ChatGPT generates a description corresponding to the state of the technology at the time it was introduced to the forum. I also limit the response length to 60 tokens to maintain consistency with the descriptors used in O*NET. Next, I offer an explicit example using the tag "speech-recognition". After that, I pass the name of an AI-related tag in {tag} and its description from Stack Overflow in {excerpt}. I use the earliest description available on Stack Overflow for each tag to ensure that the technology is described in its original form and unaffected by later AI developments.

\subsection{Automation AI exposure}\label{app: Automation AI exposure}

The methodology for constructing the index of automation AI exposure involves matching developed AI algorithms in the US market (measured by AI-related questions from Stack Overflow) with occupational requirements from O*NET. This is accomplished through the following six steps.

First, I smooth the scores of AI-related questions from 2022 using a yearly decay factor.\footnote{Questions with a negative score are excluded, as they are deemed uninformative by the Stack Overflow community.} The scores represent the net votes given by Stack Overflow members. A decay factor of 50\% is applied, meaning the impact of a question is halved for each additional year since its publication. This approach accounts for the rapid emergence of new technologies, which diminishes the relevance of older questions. The formula for calculating the smoothed score of an AI-related question for a specific year is:

\begin{equation}\label{eq: smooth_score_question}
	S_{qt} = V_{q2022} * \frac{0.5^{(t-k)}}{\sum_{k}^{2022} 0.5^{(2022-k)}}
\end{equation}

Where $S_{qt}$ is the smoothed score for question $q$ in year $t$, $V_{q2022}$ represents the votes the question gets in 2022, and $k$ is the year of publication. To illustrate the decay factor, a question published in 2020 with a score of 10 in 2022 has a smoothed score of 5.7, 2.9, and 1.4 in 2020, 2021, and 2022, respectively. This step gives a table containing a smoothed yearly score for each AI-related question between its year of publication and 2022.

In the second step, I compute the yearly AI-related tag scores. To accomplish this, I divide the smoothed scores from equation~\ref{eq: smooth_score_question} by the number of tags attached to each question, thereby avoiding double counting. These adjusted scores are then aggregated at the tag level as follows:

\begin{equation}\label{eq: score per tag}
	ST_{gt} = \sum_{q \in Q_{g}}^{} \frac{S_{qt}}{n_q}
\end{equation}

Here, $ST_{gt}$ represents the yearly score for the AI-related tag $g$ in year $t$. $Q_{g}$ denotes the set of questions tagged with $g$ and $n_q$ indicates the number of tags attached to question $q$. This method yields a yearly score for each AI-related tag from 2010 to 2022. For example, consider two questions with smoothed scores of 15 and 6 in 2010. The first question is tagged with three tags (Machine Learning, Semantic Comparison, and NLP), and the second is tagged with two tags (Machine Learning and Deep Learning). The tag Machine Learning will have a yearly score of 8 in 2010, with 5 points contributed by the first question and 3 points by the second.

Third, I construct two transition matrices to link information from Stack Overflow with data from O*NET. The first matrix is based on the names of the tags and abilities, while the second matches their descriptions. Specifically, I use Sentence-BERT embeddings to project AI-related tags and abilities into 768-dimensional vector spaces (\cite{Reimers_Gurevych_2019}).\footnote{The "all-mpnet-base-v2" model is used for sentence embeddings.} These embeddings allow me to populate the transition matrices with cosine similarity scores for each tag-ability pair. Cosine similarity scores, ranging from -1 (opposite meaning) to 1 (similar meaning), quantify the semantic similarity between names for the first matrix and descriptions for the second. Negative scores are replaced with 0, as AI is unlikely to be developed for conceptually opposite tasks. For example, the AI-related tag "Vision" has cosine similarity scores of 0.70 and 0.66 with the abilities "Near Vision" and "Far Vision", respectively. Lastly, following \textcite{autor_2024} and \textcite{prytkova_al_2024}, I retain the similarity scores that fall within the top 25\% in both transition matrices and compute an averaged transition matrix. Applying this threshold improves the match by filtering out noise from weakly matched tag-ability pairs.\footnote{In the Appendix~\ref{app: full matrices ss}, I rerun the analysis using the full matrix of sentence similarity scores. The results are qualitatively similar.}

In the fourth step, I utilize the yearly tag scores from equation (\ref{eq: score per tag}) in conjunction with the averaged transition matrix to compute the exposure to AI at the ability level:

\begin{equation}\label{eq: automation ai ability}
	A_{at} = \sum_{2010}^{t}\sum_{g=1}^{1182} ST_{gt}*C_{ag}
\end{equation}

In this equation, $A_{at}$ measures the exposure to AI for the ability $a$ in year $t$. It represents the cumulative sum of yearly tag scores, weighted by the cosine similarity measures $C_{ag}$ between the tags $g$ and the ability $a$. This weighting gives greater significance to yearly tag scores where the tag closely aligns with the ability.

Fifth, I use the yearly ability AI exposure and the importance and level scores provided by O*NET to compute the automation AI exposure index at the occupational level. The computation is as follows:

\begin{equation}\label{eq: automation ai occupation}
	AI\_auto_{ot} = \sum_{a=1}^{52} \frac{A_{at}*L_{ao}*I_{ao}}{\sum_{a=1}^{52} L_{ao}*I_{ao}}
\end{equation}

Here, $AI\_auto_{ot}$ represents the automation AI exposure for occupation $o$ in year $t$. This index is the weighted average of the abilities' AI exposure scores from equation~(\ref{eq: automation ai ability}), where the weights are the importance $I_{oa}$ and level $L_{oa}$ scores, reflecting the varying requirements for abilities across occupations.\footnote{Higher importance and level scores indicate that an ability is crucial and frequently used within an occupation, while lower scores suggest that the ability is less relevant.} To account for the differing requirements across occupations, the weights are rescaled between 0 and 1, as commonly done in the literature (\cite{felten_raj_seamans_2021, webb_2020, brynjolfsson_mitchell_rock_2018}).\footnote{Some occupations necessitate higher scores for many abilities, resulting in a larger total score. Without rescaling, this would disproportionately increase the AI exposure for these occupations compared to those with lower total scores.} Since the importance and level scores are fixed at their 2010 values, all variations in $AI\_auto_{ot}$ are driven by the changes in questions asked on Stack Overflow over time.

Finally, following \cite{acemoglu_2024}, I impute missing values using the average exposure at the higher level of aggregation and compute simple means at the 6-digit constant Standard Occupational Classification (cSOC) level.

\subsection{Augmentation AI exposure}\label{app: Augmentation AI exposure}

The construction of the index of augmentation AI exposure follows a methodology closely aligned with that used for the index of automation AI exposure. The key difference is using micro-titles for industries and occupations from the CAI instead of abilities. \textcite{autor_chin_salomons_seegmiller_2024} demonstrate that micro-titles provide insights into the goods and services produced by an occupation or industry. Moreover, they show that micro-titles can effectively be used to measure augmenting innovations.

I compute the index of augmentation AI exposure in five steps. First, I smooth the votes of AI-related questions in 2022 by applying a decay factor of 50\%. This step allows for creating a panel dataset where the votes for AI-related questions are distributed between the year of publication and 2022. The smoothed votes are calculated as follows:

\begin{equation}\label{eq: smooth_score_question_cai}
	S_{qt} = V_{q2022} * \frac{0.5^{(t-k)}}{\sum_{k}^{2022} 0.5^{(2022-k)}}
\end{equation}

Here, $S_{qt}$ is the smoothed vote score for question $q$ in year $t$, $V_{q2022}$ represents the votes the question gets in 2022, and $k$ is the year of publication. To illustrate the effect of the decay factor, consider a question published in 2019 with a score of 15 in 2022. The smoothed scores for this question would be 8, 4, 2, and 1 in 2019, 2020, 2021, and 2022, respectively.

Second, I compute the yearly tag scores by dividing the smoothed scores from equation~\ref{eq: smooth_score_question_cai} by the number of tags attached to each question. These adjusted scores are then summed at the tag level as follows:

\begin{equation}\label{eq: score per tag_cai}
	ST_{gt} = \sum_{q \in Q_{g}}^{} \frac{S_{qt}}{n_q}
\end{equation}

Where $ST_{gt}$ is the yearly score for the AI-related tag $g$ in year $t$. $Q_{g}$ is the set of questions tagged with $g$ and $n_q$ indicates the number of tags attached to question $q$. This method gives a yearly score for each AI-related tag from 2010 to 2022. For example, consider the following scenario: two questions have smoothed scores of 20 and 10 in 2010, tagged with three tags (Deep Learning, TensorFlow, and Keras) and two tags (Deep Learning and Computer Vision), respectively. The tag Deep Learning would have a yearly score of 11.7 in 2010 (6.7 points from the first question and 5 points from the second).

Third, I create four transition matrices that link AI-related tags with micro-titles for occupations and industries from CAI. The first two matrices link the occupation micro-titles to the AI-related tags based on their names (matrix 1) and descriptions (matrix 2). Matrices 3 and 4 perform the same exercise for micro-titles related to industries. I use the same Sentence-BERT model to create sentence embeddings as for the automation AI exposure measure (\cite{Reimers_Gurevych_2019}).\footnote{I use the "all-mpnet-base-v2" model.} The transition matrices are populated with cosine similarity scores, which measure the semantic closeness of each tag to a micro-title. As with the automation AI exposure measure, I replace negative cosine similarity scores with 0, retain the top 25\% of similarity scores appearing in both transition matrices for occupations and compute an averaged transition matrix. I apply the same approach for the transition matrices based on micro-titles for industries.

Third, I derive the index of augmentation AI exposure for micro-industries by applying the following equation:

\begin{equation}\label{eq: augmentation ai industry}
	I_{it} = \sum_{2010}^{t}\sum_{g=1}^{1182} ST_{gt}*C_{ig} 
\end{equation}

Where $I_{it}$ gives the augmentation AI exposure for micro-industry $i$ in year $t$. $ST_{gt}$ is the yearly-tags scores and comes from equation~(\ref{eq: score per tag_cai}). $C_{ig}$ is the cosine similarity measure between micro-industry $i$ and tag $g$. 

Similarly, the index of augmentation AI exposure for micro-occupations is given by:

\begin{equation}\label{eq: augmentation ai occupation}
	O_{ot} = \sum_{2010}^{t}\sum_{g=1}^{1182} ST_{gt}*C_{og}
\end{equation}

Here, $O_{ot}$ is the exposure to AI that complements micro-occupation $o$ and $C_{og}$ is the cosine similarity measure between micro-occupation $i$ and tag $g$.

Finally, I compute the simple mean at 6-digit cSOC for the augmentation AI exposure for micro-occupations and at 4-digit cNAICS for the augmentation AI exposure for micro-industries, and I take the average of both scores as follows:

\begin{equation}\label{eq: augmentation ai occupation-indutry}
	AI\_augm_{oit} = \frac{I_{it}+O_{ot}}{2} 
\end{equation}

The index of augmentation AI exposure at the occupational*industry level is given by $AI\_augm_{oit}$, where $I_{it}$ is given by equation~(\ref{eq: augmentation ai industry}) and $O_{ot}$ by equation~(\ref{eq: augmentation ai occupation}).

\section{Abilities and occupational AI exposure}\label{app: Abilities and occupational AI exposure}

\setcounter{table}{0}
\renewcommand{\thetable}{B\arabic{table}}

\setcounter{figure}{0}
\renewcommand{\thefigure}{B\arabic{figure}}

\begin{figure}[H]
	\centering
	\caption{AI exposure for abilities, 2022}
	\label{fig: AI exposure abilities}
	\includegraphics[scale=0.7]{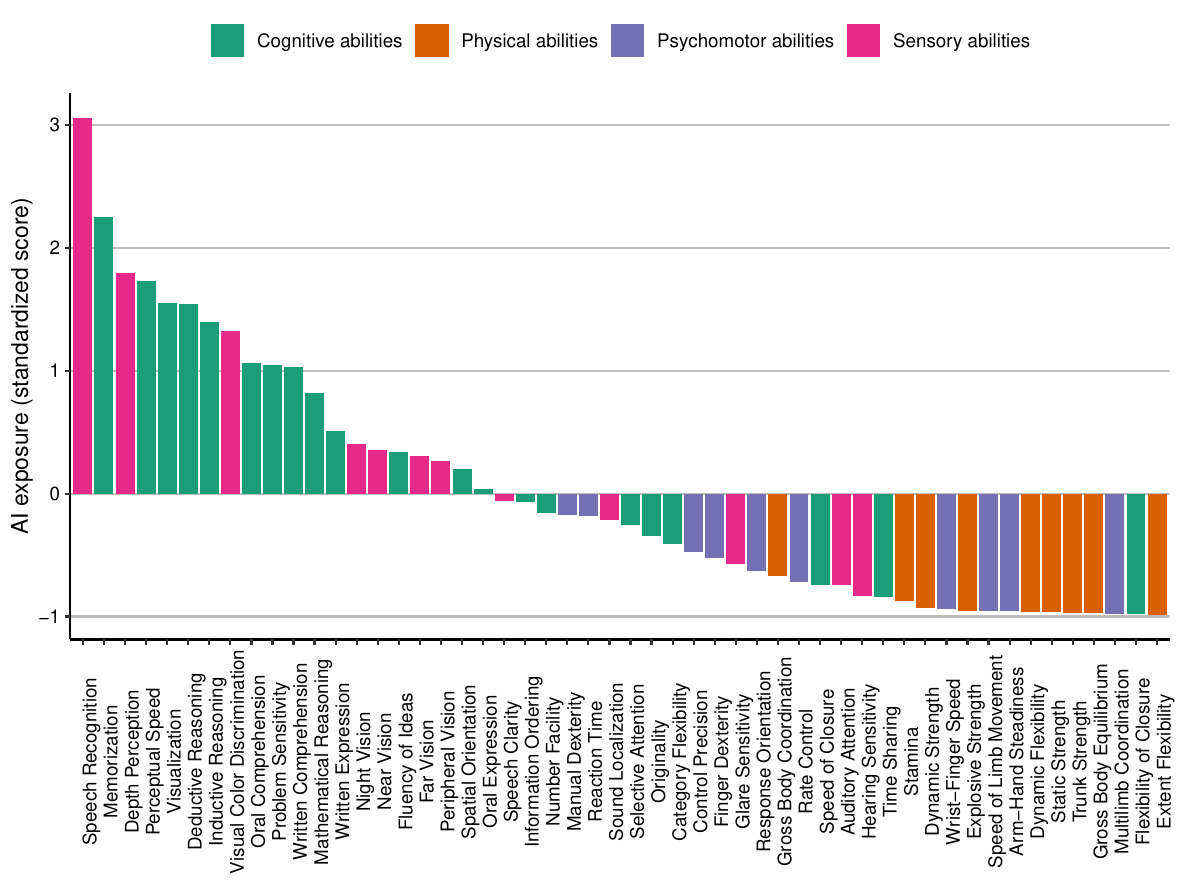}
	\caption*{\footnotesize{Note: The y-axis measures the standardized scores of AI exposure for abilities obtained from equation~(\ref{eq: automation ai ability}). Abilities are ranked in decreasing order of the score.}}
\end{figure}

Figure \ref{fig: AI exposure abilities} presents AI automation exposure at the ability level for 2022, as determined by equation~\ref{eq: automation ai ability}. The figure highlights a clear distinction between cognitive and sensory abilities versus physical and psychomotor abilities. Cognitive and sensory abilities are significantly more exposed to AI, confirming findings in prior research (\cite{felten_raj_seamans_2021}). Among the most exposed abilities are those related to vision, language understanding, and the ability to combine information and extract patterns—areas where AI algorithms have demonstrated strong capabilities (\cite{stanford_ai_2024}).

In contrast, the least exposed abilities include psychomotor abilities (e.g., multilimb coordination) and physical abilities (e.g., extent flexibility and gross body equilibrium). These abilities are less vulnerable to AI automation due to their dependence on physical and psychomotor inputs, which remains challenging for AI to replicate.

Table~\ref{tbl: highest lowest automation AI exposure occupations} presents the most and least exposed occupations to AI automation. The most exposed occupations are predominantly white-collar roles requiring advanced educational qualifications and high cognitive and sensory abilities. Among these, "Real Estate Brokers", "Proofreaders and Copy Markers", and "Foreign Language and Literature Teachers, Postsecondary" exhibit the highest exposure scores. In contrast, the 20 least exposed occupations are exclusively blue-collar jobs that demand substantial physical and psychomotor abilities. Occupations such as "Dancers", "Reinforcing Iron and Rebar Workers", and "Fallers" have the lowest exposure scores. These findings are consistent with those of \textcite{felten_raj_seamans_2018, felten_raj_seamans_2021}, who identified similar patterns in AI exposure across various occupations.

\begin{table}[H]
	\caption{Occupations with the highest and lowest automation AI exposure, 2022}
	\label{tbl: highest lowest automation AI exposure occupations}
	\resizebox{\textwidth}{!}{
		\centering
		\begin{tabular}{|l|l|c|}
			\hline
			Rank & Occupation & Automation AI score \\ 
			\hline
			1 & Real Estate Brokers & 1.74 \\ 
			2 & Proofreaders and Copy Markers & 1.74 \\ 
			3 & Foreign Language and Literature Teachers, Postsecondary & 1.63 \\ 
			4 & Law Teachers, Postsecondary & 1.63 \\ 
			5 & Judges, Magistrate Judges, and Magistrates & 1.62 \\ 
			6 & Psychology Teachers, Postsecondary & 1.54 \\ 
			7 & Forestry and Conservation Science Teachers, Postsecondary & 1.52 \\ 
			8 & Sociologists & 1.52 \\ 
			9 & Door-to-Door Sales Workers, News and Street Vendors, and Related Workers & 1.51 \\ 
			10 & Arbitrators, Mediators, and Conciliators & 1.50 \\ 
			11 & Education Teachers, Postsecondary & 1.50 \\ 
			12 & Procurement Clerks & 1.49 \\ 
			13 & Sociology Teachers, Postsecondary & 1.47 \\ 
			14 & Title Examiners, Abstractors, and Searchers & 1.46 \\ 
			15 & Economists & 1.46 \\ 
			16 & Loan Officers & 1.45 \\ 
			17 & Telemarketers & 1.45 \\ 
			18 & Communications Teachers, Postsecondary & 1.45 \\ 
			19 & Philosophy and Religion Teachers, Postsecondary & 1.43 \\ 
			20 & Social Work Teachers, Postsecondary & 1.43 \\ 
			…    & … & … \\
			746 & Dining Room and Cafeteria Attendants and Bartender Helpers & -1.69 \\ 
			747 & Food Preparation and Serving Related Workers, All Other & -1.69 \\ 
			748 & Derrick Operators, Oil and Gas & -1.71 \\ 
			749 & Landscaping and Groundskeeping Workers & -1.71 \\ 
			750 & Structural Iron and Steel Workers & -1.77 \\ 
			751 & Rock Splitters, Quarry & -1.78 \\ 
			752 & Helpers--Roofers & -1.78 \\ 
			753 & Dishwashers & -1.80 \\ 
			754 & Foundry Mold and Coremakers & -1.81 \\ 
			755 & Slaughterers and Meat Packers & -1.82 \\ 
			756 & Helpers--Painters, Paperhangers, Plasterers, and Stucco Masons & -1.83 \\ 
			757 & Fence Erectors & -1.84 \\ 
			758 & Tire Builders & -1.85 \\ 
			759 & Helpers--Brickmasons, Blockmasons, Stonemasons, and Tile and Marble Setters & -1.90 \\ 
			760 & Roof Bolters, Mining & -1.92 \\ 
			761 & Exercise Trainers and Group Fitness Instructors & -1.92 \\ 
			762 & Pressers, Textile, Garment, and Related Materials & -2.01 \\ 
			763 & Fallers & -2.02 \\ 
			764 & Reinforcing Iron and Rebar Workers & -2.08 \\ 
			765 & Dancers & -2.35 \\   
			\hline
		\end{tabular}
	}
	\caption*{\footnotesize{Note: Occupations are ranked by their score of automation AI exposure at 6-digit constant Standard Occupational Classification (cSOC) level and computed following equation (\ref{eq: automation ai occupation}). AI score is standardized to have a mean of 0 and a standard deviation of 1.}}
\end{table}

Table \ref{tbl: highest lowest augmentation AI exposure occupations} presents the occupations most and least exposed to augmentation AI. Among the most exposed occupations, those related to computers are highly represented, including "Computer and Information Research Scientists", "Computer Programmers", and "Computer Systems Analysts". In contrast, the least exposed occupations are more diverse and include clerk occupations, such as "New Accounts Clerks", as well as specialist physicians like "Oral and Maxillofacial Surgeons", "Orthodontists", and "Dental Hygienists".

\begin{table}[H]
	\caption{Occupations with the highest and lowest augmentation AI exposure, 2022}
	\label{tbl: highest lowest augmentation AI exposure occupations}
	\centering
	\resizebox{.9\textwidth}{!}{
		\begin{tabular}{|l|l|c|}
			\hline
			Rank & Occupation & Augmentation AI score \\ 
			\hline
			1 & Computer and Information Research Scientists & 8.00 \\ 
			2 & Computer Programmers & 5.23 \\ 
			3 & Computer Systems Analysts & 5.10 \\ 
			4 & Veterinary Technologists and Technicians & 3.66 \\ 
			5 & Computer Hardware Engineers & 3.66 \\ 
			6 & Information Security Analysts, Web Developers, and Computer Network Architects & 3.42 \\ 
			7 & Industrial-Organizational Psychologists & 3.42 \\ 
			8 & Environmental Engineering Technologists and Technicians & 3.42 \\ 
			9 & Computer Science Teachers, Postsecondary & 3.40 \\ 
			10 & Software Developers and Software Quality Assurance Analysts and Testers & 3.35 \\ 
			11 & Civil Engineering Technologists and Technicians & 3.20 \\ 
			12 & Mathematicians & 3.07 \\ 
			13 & Statisticians & 3.03 \\ 
			14 & Computer support specialist & 2.85 \\ 
			15 & Operations Research Analysts & 2.63 \\ 
			16 & Forensic Science Technicians & 2.61 \\ 
			17 & Agricultural Engineers & 2.60 \\ 
			18 & Urban and Regional Planners & 2.52 \\ 
			19 & Sales Engineers & 2.48 \\ 
			20 & Network and Computer Systems Administrators & 2.47 \\ 
			…    & … & … \\
			742 & Religious Workers, All Other & -1.29 \\ 
			741 & Residential Advisors & -1.29 \\ 
			742 & Property, Real Estate, and Community Association Managers & -1.32 \\ 
			743 & Hosts and Hostesses, Restaurant, Lounge, and Coffee Shop & -1.33 \\ 
			744 & Postmasters and Mail Superintendents & -1.33 \\ 
			745 & Chiropractors & -1.37 \\ 
			746 & Physician Assistants & -1.39 \\ 
			747 & New Accounts Clerks & -1.39 \\ 
			748 & Podiatrists & -1.40 \\ 
			749 & Obstetricians and Gynecologists & -1.42 \\ 
			750 & Family Medicine Physicians & -1.48 \\ 
			751 & Tellers & -1.49 \\ 
			752 & Dentists, All Other Specialists & -1.56 \\ 
			753 & Clergy & -1.56 \\ 
			754 & Oral and Maxillofacial Surgeons & -1.57 \\ 
			755 & Prosthodontists & -1.58 \\ 
			756 & Dentists, General & -1.59 \\ 
			757 & Orthodontists & -1.61 \\ 
			758 & Lodging Managers & -1.63 \\ 
			759 & Dental Hygienists & -1.73 \\  
			\hline
		\end{tabular}
	}
	\caption*{\footnotesize{Note: The table shows the weighted average exposure at the occupational level, using employment size in industries as weight. Occupations are ranked by their score of augmentation AI exposure at 6-digit constant Standard Occupational Classification (cSOC) level and computed following equation (\ref{eq: augmentation ai occupation-indutry}).  AI score is standardized to have a mean of 0 and a standard deviation of 1.}}
\end{table}

\section{Descriptive statistics}\label{app: Descriptive statistics}

\setcounter{table}{0}
\renewcommand{\thetable}{C\arabic{table}}

\setcounter{figure}{0}
\renewcommand{\thefigure}{C\arabic{figure}}

\begin{table}[H]
	\centering
	\caption{Descriptive statistics}
	\label{tbl: Descriptive statistics}
	\begin{tabular}[t]{lcccc}
		\toprule
		& Mean & SD & Min & Max\\
		\midrule
		Automation AI exposure & 0.00 & 0.99 & -2.03 & 2.27 \\ 
		Augmentation AI exposure & -0.16 & 1.00 & -1.57 & 11.27 \\ 
		Share new work & 0.02 & 0.04 & 0.00 & 2.33 \\ 
		Hourly wages (log) & 3.19 & 0.51 & 2.19 & 5.57 \\ 
		Employment size (log) & 10.99 & 2.34 & 3.40 & 15.21 \\ 
		Female & 0.49 & 0.19 & 0.05 & 0.90 \\ 
		Male & 0.51 & 0.19 & 0.10 & 0.95 \\ 
		Aged 14-34 & 0.35 & 0.11 & 0.11 & 0.63 \\ 
		Aged 35-54 & 0.42 & 0.07 & 0.27 & 0.58 \\ 
		Aged 55-99 & 0.23 & 0.05 & 0.10 & 0.54 \\ 
		Race: White & 0.77 & 0.07 & 0.46 & 0.96 \\ 
		Race: Black & 0.13 & 0.06 & 0.01 & 0.43 \\ 
		Race: Asian & 0.07 & 0.04 & 0.00 & 0.26 \\ 
		Race: Others & 0.03 & 0.01 & 0.00 & 0.16 \\ 
		Ethnicity: not Hispanic or Latino & 0.84 & 0.05 & 0.47 & 0.97 \\ 
		Ethnicity: Hispanic or Latino & 0.16 & 0.05 & 0.03 & 0.53 \\ 
		Education: High school or lower & 0.36 & 0.08 & 0.17 & 0.58 \\ 
		Education: Some College or Associate degree & 0.27 & 0.04 & 0.18 & 0.35 \\ 
		Education: Bachelor's degree or advanced degree & 0.24 & 0.11 & 0.10 & 0.57 \\ 
		Education: not available & 0.13 & 0.10 & 0.01 & 0.39 \\ 
		Imports per capita (log) & 0.97 & 2.93 & 0.00 & 14.98 \\  
		\bottomrule
	\end{tabular}
	\caption*{\footnotesize{Note: Automation and augmentation AI exposures are standardized. All statistics are weighted by 2015 employment size.}}
\end{table}

\section{Instrumental variable}\label{app: Instrumental variable}

\setcounter{table}{0}
\renewcommand{\thetable}{D\arabic{table}}

\setcounter{figure}{0}
\renewcommand{\thefigure}{D\arabic{figure}}

\begin{figure}[H]
	\centering
	\caption{Top 15 contributors to the IV countries group}
	\label{fig: Top 15 contributors to the IV countries group}
	\includegraphics[scale=0.7]{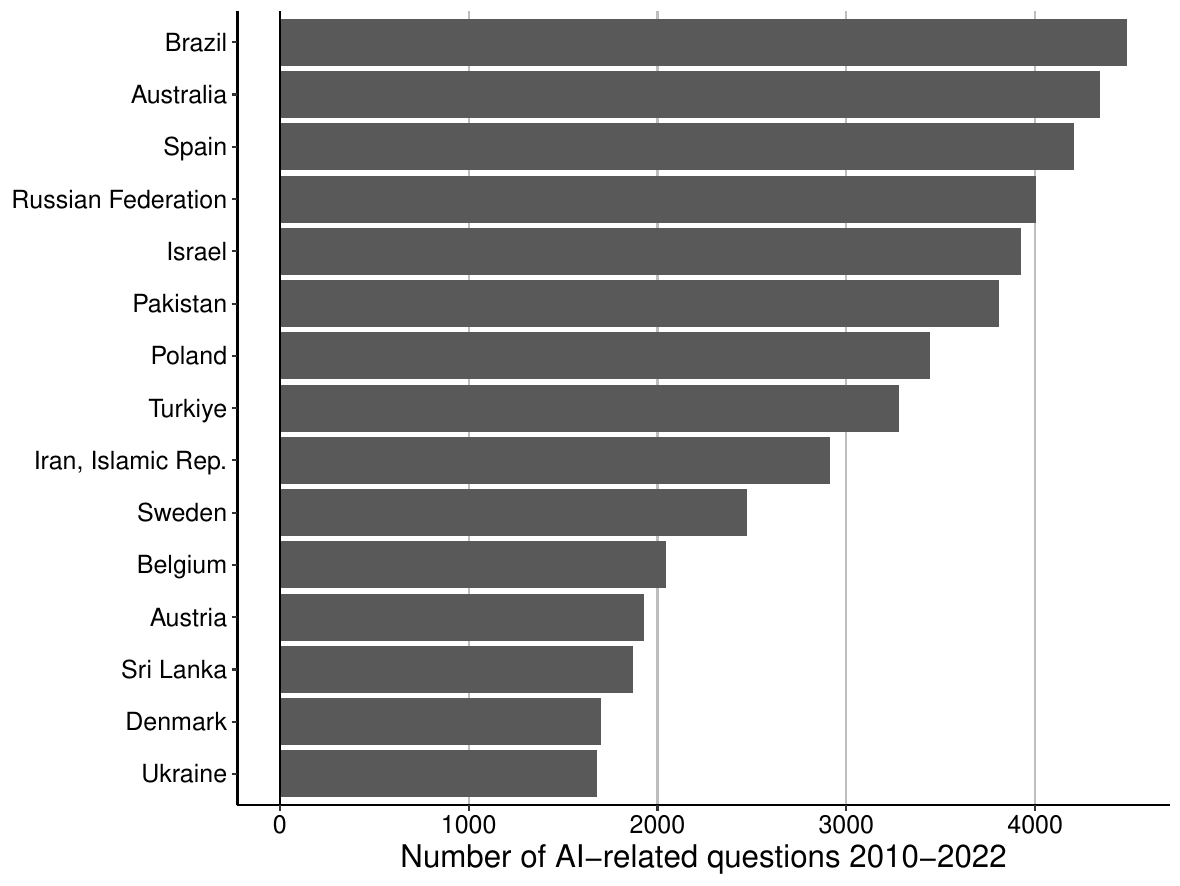}
	\caption*{\footnotesize{Note: The figure shows the top 15 countries regarding AI-related questions for the IV countries group.}}
\end{figure}

\begin{figure}[H]
	\centering
	\caption{Yearly AI-related questions posted on Stack Overflow, by country of residence 2010-2022}
	\label{fig: Question on SO for US and IV}
	\includegraphics[scale=0.55]{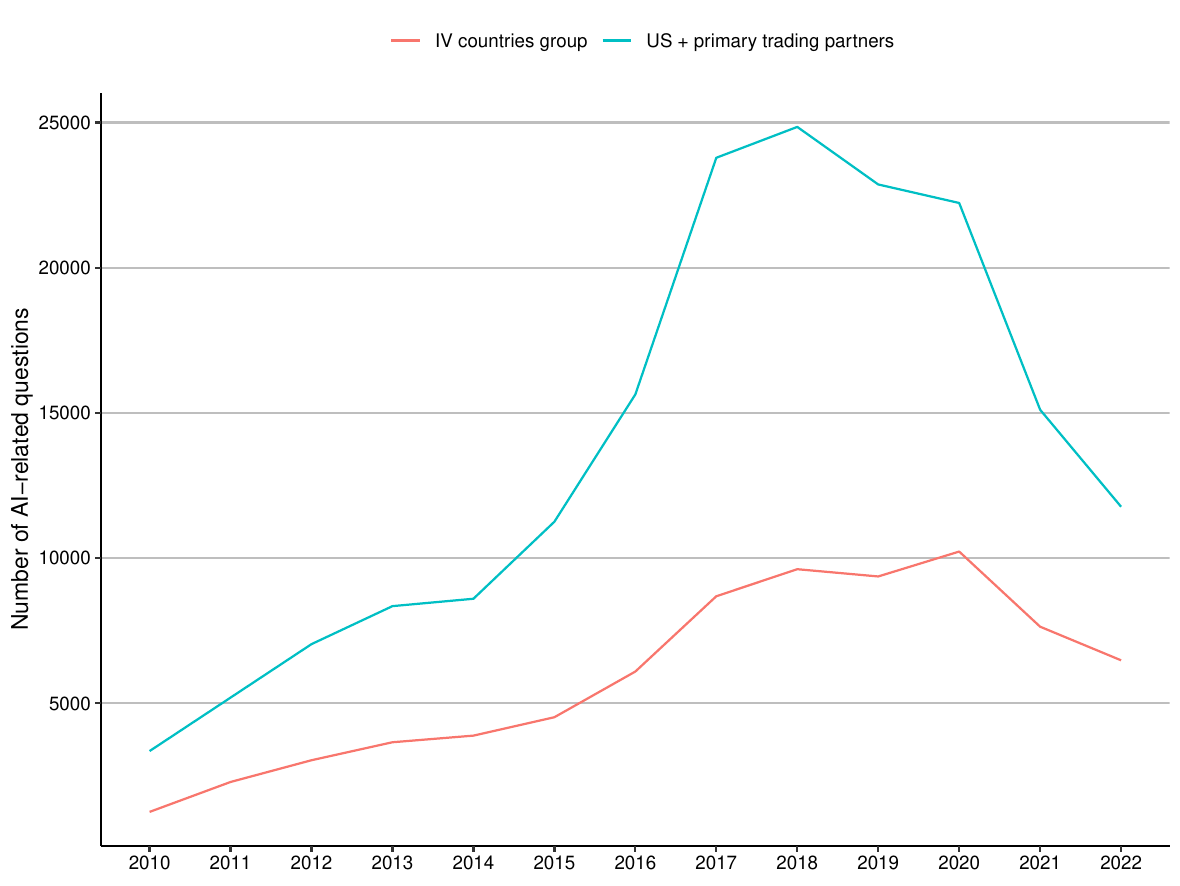}
	\caption*{\footnotesize{Note: This figure displays the yearly AI-related questions posted on Stack Overflow by members residing in the US and its primary trading partners compared to those in the IV countries group. The IV countries group includes 134 countries with no significant economic relationships with the United States.}}
\end{figure}

\section{Validation AI exposure indices}\label{app: Validation AI exposure indices}

\setcounter{table}{0}
\renewcommand{\thetable}{E\arabic{table}}

\setcounter{figure}{0}
\renewcommand{\thefigure}{E\arabic{figure}}

\begin{figure}[H]
	\centering
	\caption{Automation AI exposure and percentage firms adopting AI for task automation by broad industry}
	\label{fig: automation AI VS indus automation}
	\includegraphics[scale=0.7]{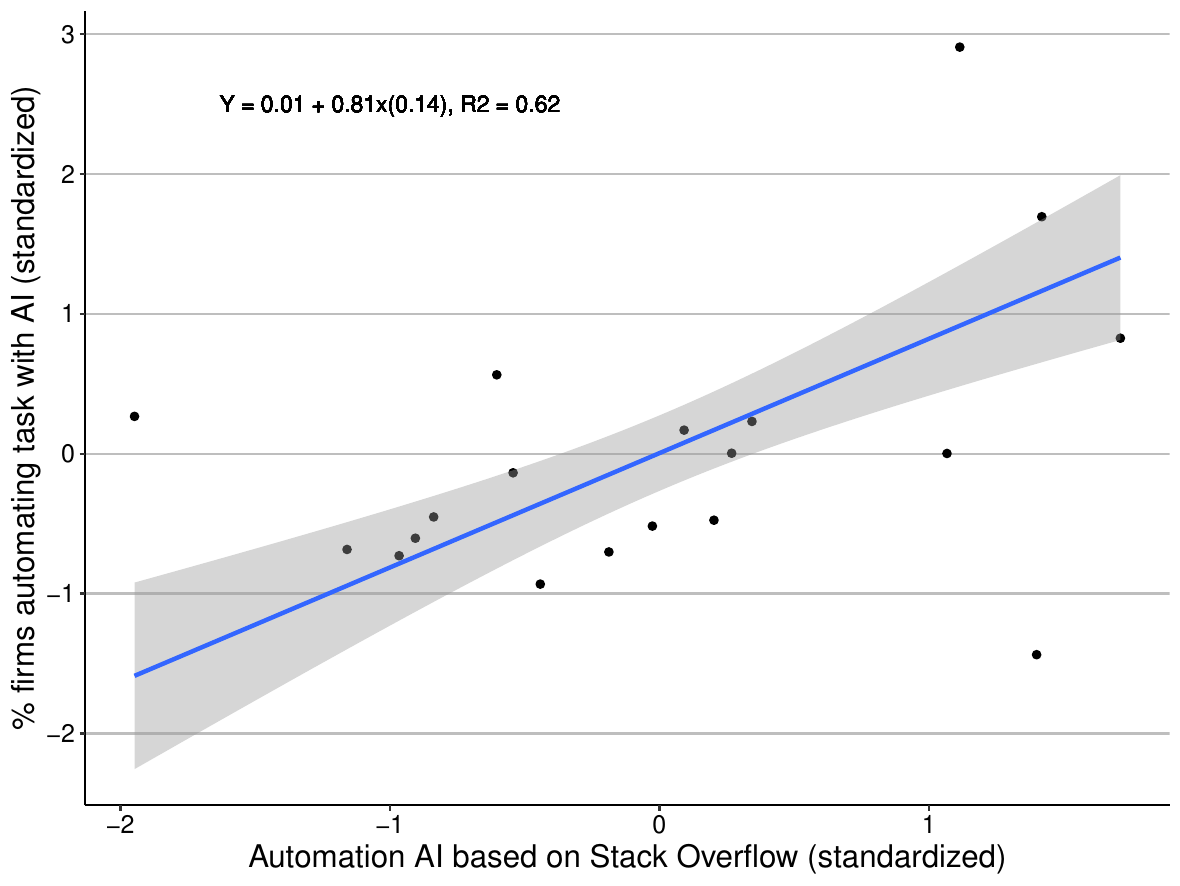}
	\caption*{\footnotesize{Note: The figure illustrates the correlation between the automation AI exposure measure developed in this study (x-axis) and self-reported AI adoption for task automation by firms at the industry level (y-axis) (n = 19). Automation AI exposure is the weighted average exposure at the industry level, using employment size in occupations as weight. Firms' AI adoption data are sourced from the 2019 Annual Business Survey \citep{ncses_2019}. The blue line represents the OLS regression line, weighted by the number of firms in each industry.}}
\end{figure}

Figure~\ref{fig: automation AI VS indus automation} illustrates the correlation between the automation AI exposure measure developed in this study and the share of firms adopting AI for task automation across broad industries (n = 19). Data on AI adoption are sourced from the 2019 Annual Business Survey \citep{ncses_2019}, a nationally representative US survey covering over 4.8 million firms. The figure reveals a strong positive correlation between the two measures, supporting the validity of the automation AI exposure index.

\begin{figure}[H]
	\centering
	\caption{Strong dispersion between automation AI and augmentation AI exposure, 2022}
	\label{fig: Automation AI VS augmentation AI}
	\includegraphics[scale=0.7]{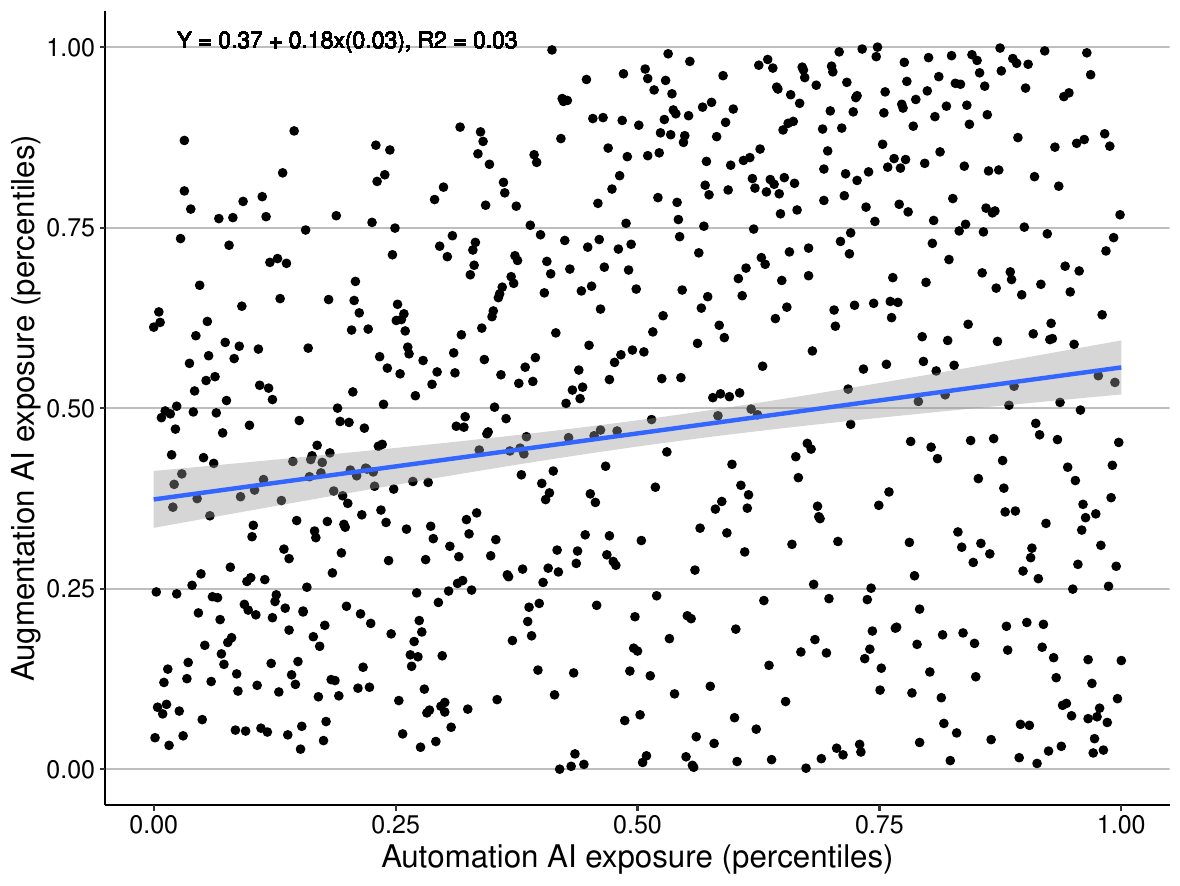}
	\caption*{\footnotesize{Note: The figure shows the relationship between automation AI exposure (x-axis) and augmentation AI exposure (y-axis) at the occupational level (6-digit). Augmentation AI exposure is the weighted average exposure at the occupational level, using employment size in industries as weight. The blue line represents the regression line weighted by the employment size. The regression line has a slope of \num{0.12} (SE $=$ \num{0.03}) and an intercept of \num{0.36} with R2 = 0.01.}}
\end{figure}

Figure~\ref{fig: Automation AI VS augmentation AI} presents a scatterplot with automation AI exposure on the x-axis and augmentation AI exposure on the y-axis, both in percentile ranks. The figure shows a weak positive relationship between the two measures. This finding is significant for this study, as it suggests that automation AI exposure and augmentation AI exposure pertain to distinct occupations, confirming that the two indices measure different aspects of AI development.\footnote{For comparison, \textcite{autor_chin_salomons_seegmiller_2024} present a similar figure depicting the relationship between exposure to automation and augmentation innovations at the occupational level from 1980 to 2018, where they observe a much stronger positive relationship.}

Table~\ref{tbl: comparison AI indices} presents the results of estimating the relationships between the AI exposure indices developed in this study and other AI indices proposed in the literature, using OLS estimators. While indices proposed in the literature claim to be agnostic about whether they measure automation or augmentation, their methodologies align them more with automation measures. The indices are converted into percentile ranks to facilitate interpretation.

Columns 1 to 3 use the automation AI exposure index as the dependent variable. Automation AI exposure appears closely aligned with the index suggested by \textcite{felten_raj_seamans_2021}, with a point estimate of 0.97 (Column 1). This result is unsurprising since the methodologies for producing these two indices are similar, relying on abilities to describe the content of occupations. Additionally, \textcite{felten_raj_seamans_2021} focus on 10 AI applications that have experienced the fastest growth since 2010 and are believed to be more likely used in the medium term. Significant overlap exists between their AI applications and the AI-related tags on Stack Overflow.

The correlation with the index from \textcite{brynjolfsson_mitchell_rock_2018} is statistically significant but relatively weak (point estimate of 0.25) (Column 2). This can be explained by the nature of their index, which focuses solely on machine learning, a subfield of AI, whereas this study employs a broader concept of AI.

The relationship between the measure of automation AI exposure and the index from \textcite{webb_2020} is not statistically significant (Column 3). This discrepancy may stem from the definition of AI adopted by \textcite{webb_2020} and the reliance on patents to measure AI exposure. \textcite{webb_2020} define AI as supervised learning and reinforcement learning algorithms, subfields of machine learning. Furthermore, patents may reflect only a limited aspect of AI adoption due to varied protection strategies. AI systems are often protected as trade secrets, and protecting them under copyright and patent laws presents challenges (\cite{Foss-Solbrekk_2021, Hattenbach_Snyder_2018, Shuijing_Tao_2019}). Additionally, AI algorithms might be published as open source.

Columns 4 to 6 present the relationships when the dependent variable is the augmentation AI exposure measure. The association with \textcite{felten_raj_seamans_2021} and \textcite{brynjolfsson_mitchell_rock_2018} is much weaker than for automation AI exposure and is not significant for the latter (Columns 4 and 5, respectively). These results indicate that augmentation AI exposure captures a different aspect of AI exposure than these previous indices.

Regarding \textcite{webb_2020}, the coefficient is 0.25 and is significantly different from zero. This result suggests that firms might be more likely to patent AI algorithms that complement labor rather than replace it.

\begin{table}[H]
	\centering
	\caption{Relationships with previous AI exposure indices}
	\label{tbl: comparison AI indices}
	\resizebox{\textwidth}{!}{%
		\begin{tabular}[t]{lcccccc}
			\toprule
			\textit{Dependent variable:} & \multicolumn{3}{c}{Automation AI exposure} & \multicolumn{3}{c}{Augmentation AI exposure} \\
			& (1) & (2) & (3) & (4) & (5) & (6)\\
			\midrule
			\textcite{felten_raj_seamans_2021}  & \num{0.973}*** &                 &                 & \num{0.252}*** &                 &                 \\
			& (\num{0.009})  &                 &                 & (\num{0.036})  &                 &                 \\
			\textcite{brynjolfsson_mitchell_rock_2018} &                 & \num{0.254}*** &                 &                 & \num{0.036}    &                 \\
			&                 & (\num{0.036})  &                 &                 & (\num{0.037})  &                 \\
			\textcite{webb_2020} &                 &                 & \num{-0.058}   &                 &                 & \num{0.257}*** \\
			&                 &                 & (\num{0.037})  &                 &                 & (\num{0.036})  \\
			Intercept & \num{0.014}*** & \num{0.373}*** & \num{0.529}*** & \num{0.374}*** & \num{0.482}*** & \num{0.372}*** \\
			& (\num{0.005})  & (\num{0.021})  & (\num{0.021})  & (\num{0.021})  & (\num{0.021})  & (\num{0.021})  \\
			\midrule
			$R^2 adj.$ & \num{0.946}    & \num{0.063}    & \num{0.002}    & \num{0.062}    & \num{0.000}    & \num{0.065}\\
			Observations & \num{723} & \num{723} & \num{723} & \num{723} & \num{723} & \num{723}\\
			\bottomrule
		\end{tabular}
	}
	\caption*{\footnotesize{Note: Augmentation AI exposure is the weighted average exposure at the occupational level, using employment size in industries as weight. Indices are converted into percentile ranks. The indices for automation AI exposure and augmentation AI exposure are for 2022. Columns 1 to 3 use automation AI occupation for the dependent variable, while Columns 4 to 6 use augmentation AI exposure. The table shows the results using OLS estimators. $\star\star\star$ Significant at the 1\% level; $\star\star$ significant at the 5\% level; $\star$ significant at the 10\% level.}}
\end{table}

\section{First Stage 2SLS estimators}\label{app: IV first-stage}

\setcounter{table}{0}
\renewcommand{\thetable}{F\arabic{table}}

\setcounter{figure}{0}
\renewcommand{\thefigure}{F\arabic{figure}}

\begin{table}[H]
	\centering
	\caption{First-stage estimates for cumulative share of new work and mean hourly wages}
	\label{tbl: first_stage new work wages}
	\begin{tabular}[t]{lcccc}
		\toprule
		\textit{Dependent variable:} & \multicolumn{2}{c}{Automation AI exposure} & \multicolumn{2}{c}{Augmentation AI exposure} \\
		& (1) & (2) & (3) & (4) \\
		\midrule
		Automation AI IV   & \num{0.865}*** & \num{0.865}*** &                 & \num{-0.091}*** \\
		& (\num{0.007})  & (\num{0.007})  &                 & (\num{0.016})   \\
		Augmentation AI IV &                 & \num{-0.002}   & \num{0.627}*** & \num{0.629}***  \\
		&                 & (\num{0.002})  & (\num{0.012})  & (\num{0.012})   \\
		
		&  &  &  &  \\
		F-stat. & \num{1694}  & \num{1583} & \num{247} & \num{244}\\
		\textit{Covariates included} & \checkmark & \checkmark & \checkmark & \checkmark \\
		\textit{Fixed effects:}  &  &  &  & \\
		cNAICS*cSOC & \checkmark & \checkmark & \checkmark & \checkmark \\
		cNAICS*year (3-digit) & \checkmark & \checkmark & \checkmark & \checkmark \\
		Observations & \num{202695} & \num{202695} & \num{202695} & \num{202695}\\
		\bottomrule
	\end{tabular}
	\caption*{\footnotesize{Note: The dependent variable is exposure to automation AI (Columns 1 and 2) and augmentation AI (Columns 3 and 4). The instrumental variable is the five-year lagged AI exposures from countries with no significant economic ties to the US. All variables have been standardized. Covariates included are: employment size (log), trade per capita and demographic characteristics (age, gender, education, ethnicity, and race) for industry (4-digit). Observations are weighted by 2015 employment size. Standard errors are reported in brackets and are clustered at the occupation-industry cell. $\star\star\star$ Significant at the 1\% level; $\star\star$ Significant at the 5\% level; $\star$ Significant at the 10\% level.}}
\end{table}

\begin{table}[H]
	\centering
	\caption{First-stage estimates for employment}
	\label{tbl: first_stage employment}
	\begin{tabular}[t]{lcccc}
		\toprule
		\textit{Dependent variable:} & \multicolumn{2}{c}{Automation AI exposure} & \multicolumn{2}{c}{Augmentation AI exposure} \\
		& (1) & (2) & (3) & (4) \\
		\midrule
		Automation AI IV   & \num{0.859}*** & \num{0.859}*** &                 & \num{-0.091}*** \\
		& (\num{0.007})  & (\num{0.007})  &                 & (\num{0.016})   \\
		Augmentation AI IV &                 & \num{-0.002}   & \num{0.628}*** & \num{0.630}***  \\
		&                 & (\num{0.002})  & (\num{0.012})  & (\num{0.012})   \\
		
		&  &  &  &  \\
		F-stat. & \num{1573}  & \num{1466} & \num{248} & \num{246}\\
		\textit{Covariates included} & \checkmark & \checkmark & \checkmark & \checkmark \\
		\textit{Fixed effects:}  &  &  &  & \\
		cNAICS*cSOC & \checkmark & \checkmark & \checkmark & \checkmark \\
		cNAICS*year (3-digit) & \checkmark & \checkmark & \checkmark & \checkmark \\
		Observations & \num{202695} & \num{202695} & \num{202695} & \num{202695}\\
		\bottomrule
	\end{tabular}
	\caption*{\footnotesize{Note: The dependent variable is exposure to automation AI (Columns 1 and 2) and augmentation AI (Columns 3 and 4). The instrumental variable is the five-year lagged AI exposures from countries with no significant economic ties to the US. All variables have been standardized. Covariates included are: mean hourly wages (log), trade per capita and demographic characteristics (age, gender, education, ethnicity, and race) for industry (4-digit). Observations are weighted by 2015 employment size. Standard errors are reported in brackets and are clustered at the occupation-industry cell. $\star\star\star$ Significant at the 1\% level; $\star\star$ Significant at the 5\% level; $\star$ Significant at the 10\% level.}}
\end{table}

\section{Effect of AI exposure by skill requirements}\label{app: ai impact education}

\setcounter{table}{0}
\renewcommand{\thetable}{G\arabic{table}}

\setcounter{figure}{0}
\renewcommand{\thefigure}{G\arabic{figure}}

\begin{table}[H]
	\centering
	\caption{Effect of AI exposure for low-skilled occupations - OLS estimators}
	\label{app: ai impact education low skilled}
	\resizebox{\textwidth}{!}{%
		\begin{tabular}[t]{lccc}
			\toprule
			
			\textit{Dependent variable:} & \makecell{Share New Work} & \makecell{Employment (log)} & \makecell{Hourly Wages (log)} \\
			& (1) & (2) & (3) \\
			
			\midrule
			Automation AI   & \num{-0.008}** & \num{-0.101}*** & \num{-0.054}*** \\
			& (\num{0.004})  & (\num{0.039})   & (\num{0.006})   \\
			Augmentation AI & \num{-0.013}** & \num{0.138}***  & \num{0.001}     \\
			& (\num{0.006})  & (\num{0.033})   & (\num{0.009})   \\
			
			\addlinespace
			$R^2 \ adj.$ & \num{0.710}    & \num{0.996}     & \num{0.984}  \\
			$R^2 \ within \ adj.$ & \num{0.021}    & \num{0.031}     & \num{0.048}\\
			
			\addlinespace
			\textit{Covariates included} & \checkmark & \checkmark & \checkmark \\
			\textit{Fixed effects:}  &  &  & \\
			cNAICS*cSOC & \checkmark & \checkmark & \checkmark \\
			cNAICS*year (3-digit) & \checkmark & \checkmark & \checkmark \\
			
			\midrule
			
			Observations & \num{73722} & \num{53391} & \num{66277}\\
			Unique cSOC (6-digit) & \num{288} & \num{182} & \num{192} \\
			Unique cNAICS (4-digit) & \num{220} & \num{220} & \num{220} \\
			
			\midrule
			\textit{Mean outcome} & \num{0.02} & \num{10.2} & \num{3.4} \\
			\textit{SD outcome} & \num{0.04} & \num{1.8} & \num{0.4} \\
						
			\bottomrule
		\end{tabular}
	}
	
		\caption*{\footnotesize{Note: The table presents the outputs for regressions (\ref{eq: impact ai on new work}) (Column 1) and (\ref{eq: impact ai on wages and employment}) (Column 2 and 3). In Columns 1, the dependent variable is the cumulative share of new work; in Column 2, it is the logarithm of employment size; and in Column 3, it is the logarithm of mean hourly wages. The sample is composed of occupations requiring a low level of skill according to O*NET job zones ("Little or No Preparation Needed" or "Some Preparation Needed"). Covariates included are: employment size (log, for Columns 1 and 3), mean hourly wages (log, for Column 2), trade per capita and demographic characteristics (age, gender, ethnicity, and race) for industry (4-digit). The results are estimated with OLS estimators. The estimations are weighted by 2015 employment size. Standard errors are reported in brackets and are clustered at the occupation-industry level. $\star\star\star$ Significant at the 1\% level; $\star\star$ significant at the 5\% level; $\star$ significant at the 10\% level.}}
\end{table}

\begin{table}[H]
	\centering
	\caption{Effect of AI exposure for middle-skilled occupations - OLS estimators}
	\label{app: ai impact education middle skilled}
	\resizebox{\textwidth}{!}{%
		\begin{tabular}[t]{lccc}
			\toprule
			
			\textit{Dependent variable:} & \makecell{Share New Work} & \makecell{Employment (log)} & \makecell{Hourly Wages (log)} \\
			& (1) & (2) & (3) \\
			
			\midrule
			Automation AI   & \num{-0.012}*  & \num{-0.010}  & \num{-0.051}*** \\
			& (\num{0.007})  & (\num{0.030}) & (\num{0.007})   \\
			Augmentation AI & \num{0.022}*** & \num{0.035}** & \num{0.012}**   \\
			& (\num{0.005})  & (\num{0.014}) & (\num{0.005})   \\
			
			\addlinespace
			$R^2 \ adj.$ & \num{0.772}    & \num{0.993}   & \num{0.988}  \\
			$R^2 \ within \ adj.$ & \num{0.092}    & \num{0.051}   & \num{0.079}\\
			
			\addlinespace
			\textit{Covariates included} & \checkmark & \checkmark & \checkmark \\
			\textit{Fixed effects:}  &  &  & \\
			cNAICS*cSOC & \checkmark & \checkmark & \checkmark \\
			cNAICS*year (3-digit) & \checkmark & \checkmark & \checkmark \\
			
			\midrule
			
			Observations & \num{73722} & \num{53391} & \num{66277}\\
			Unique cSOC (6-digit) & \num{288} & \num{182} & \num{192} \\
			Unique cNAICS (4-digit) & \num{220} & \num{220} & \num{220} \\
			
			\midrule
			Mean outcome & \num{0.03} & \num{10.2} & \num{3.9}\\
			SD outcome & \num{0.05} & \num{2.1} & \num{0.4} \\
						
			\bottomrule
		\end{tabular}
	}
	\caption*{\footnotesize{Note: The table presents the outputs for regressions (\ref{eq: impact ai on new work}) (Column 1) and (\ref{eq: impact ai on wages and employment}) (Column 2 and 3). In Columns 1, the dependent variable is the cumulative share of new work; in Column 2, it is the logarithm of employment size; and in Column 3, it is the logarithm of mean hourly wages. The sample is composed of occupations requiring a middle level of skill according to O*NET job zones ("Medium Preparation Needed"). Covariates included are: employment size (log, for Columns 1 and 3), mean hourly wages (log, for Column 2), trade per capita and demographic characteristics (age, gender, ethnicity, and race) for industry (4-digit). The results are estimated with OLS estimators. The estimations are weighted by 2015 employment size. Standard errors are reported in brackets and are clustered at the occupation-industry level. $\star\star\star$ Significant at the 1\% level; $\star\star$ significant at the 5\% level; $\star$ significant at the 10\% level.}}
\end{table}

\begin{table}[H]
	\centering
	\caption{Effect of AI exposure for high-skilled occupations - OLS estimators}
	\label{app: ai impact education high skilled}
	\resizebox{\textwidth}{!}{%
		\begin{tabular}[t]{lccc}
			\toprule
			
			\textit{Dependent variable:} & \makecell{Share New Work} & \makecell{Employment (log)} & \makecell{Hourly Wages (log)} \\
			& (1) & (2) & (3) \\
			
			\midrule
			Automation AI   & \num{0.017}    & \num{0.222}*** & \num{-0.032}   \\
			& (\num{0.013})  & (\num{0.073})  & (\num{0.022})  \\
			Augmentation AI & \num{0.016}*** & \num{-0.038}*  & \num{0.007}*** \\
			& (\num{0.002})  & (\num{0.022})  & (\num{0.002})  \\
			
			\addlinespace
			$R^2 \ adj.$ & \num{0.682}    & \num{0.994}    & \num{0.986}  \\
			$R^2 \ within \ adj.$ & \num{0.046}    & \num{0.026}    & \num{0.017}\\
			
			\addlinespace
			\textit{Covariates included} & \checkmark & \checkmark & \checkmark \\
			\textit{Fixed effects:}  &  &  & \\
			cNAICS*cSOC & \checkmark & \checkmark & \checkmark \\
			cNAICS*year (3-digit) & \checkmark & \checkmark & \checkmark \\
			
			\midrule
			
			Observations & \num{73722} & \num{53391} & \num{66277}\\
			Unique cSOC (6-digit) & \num{288} & \num{182} & \num{192} \\
			Unique cNAICS (4-digit) & \num{220} & \num{220} & \num{220} \\
			
			\midrule
			Mean outcome & \num{0.02} & \num{10.2} & \num{3.4} \\
			SD outcome & \num{0.04} & \num{1.8} & \num{0.4} \\
			
			\midrule
			Observations & \num{73722} & \num{53391} & \num{66277} \\
			
			\bottomrule
		\end{tabular}
	}
	
		\caption*{\footnotesize{Note: The table presents the outputs for regressions (\ref{eq: impact ai on new work}) (Column 1) and (\ref{eq: impact ai on wages and employment}) (Column 2 and 3). In Columns 1, the dependent variable is the cumulative share of new work; in Column 2, it is the logarithm of employment size; and in Column 3, it is the logarithm of mean hourly wages. The sample is composed of occupations requiring a high level of skill according to O*NET job zones ("Considerable Preparation Needed" or "Extensive Preparation Needed"). Covariates included are: employment size (log, for Columns 1 and 3), mean hourly wages (log, for Column 2), trade per capita and demographic characteristics (age, gender, ethnicity, and race) for industry (4-digit). The estimations are weighted by 2015 employment size. Standard errors are reported in brackets and are clustered at the occupation-industry level. $\star\star\star$ Significant at the 1\% level; $\star\star$ significant at the 5\% level; $\star$ significant at the 10\% level.}}
\end{table}

\section{IV countries group using high-income countries}\label{app: IV_high_income_countries}

\setcounter{table}{0}
\renewcommand{\thetable}{H\arabic{table}}

\setcounter{figure}{0}
\renewcommand{\thefigure}{H\arabic{figure}}

\begin{table}[H]
	\centering
	\caption{Effect of AI exposure on the cumulative share of new work - 2SLS estimators}
	\label{tbl: IV_high_income_countries new work}
	\resizebox{\textwidth}{!}{%
		\begin{tabular}[t]{lcccccc}
			\toprule
			& (1) & (2) & (3) & (4) & (5) & (6)\\
			\midrule
			Automation AI   & \num{-0.004}  & \num{-0.003}  &                 &                 & \num{-0.004}   & \num{-0.002}   \\
			& (\num{0.003}) & (\num{0.003}) &                 &                 & (\num{0.003})  & (\num{0.003})  \\
			Augmentation AI &                &                & \num{0.017}*** & \num{0.022}*** & \num{0.017}*** & \num{0.022}*** \\
			&                &                & (\num{0.002})  & (\num{0.002})  & (\num{0.002})  & (\num{0.002})  \\
			\addlinespace
			F-Stat (auto) & \num{15380} & \num{1975} &  &  & \num{15950} & \num{1834} \\
			F-Stat (augm) &  &  & \num{1675} & \num{242} & \num{1854} & \num{239} \\
			
			&  &  &  &  &  & \\
			
			\textit{Covariates included} & \checkmark & \checkmark & \checkmark & \checkmark & \checkmark & \checkmark\\
			\textit{Fixed effects:}  &  &  &  &  &  & \\
			cNAICS*cSOC & \checkmark & \checkmark & \checkmark & \checkmark & \checkmark & \checkmark \\
			cNAICS*year (3-digit) & & \checkmark & & \checkmark & & \checkmark \\
			
			\midrule
			Observations & \num{202695} & \num{202695} & \num{202695} & \num{202695} & \num{202695} & \num{202695}\\
			Unique cSOC (6-digit) & \num{702} & \num{702} & \num{702} & \num{702} & \num{702} & \num{702}\\
			Unique cNAICS (4-digit) & \num{220} & \num{220} & \num{220} & \num{220} & \num{220} & \num{220}\\
			
			\midrule
			Mean outcome & \num{0.02} & \num{0.02} & \num{0.02} & \num{0.02} & \num{0.02} & \num{0.02}\\
			SD outcome & \num{0.04} & \num{0.04} & \num{0.04} & \num{0.04} & \num{0.04} & \num{0.04}\\
			\bottomrule
		\end{tabular}
	}
	\caption*{\footnotesize{Note: The table presents the outputs for the regressions (\ref{eq: impact ai on new work}) when the dependent variable is the cumulative share of new work and 2SLS estimators is applied. Only high income countries in 2010 have been included in the IV countries group (n = 43). Augmentation AI exposure and automation AI exposure are standardized to have a mean of 0 and a standard deviation equal to 1. Covariates included are: employment size (log), trade per capita and demographic characteristics (age, gender, education, ethnicity, and race) for industry (4-digit). The estimations are weighted by 2015 employment size. Standard errors are reported in brackets and are clustered at the occupation-industry cell. $\star\star\star$ Significant at the 1\% level; $\star\star$ significant at the 5\% level; $\star$ significant at the 10\% level.}}
\end{table}

\begin{table}[H]
	\centering
	\caption{Effect of AI exposure on employment (log) - 2SLS estimators}
	\label{tbl: IV_high_income_countries employment}
	\resizebox{\textwidth}{!}{%
		\begin{tabular}[t]{lcccccc}
			\toprule
			& (1) & (2) & (3) & (4) & (5) & (6)\\
			\midrule
			Automation AI   & \num{0.026}   & \num{0.016}   &                &                & \num{0.026}   & \num{0.017}   \\
			& (\num{0.031}) & (\num{0.025}) &                &                & (\num{0.031}) & (\num{0.025}) \\
			Augmentation AI &                &                & \num{0.026}   & \num{0.035}*  & \num{0.025}   & \num{0.035}*  \\
			&                &                & (\num{0.018}) & (\num{0.019}) & (\num{0.018}) & (\num{0.019}) \\
			\addlinespace
			F-Stat (auto) & \num{19150} & \num{1776} &  &  & \num{19350} & \num{1650} \\
			F-Stat (augm) &  &  & \num{1967} & \num{242} & \num{2173} & \num{241} \\
			
			&  &  &  &  &  & \\
			
			\textit{Covariates included} & \checkmark & \checkmark & \checkmark & \checkmark & \checkmark & \checkmark\\
			\textit{Fixed effects:}  &  &  &  &  &  & \\
			cNAICS*cSOC & \checkmark & \checkmark & \checkmark & \checkmark & \checkmark & \checkmark \\
			cNAICS*year (3-digit) & & \checkmark & & \checkmark & & \checkmark \\
			
			\midrule
			Observations & \num{202695} & \num{202695} & \num{202695} & \num{202695} & \num{202695} & \num{202695}\\
			Unique cSOC (6-digit) & \num{702} & \num{702} & \num{702} & \num{702} & \num{702} & \num{702}\\
			Unique cNAICS (4-digit) & \num{220} & \num{220} & \num{220} & \num{220} & \num{220} & \num{220}\\
			
			\midrule
			Mean outcome & \num{11.0} & \num{11.0} & \num{11.0} & \num{11.0} & \num{11.0} & \num{11.0}\\
			SD outcome & \num{2.3} & \num{2.3} & \num{2.3} & \num{2.3} & \num{2.3} & \num{2.3}\\
			\bottomrule
		\end{tabular}
	}
	\caption*{\footnotesize{Note: The table presents the outputs for the regressions (\ref{eq: impact ai on wages and employment}) when the dependent variable is the employment size in log and 2SLS estimators is applied. Only high income countries in 2010 have been included in the IV countries group (n = 43). Augmentation AI exposure and automation AI exposure are standardized to have a mean of 0 and a standard deviation equal to 1. Covariates included are: mean hourly wages (log), trade per capita and demographic characteristics (age, gender, education, ethnicity, and race) for industry (4-digit). The estimations are weighted by 2015 employment size. Standard errors are reported in brackets and are clustered at the occupation-industry cell. $\star\star\star$ Significant at the 1\% level; $\star\star$ significant at the 5\% level; $\star$ significant at the 10\% level.}}
\end{table}

\begin{table}[H]
	\centering
	\caption{Effect of AI exposure on mean hourly wages (log) - 2SLS estimators}
	\label{tbl: IV_high_income_countries wages}
	\resizebox{\textwidth}{!}{%
		\begin{tabular}[t]{lcccccc}
			\toprule
			& (1) & (2) & (3) & (4) & (5) & (6)\\
			\midrule
			Automation AI   & \num{-0.085}*** & \num{-0.080}*** &                  &                & \num{-0.084}*** & \num{-0.080}*** \\
			& (\num{0.006})   & (\num{0.005})   &                  &                & (\num{0.006})   & (\num{0.005})   \\
			Augmentation AI &                  &                  & \num{-0.010}*** & \num{-0.002}  & \num{-0.007}*** & \num{-0.001}    \\
			&                  &                  & (\num{0.003})   & (\num{0.003}) & (\num{0.003})   & (\num{0.003})   \\
			\addlinespace
			F-Stat (auto) & \num{15380} & \num{1975} &  &  & \num{15950} & \num{1834} \\
			F-Stat (augm) &  &  & \num{1675} & \num{242} & \num{1854} & \num{239} \\
			
			&  &  &  &  &  & \\
			
			\textit{Covariates included} & \checkmark & \checkmark & \checkmark & \checkmark & \checkmark & \checkmark\\
			\textit{Fixed effects:}  &  &  &  &  &  & \\
			cNAICS*cSOC & \checkmark & \checkmark & \checkmark & \checkmark & \checkmark & \checkmark \\
			cNAICS*year (3-digit) & & \checkmark & & \checkmark & & \checkmark \\
			
			\midrule
			Observations & \num{202695} & \num{202695} & \num{202695} & \num{202695} & \num{202695} & \num{202695}\\
			Unique cSOC (6-digit) & \num{702} & \num{702} & \num{702} & \num{702} & \num{702} & \num{702}\\
			Unique cNAICS (4-digit) & \num{220} & \num{220} & \num{220} & \num{220} & \num{220} & \num{220}\\
			
			\midrule
			Mean outcome & \num{3.2} & \num{3.2} & \num{3.2} & \num{3.2} & \num{3.2} & \num{3.2}\\
			SD outcome & \num{0.5} & \num{0.5} & \num{0.5} & \num{0.5} & \num{0.5} & \num{0.5}\\
			\bottomrule
		\end{tabular}
	}
	\caption*{\footnotesize{Note: The table presents the outputs for the regressions (\ref{eq: impact ai on wages and employment}) when the dependent variable is mean hourly wages in log and 2SLS estimators is applied. Only high income countries in 2010 have been included in the IV countries group (n = 43). Augmentation AI exposure and automation AI exposure are standardized to have a mean of 0 and a standard deviation equal to 1. Covariates included are: employment size (log), trade per capita and demographic characteristics (age, gender, education, ethnicity, and race) for industry (4-digit). The estimations are weighted by 2015 employment size. Standard errors are reported in brackets and are clustered at the occupation-industry cell. $\star\star\star$ Significant at the 1\% level; $\star\star$ significant at the 5\% level; $\star$ significant at the 10\% level.}}
\end{table}

\section{Using full matrices of sentence similarity scores}\label{app: full matrices ss}

\setcounter{table}{0}
\renewcommand{\thetable}{I\arabic{table}}

\setcounter{figure}{0}
\renewcommand{\thefigure}{I\arabic{figure}}

\begin{table}[H]
	\centering
	\caption{Effect of AI exposure on the cumulative share of new work}
	\label{tbl: full matrices ss new work}
	\resizebox{\textwidth}{!}{%
		\begin{tabular}[t]{lcccccc}
			\toprule
			& (1) & (2) & (3) & (4) & (5) & (6)\\
			\midrule
			&  &  &  &  &  & \\
			& \multicolumn{6}{c}{\textit{Panel A: OLS estimators}} \\
			\cline{2-7}
			&  &  &  &  &  & \\
			Automation AI   & \num{-0.004}  & \num{-0.001}  &                 &                 & \num{-0.005}   & \num{-0.001}   \\
			& (\num{0.006}) & (\num{0.006}) &                 &                 & (\num{0.006})  & (\num{0.006})  \\
			Augmentation AI &                &                & \num{0.016}*** & \num{0.023}*** & \num{0.016}*** & \num{0.023}*** \\
			&                &                & (\num{0.004})  & (\num{0.004})  & (\num{0.004})  & (\num{0.004})  \\
			$R^2 \ adj.$ & \num{0.644}   & \num{0.651}   & \num{0.647}    & \num{0.657}    & \num{0.647}    & \num{0.657}    \\
			$R^2 \ within \ adj.$ & \num{0.024}   & \num{0.007}   & \num{0.034}    & \num{0.024}    & \num{0.034}    & \num{0.024}    \\
			
			&  &  &  &  &  & \\
			
			& \multicolumn{6}{c}{\textit{Panel B: 2SLS estimators}} \\
			\cline{2-7}
			&  &  &  &  &  & \\
			Automation AI   & \num{-0.007}  & \num{-0.005}  &                 &                 & \num{-0.007}   & \num{-0.005}   \\
			& (\num{0.007}) & (\num{0.007}) &                 &                 & (\num{0.007})  & (\num{0.007})  \\
			Augmentation AI &                &                & \num{0.025}*** & \num{0.033}*** & \num{0.026}*** & \num{0.033}*** \\
			&                &                & (\num{0.004})  & (\num{0.004})  & (\num{0.004})  & (\num{0.004})  \\
			\addlinespace
			F-Stat (auto) & \num{13120} & \num{2083} &  &  & \num{12390} & \num{1996} \\
			F-Stat (augm) &  &  & \num{7059} & \num{321} & \num{6664} & \num{323} \\
			
			&  &  &  &  &  & \\
			
			\textit{Covariates included} & \checkmark & \checkmark & \checkmark & \checkmark & \checkmark & \checkmark\\
			\textit{Fixed effects:}  &  &  &  &  &  & \\
			cNAICS*cSOC & \checkmark & \checkmark & \checkmark & \checkmark & \checkmark & \checkmark \\
			cNAICS*year (3-digit) & & \checkmark & & \checkmark & & \checkmark \\
			
			\midrule
			Observations & \num{202695} & \num{202695} & \num{202695} & \num{202695} & \num{202695} & \num{202695}\\
			Unique cSOC (6-digit) & \num{702} & \num{702} & \num{702} & \num{702} & \num{702} & \num{702}\\
			Unique cNAICS (4-digit) & \num{220} & \num{220} & \num{220} & \num{220} & \num{220} & \num{220}\\
			
			\midrule
			Mean outcome & \num{0.02} & \num{0.02} & \num{0.02} & \num{0.02} & \num{0.02} & \num{0.02}\\
			SD outcome & \num{0.04} & \num{0.04} & \num{0.04} & \num{0.04} & \num{0.04} & \num{0.04}\\
			\bottomrule
		\end{tabular}
	}
	\caption*{\footnotesize{Note: The table presents the outputs for the regressions (\ref{eq: impact ai on new work}) when the dependent variable is the cumulative share of new work. Panel A shows the output when OLS estimators are used, whereas Panel B is when 2SLS estimators is applied. Augmentation AI exposure and automation AI exposure are standardized to have a mean of 0 and a standard deviation equal to 1. Covariates included are: employment size (log), trade per capita and demographic characteristics (age, gender, education, ethnicity, and race) for industry (4-digit). The estimations are weighted by 2015 employment size. Standard errors are reported in brackets and are clustered at the occupation-industry cell. $\star\star\star$ Significant at the 1\% level; $\star\star$ significant at the 5\% level; $\star$ significant at the 10\% level.}}
\end{table}

\begin{table}[H]
	\centering
	\caption{Effect of AI exposure on employment (log)}
	\label{tbl: full matrices ss employment}
	\resizebox{\textwidth}{!}{%
		\begin{tabular}[t]{lcccccc}
			\toprule
			& (1) & (2) & (3) & (4) & (5) & (6)\\
			\midrule
			&  &  &  &  &  & \\
			& \multicolumn{6}{c}{\textit{Panel A: OLS estimators}} \\
			\cline{2-7}
			&  &  &  &  &  & \\
			Automation AI   & \num{0.083}   & \num{0.065}   &                &                 & \num{0.081}   & \num{0.066}    \\
			& (\num{0.062}) & (\num{0.053}) &                &                 & (\num{0.063}) & (\num{0.053})  \\
			Augmentation AI &                &                & \num{0.055}** & \num{0.087}*** & \num{0.054}** & \num{0.087}*** \\
			&                &                & (\num{0.024}) & (\num{0.025})  & (\num{0.024}) & (\num{0.025})  \\
			$R^2 \ adj.$ & \num{0.994}   & \num{0.995}   & \num{0.994}   & \num{0.995}    & \num{0.994}   & \num{0.995}    \\
			$R^2 \ within \ adj.$ & \num{0.052}   & \num{0.026}   & \num{0.054}   & \num{0.030}    & \num{0.054}   & \num{0.030}    \\
			
			&  &  &  &  &  & \\
			
			& \multicolumn{6}{c}{\textit{Panel B: 2SLS estimators}} \\
			\cline{2-7}
			&  &  &  &  &  & \\
			Automation AI   & \num{0.046}   & \num{0.031}   &                &                & \num{0.045}   & \num{0.030}   \\
			& (\num{0.067}) & (\num{0.057}) &                &                & (\num{0.067}) & (\num{0.057}) \\
			Augmentation AI &                &                & \num{0.046}   & \num{0.074}** & \num{0.045}   & \num{0.073}** \\
			&                &                & (\num{0.028}) & (\num{0.031}) & (\num{0.028}) & (\num{0.031}) \\
			\addlinespace
			F-Stat (auto) & \num{17110} & \num{2002} &  &  & \num{17380} & \num{1913} \\
			F-Stat (augm) &  &  & \num{10020} & \num{323} & \num{9400} & \num{316} \\
			
			&  &  &  &  &  & \\
			
			\textit{Covariates included} & \checkmark & \checkmark & \checkmark & \checkmark & \checkmark & \checkmark\\
			\textit{Fixed effects:}  &  &  &  &  &  & \\
			cNAICS*cSOC & \checkmark & \checkmark & \checkmark & \checkmark & \checkmark & \checkmark \\
			cNAICS*year (3-digit) & & \checkmark & & \checkmark & & \checkmark \\
			
			\midrule
			Observations & \num{202695} & \num{202695} & \num{202695} & \num{202695} & \num{202695} & \num{202695}\\
			Unique cSOC (6-digit) & \num{702} & \num{702} & \num{702} & \num{702} & \num{702} & \num{702}\\
			Unique cNAICS (4-digit) & \num{220} & \num{220} & \num{220} & \num{220} & \num{220} & \num{220}\\
			
			\midrule
			Mean outcome & \num{11.0} & \num{11.0} & \num{11.0} & \num{11.0} & \num{11.0} & \num{11.0}\\
			SD outcome & \num{2.3} & \num{2.3} & \num{2.3} & \num{2.3} & \num{2.3} & \num{2.3}\\
			\bottomrule
		\end{tabular}
	}
	\caption*{\footnotesize{Note: The table presents the outputs for the regressions (\ref{eq: impact ai on wages and employment}) when the dependent variable is the employment size in log. Panel A shows the output when OLS estimators are used, whereas Panel B is when 2SLS estimators is applied. Augmentation AI exposure and automation AI exposure are standardized to have a mean of 0 and a standard deviation equal to 1. Covariates included are: mean hourly wages (log), trade per capita and demographic characteristics (age, gender, education, ethnicity, and race) for industry (4-digit). The estimations are weighted by 2015 employment size. Standard errors are reported in brackets and are clustered at the occupation-industry cell. $\star\star\star$ Significant at the 1\% level; $\star\star$ significant at the 5\% level; $\star$ significant at the 10\% level.}}
\end{table}

\begin{table}[H]
	\centering
	\caption{Effect of AI exposure on hourly wages (log)}
	\label{tbl: full matrices ss wages}
	\resizebox{\textwidth}{!}{%
		\begin{tabular}[t]{lcccccc}
			\toprule
			& (1) & (2) & (3) & (4) & (5) & (6)\\
			\midrule
			&  &  &  &  &  & \\
			& \multicolumn{6}{c}{\textit{Panel A: OLS estimators}} \\
			\cline{2-7}
			&  &  &  &  &  & \\
			Automation AI   & \num{-0.185}*** & \num{-0.178}*** &                &                & \num{-0.185}*** & \num{-0.178}*** \\
			& (\num{0.012})   & (\num{0.011})   &                &                & (\num{0.012})   & (\num{0.011})   \\
			Augmentation AI &                  &                  & \num{0.003}   & \num{0.007}   & \num{0.004}     & \num{0.006}     \\
			&                  &                  & (\num{0.005}) & (\num{0.005}) & (\num{0.004})   & (\num{0.005})   \\
			$R^2 \ adj.$ & \num{0.994}     & \num{0.994}     & \num{0.993}   & \num{0.994}   & \num{0.994}     & \num{0.994}     \\
			$R^2 \ within \ adj.$ & \num{0.156}     & \num{0.081}     & \num{0.093}   & \num{0.020}   & \num{0.156}     & \num{0.081}     \\
			
			&  &  &  &  &  & \\
			
			& \multicolumn{6}{c}{\textit{Panel B: 2SLS estimators}} \\
			\cline{2-7}
			&  &  &  &  &  & \\
			Automation AI   & \num{-0.187}*** & \num{-0.176}*** &                &                & \num{-0.187}*** & \num{-0.176}*** \\
			& (\num{0.013})   & (\num{0.012})   &                &                & (\num{0.013})   & (\num{0.012})   \\
			Augmentation AI &                  &                  & \num{-0.009}* & \num{-0.001}  & \num{-0.003}    & \num{0.004}     \\
			&                  &                  & (\num{0.005}) & (\num{0.005}) & (\num{0.005})   & (\num{0.005})   \\
			\addlinespace
			F-Stat (auto) & \num{13120} & \num{2083} &  &  & \num{12390} & \num{1996} \\
			F-Stat (augm) &  &  & \num{7059} & \num{321} & \num{6664} & \num{323} \\
			
			&  &  &  &  &  & \\
			
			\textit{Covariates included} & \checkmark & \checkmark & \checkmark & \checkmark & \checkmark & \checkmark\\
			\textit{Fixed effects:}  &  &  &  &  &  & \\
			cNAICS*cSOC & \checkmark & \checkmark & \checkmark & \checkmark & \checkmark & \checkmark \\
			cNAICS*year (3-digit) & & \checkmark & & \checkmark & & \checkmark \\
			
			\midrule
			Observations & \num{202695} & \num{202695} & \num{202695} & \num{202695} & \num{202695} & \num{202695}\\
			Unique cSOC (6-digit) & \num{702} & \num{702} & \num{702} & \num{702} & \num{702} & \num{702}\\
			Unique cNAICS (4-digit) & \num{220} & \num{220} & \num{220} & \num{220} & \num{220} & \num{220}\\
			
			\midrule
			Mean outcome & \num{3.2} & \num{3.2} & \num{3.2} & \num{3.2} & \num{3.2} & \num{3.2}\\
			SD outcome & \num{0.5} & \num{0.5} & \num{0.5} & \num{0.5} & \num{0.5} & \num{0.5}\\
			\bottomrule
		\end{tabular}
	}
	\caption*{\footnotesize{Note: The table presents the outputs for the regressions (\ref{eq: impact ai on wages and employment}) when the dependent variable is the mean hourly wages in log. Panel A shows the output when OLS estimators are used, whereas Panel B is when 2SLS estimators is applied. Augmentation AI exposure and automation AI exposure are standardized to have a mean of 0 and a standard deviation equal to 1. Covariates included are: employment size (log), trade per capita and demographic characteristics (age, gender, education, ethnicity, and race) for industry (4-digit). The estimations are weighted by 2015 employment size. Standard errors are reported in brackets and are clustered at the occupation-industry cell. $\star\star\star$ Significant at the 1\% level; $\star\star$ significant at the 5\% level; $\star$ significant at the 10\% level.}}
\end{table}

\section{Tags' technical descriptions from Stack Overflow}\label{app: first_excerpt}

\setcounter{table}{0}
\renewcommand{\thetable}{J\arabic{table}}

\setcounter{figure}{0}
\renewcommand{\thefigure}{J\arabic{figure}}

\begin{table}[H]
	\centering
	\caption{Effect of AI exposure on the cumulative share of new work}
	\label{tbl: app first excerpt new work}
	\resizebox{\textwidth}{!}{%
		\begin{tabular}[t]{lcccccc}
			\toprule
			& (1) & (2) & (3) & (4) & (5) & (6)\\
			\midrule
			&  &  &  &  &  & \\
			& \multicolumn{6}{c}{\textit{Panel A: OLS estimators}} \\
			\cline{2-7}
			&  &  &  &  &  & \\
			Automation AI   & \num{-0.002}  & \num{0.000}   &                 &                 & \num{-0.002}   & \num{0.001}    \\
			& (\num{0.004}) & (\num{0.003}) &                 &                 & (\num{0.004})  & (\num{0.003})  \\
			Augmentation AI &                &                & \num{0.012}*** & \num{0.017}*** & \num{0.012}*** & \num{0.017}*** \\
			&                &                & (\num{0.002})  & (\num{0.002})  & (\num{0.002})  & (\num{0.002})  \\
			$R^2 \ adj.$ & \num{0.644}   & \num{0.651}   & \num{0.649}    & \num{0.660}    & \num{0.649}    & \num{0.660}\\
			$R^2 \ within \ adj.$ & \num{0.024}   & \num{0.007}   & \num{0.039}    & \num{0.031}    & \num{0.039}    & \num{0.031}\\
			
			&  &  &  &  &  & \\
			
			& \multicolumn{6}{c}{\textit{Panel B: 2SLS estimators}} \\
			\cline{2-7}
			&  &  &  &  &  & \\
			Automation AI   & \num{-0.003}  & \num{-0.001}  &                 &                 & \num{-0.004}   & \num{-0.002}   \\
			& (\num{0.004}) & (\num{0.004}) &                 &                 & (\num{0.004})  & (\num{0.004})  \\
			Augmentation AI &                &                & \num{0.019}*** & \num{0.024}*** & \num{0.019}*** & \num{0.024}*** \\
			&                &                & (\num{0.002})  & (\num{0.002})  & (\num{0.002})  & (\num{0.002})  \\
			\addlinespace
			F-Stat (auto) & \num{17840} & \num{854} &  &  & \num{16620} & \num{1832} \\
			F-Stat (augm) &  &  & \num{1651} & \num{295} & \num{801} & \num{281} \\
			
			&  &  &  &  &  & \\
			
			\textit{Covariates included} & \checkmark & \checkmark & \checkmark & \checkmark & \checkmark & \checkmark\\
			\textit{Fixed effects:}  &  &  &  &  &  & \\
			cNAICS*cSOC & \checkmark & \checkmark & \checkmark & \checkmark & \checkmark & \checkmark \\
			cNAICS*year (3-digit) & & \checkmark & & \checkmark & & \checkmark \\
			
			\midrule
			Observations & \num{202695} & \num{202695} & \num{202695} & \num{202695} & \num{202695} & \num{202695}\\
			Unique cSOC (6-digit) & \num{702} & \num{702} & \num{702} & \num{702} & \num{702} & \num{702}\\
			Unique cNAICS (4-digit) & \num{220} & \num{220} & \num{220} & \num{220} & \num{220} & \num{220}\\
			
			\midrule
			Mean outcome & \num{0.02} & \num{0.02} & \num{0.02} & \num{0.02} & \num{0.02} & \num{0.02}\\
			SD outcome & \num{0.04} & \num{0.04} & \num{0.04} & \num{0.04} & \num{0.04} & \num{0.04}\\
			\bottomrule
		\end{tabular}
	}
	\caption*{\footnotesize{Note: The table presents the outputs for the regressions (\ref{eq: impact ai on new work}) when the dependent variable is the cumulative share of new work. Panel A shows the output when OLS estimators are used, whereas Panel B is when 2SLS estimators is applied. Augmentation AI exposure and automation AI exposure are standardized to have a mean of 0 and a standard deviation equal to 1. Covariates included are: employment size (log), trade per capita and demographic characteristics (age, gender, education, ethnicity, and race) for industry (4-digit). The estimations are weighted by 2015 employment size. Standard errors are reported in brackets and are clustered at the occupation-industry cell. $\star\star\star$ Significant at the 1\% level; $\star\star$ significant at the 5\% level; $\star$ significant at the 10\% level.}}
\end{table}

\begin{table}[H]
	\centering
	\caption{Effect of AI exposure on employment (log)}
	\label{tbl: app first excerpt employment}
	\resizebox{\textwidth}{!}{%
		\begin{tabular}[t]{lcccccc}
			\toprule
			& (1) & (2) & (3) & (4) & (5) & (6)\\
			\midrule
			&  &  &  &  &  & \\
			& \multicolumn{6}{c}{\textit{Panel A: OLS estimators}} \\
			\cline{2-7}
			&  &  &  &  &  & \\
			Automation AI   & \num{0.064}*  & \num{0.053}*  &                &                & \num{0.063}*  & \num{0.054}*  \\
			& (\num{0.038}) & (\num{0.032}) &                &                & (\num{0.038}) & (\num{0.031}) \\
			Augmentation AI &                &                & \num{0.021}   & \num{0.036}** & \num{0.021}   & \num{0.036}** \\
			&                &                & (\num{0.015}) & (\num{0.016}) & (\num{0.015}) & (\num{0.016}) \\
			$R^2 \ adj.$          & \num{0.994}   & \num{0.995}   & \num{0.994}   & \num{0.995}   & \num{0.994}   & \num{0.995}   \\
			$R^2 \ within \ adj.$ & \num{0.053}   & \num{0.027}   & \num{0.052}   & \num{0.028}   & \num{0.054}   & \num{0.029}   \\
			
			&  &  &  &  &  & \\
			
			& \multicolumn{6}{c}{\textit{Panel B: 2SLS estimators}} \\
			\cline{2-7}
			&  &  &  &  &  & \\
			Automation AI   & \num{0.065}*  & \num{0.060}*  &                &                & \num{0.064}*  & \num{0.059}*  \\
			& (\num{0.039}) & (\num{0.032}) &                &                & (\num{0.039}) & (\num{0.032}) \\
			Augmentation AI &                &                & \num{0.009}   & \num{0.025}   & \num{0.008}   & \num{0.024}   \\
			&                &                & (\num{0.018}) & (\num{0.019}) & (\num{0.018}) & (\num{0.019}) \\
			\addlinespace
			F-Stat (auto) & \num{22100} & \num{871} &  &  & \num{20220} & \num{2318} \\
			F-Stat (augm) &  &  & \num{1987} & \num{299} & \num{818} & \num{285} \\
			
			&  &  &  &  &  & \\
			
			\textit{Covariates included} & \checkmark & \checkmark & \checkmark & \checkmark & \checkmark & \checkmark\\
			\textit{Fixed effects:}  &  &  &  &  &  & \\
			cNAICS*cSOC & \checkmark & \checkmark & \checkmark & \checkmark & \checkmark & \checkmark \\
			cNAICS*year (3-digit) & & \checkmark & & \checkmark & & \checkmark \\
			
			\midrule
			Observations & \num{202695} & \num{202695} & \num{202695} & \num{202695} & \num{202695} & \num{202695}\\
			Unique cSOC (6-digit) & \num{702} & \num{702} & \num{702} & \num{702} & \num{702} & \num{702}\\
			Unique cNAICS (4-digit) & \num{220} & \num{220} & \num{220} & \num{220} & \num{220} & \num{220}\\
			
			\midrule
			Mean outcome & \num{11.0} & \num{11.0} & \num{11.0} & \num{11.0} & \num{11.0} & \num{11.0}\\
			SD outcome & \num{2.3} & \num{2.3} & \num{2.3} & \num{2.3} & \num{2.3} & \num{2.3}\\
			\bottomrule
		\end{tabular}
	}
	\caption*{\footnotesize{Note: The table presents the outputs for the regressions (\ref{eq: impact ai on wages and employment}) when the dependent variable is the employment size in log. Panel A shows the output when OLS estimators are used, whereas Panel B is when 2SLS estimators is applied. Augmentation AI exposure and automation AI exposure are standardized to have a mean of 0 and a standard deviation equal to 1. Covariates included are: mean hourly wages (log), trade per capita and demographic characteristics (age, gender, education, ethnicity, and race) for industry (4-digit). The estimations are weighted by 2015 employment size. Standard errors are reported in brackets and are clustered at the occupation-industry cell. $\star\star\star$ Significant at the 1\% level; $\star\star$ significant at the 5\% level; $\star$ significant at the 10\% level.}}
\end{table}

\begin{table}[H]
	\centering
	\caption{Effect of AI exposure on hourly wages (log)}
	\label{tbl: app first excerpt wages}
	\resizebox{\textwidth}{!}{%
		\begin{tabular}[t]{lcccccc}
			\toprule
			& (1) & (2) & (3) & (4) & (5) & (6)\\
			\midrule
			&  &  &  &  &  & \\
			& \multicolumn{6}{c}{\textit{Panel A: OLS estimators}} \\
			\cline{2-7}
			&  &  &  &  &  & \\
			Automation AI   & \num{-0.110}*** & \num{-0.106}*** &                &                & \num{-0.110}*** & \num{-0.106}*** \\
			& (\num{0.007})   & (\num{0.007})   &                &                & (\num{0.007})   & (\num{0.007})   \\
			Augmentation AI &                  &                  & \num{-0.003}  & \num{0.000}   & \num{-0.003}    & \num{-0.001}    \\
			&                  &                  & (\num{0.002}) & (\num{0.003}) & (\num{0.002})   & (\num{0.002})   \\
			$R^2 \ adj.$ & \num{0.994}     & \num{0.994}     & \num{0.993}   & \num{0.994}   & \num{0.994}     & \num{0.994}     \\
			$R^2 \ within \ adj.$ & \num{0.159}     & \num{0.083}     & \num{0.093}   & \num{0.019}   & \num{0.159}     & \num{0.083}     \\
			
			&  &  &  &  &  & \\
			
			& \multicolumn{6}{c}{\textit{Panel B: 2SLS estimators}} \\
			\cline{2-7}
			&  &  &  &  &  & \\
			Automation AI   & \num{-0.117}*** & \num{-0.108}*** &                  &                & \num{-0.117}*** & \num{-0.108}*** \\
			& (\num{0.008})   & (\num{0.007})   &                  &                & (\num{0.008})   & (\num{0.007})   \\
			Augmentation AI &                  &                  & \num{-0.010}*** & \num{-0.002}  & \num{-0.007}*** & \num{-0.001}    \\
			&                  &                  & (\num{0.003})   & (\num{0.003}) & (\num{0.003})   & (\num{0.002})   \\
			\addlinespace
			F-Stat (auto) & \num{17840} & \num{854} &  &  & \num{16620} & \num{1832} \\
			F-Stat (augm) &  &  & \num{1651} & \num{295} & \num{801} & \num{281} \\
			
			&  &  &  &  &  & \\
			
			\textit{Covariates included} & \checkmark & \checkmark & \checkmark & \checkmark & \checkmark & \checkmark\\
			\textit{Fixed effects:}  &  &  &  &  &  & \\
			cNAICS*cSOC & \checkmark & \checkmark & \checkmark & \checkmark & \checkmark & \checkmark \\
			cNAICS*year (3-digit) & & \checkmark & & \checkmark & & \checkmark \\
			
			\midrule
			Observations & \num{202695} & \num{202695} & \num{202695} & \num{202695} & \num{202695} & \num{202695}\\
			Unique cSOC (6-digit) & \num{702} & \num{702} & \num{702} & \num{702} & \num{702} & \num{702}\\
			Unique cNAICS (4-digit) & \num{220} & \num{220} & \num{220} & \num{220} & \num{220} & \num{220}\\
			
			\midrule
			Mean outcome & \num{3.2} & \num{3.2} & \num{3.2} & \num{3.2} & \num{3.2} & \num{3.2}\\
			SD outcome & \num{0.5} & \num{0.5} & \num{0.5} & \num{0.5} & \num{0.5} & \num{0.5}\\
			\bottomrule
		\end{tabular}
	}
	\caption*{\footnotesize{Note: The table presents the outputs for the regressions (\ref{eq: impact ai on wages and employment}) when the dependent variable is the employment size in log. Panel A shows the output when OLS estimators are used, whereas Panel B is when 2SLS estimators is applied. Augmentation AI exposure and automation AI exposure are standardized to have a mean of 0 and a standard deviation equal to 1. Covariates included are: employment size (log), trade per capita and demographic characteristics (age, gender, education, ethnicity, and race) for industry (4-digit). The estimations are weighted by 2015 employment size. Standard errors are reported in brackets and are clustered at the occupation-industry cell. $\star\star\star$ Significant at the 1\% level; $\star\star$ significant at the 5\% level; $\star$ significant at the 10\% level.}}
\end{table}

\section{Current employment size as weights}\label{app: current weights}

\setcounter{table}{0}
\renewcommand{\thetable}{K\arabic{table}}

\setcounter{figure}{0}
\renewcommand{\thefigure}{K\arabic{figure}}

\begin{table}[H]
	\centering
	\caption{Effect of AI exposure on the cumulative share of new work}
	\label{tbl: app current weights new work}
	\resizebox{\textwidth}{!}{%
		\begin{tabular}[t]{lcccccc}
			\toprule
			& (1) & (2) & (3) & (4) & (5) & (6)\\
			\midrule
			&  &  &  &  &  & \\
			& \multicolumn{6}{c}{\textit{Panel A: OLS estimators}} \\
			\cline{2-7}
			&  &  &  &  &  & \\
			Automation AI   & \num{-0.002}  & \num{-0.001}  &                 &                 & \num{-0.002}   & \num{0.000}    \\
			& (\num{0.003}) & (\num{0.003}) &                 &                 & (\num{0.003})  & (\num{0.003})  \\
			Augmentation AI &                &                & \num{0.012}*** & \num{0.016}*** & \num{0.012}*** & \num{0.016}*** \\
			&                &                & (\num{0.002})  & (\num{0.002})  & (\num{0.002})  & (\num{0.002})  \\
			$R^2 \ adj.$ & \num{0.655}   & \num{0.662}   & \num{0.660}    & \num{0.670}    & \num{0.660}    & \num{0.670}    \\
			$R^2 \ within \ adj.$ & \num{0.028}   & \num{0.009}   & \num{0.041}    & \num{0.030}    & \num{0.041}    & \num{0.030}    \\
			
			&  &  &  &  &  & \\
			
			& \multicolumn{6}{c}{\textit{Panel B: 2SLS estimators}} \\
			\cline{2-7}
			&  &  &  &  &  & \\
			Automation AI   & \num{-0.003}  & \num{-0.003}  &                 &                 & \num{-0.003}   & \num{-0.002}   \\
			& (\num{0.003}) & (\num{0.003}) &                 &                 & (\num{0.003})  & (\num{0.003})  \\
			Augmentation AI &                &                & \num{0.020}*** & \num{0.024}*** & \num{0.020}*** & \num{0.025}*** \\
			&                &                & (\num{0.003})  & (\num{0.003})  & (\num{0.003})  & (\num{0.003})  \\
			
			\addlinespace
			F-Stat (auto) & \num{12560} & \num{1657} &  &  & \num{13070} & \num{1541} \\
			F-Stat (augm) &  &  & \num{1462} & \num{208} & \num{1638} & \num{204} \\
			
			&  &  &  &  &  & \\
			
			\textit{Covariates included} & \checkmark & \checkmark & \checkmark & \checkmark & \checkmark & \checkmark\\
			\textit{Fixed effects:}  &  &  &  &  &  & \\
			cNAICS*cSOC & \checkmark & \checkmark & \checkmark & \checkmark & \checkmark & \checkmark \\
			cNAICS*year (3-digit) & & \checkmark & & \checkmark & & \checkmark \\
			
			\midrule
			Observations & \num{202695} & \num{202695} & \num{202695} & \num{202695} & \num{202695} & \num{202695}\\
			Unique cSOC (6-digit) & \num{702} & \num{702} & \num{702} & \num{702} & \num{702} & \num{702}\\
			Unique cNAICS (4-digit) & \num{220} & \num{220} & \num{220} & \num{220} & \num{220} & \num{220}\\
			
			\midrule
			Mean outcome & \num{0.02} & \num{0.02} & \num{0.02} & \num{0.02} & \num{0.02} & \num{0.02}\\
			SD outcome & \num{0.04} & \num{0.04} & \num{0.04} & \num{0.04} & \num{0.04} & \num{0.04}\\
			\bottomrule
		\end{tabular}
	}
	\caption*{\footnotesize{Note: The table presents the outputs for the regressions (\ref{eq: impact ai on new work}) when the dependent variable is the cumulative share of new work. Panel A shows the output when OLS estimators are used, whereas Panel B is when 2SLS estimators is applied. Augmentation AI exposure and automation AI exposure are standardized to have a mean of 0 and a standard deviation equal to 1. Covariates included are: employment size (log), trade per capita and demographic characteristics (age, gender, education, ethnicity, and race) for industry (4-digit). The estimations are weighted by employment size. Standard errors are reported in brackets and are clustered at the occupation-industry cell. $\star\star\star$ Significant at the 1\% level; $\star\star$ significant at the 5\% level; $\star$ significant at the 10\% level.}}
\end{table}

\begin{table}[H]
	\centering
	\caption{Effect of AI exposure on employment (log)}
	\label{tbl: app current weights employment}
	\resizebox{\textwidth}{!}{%
		\begin{tabular}[t]{lcccccc}
			\toprule
			& (1) & (2) & (3) & (4) & (5) & (6)\\
			\midrule
			&  &  &  &  &  & \\
			& \multicolumn{6}{c}{\textit{Panel A: OLS estimators}} \\
			\cline{2-7}
			&  &  &  &  &  & \\
			Automation AI   & \num{0.048}   & \num{0.039}   &                &                 & \num{0.048}   & \num{0.041}*   \\
			& (\num{0.031}) & (\num{0.024}) &                &                 & (\num{0.031}) & (\num{0.024})  \\
			Augmentation AI &                &                & \num{0.034}** & \num{0.040}*** & \num{0.034}** & \num{0.041}*** \\
			&                &                & (\num{0.015}) & (\num{0.014})  & (\num{0.015}) & (\num{0.014})  \\
			$R^2 \ adj.$ & \num{0.995}   & \num{0.995}   & \num{0.995}   & \num{0.995}    & \num{0.995}   & \num{0.995}    \\
			$R^2 \ within \ adj.$ & \num{0.060}   & \num{0.026}   & \num{0.062}   & \num{0.028}    & \num{0.063}   & \num{0.028}    \\
			
			&  &  &  &  &  & \\
			
			& \multicolumn{6}{c}{\textit{Panel B: 2SLS estimators}} \\
			\cline{2-7}
			&  &  &  &  &  & \\
			Automation AI   & \num{0.038}   & \num{0.028}   &                &                & \num{0.038}   & \num{0.030}   \\
			& (\num{0.032}) & (\num{0.025}) &                &                & (\num{0.032}) & (\num{0.024}) \\
			Augmentation AI &                &                & \num{0.037}** & \num{0.038}** & \num{0.035}** & \num{0.037}** \\
			&                &                & (\num{0.015}) & (\num{0.016}) & (\num{0.015}) & (\num{0.016}) \\
			\addlinespace
			F-Stat (auto) & \num{15000} & \num{1529} &  &  & \num{15480} & \num{1425} \\
			F-Stat (augm) &  &  & \num{1770} & \num{209} & \num{2003} & \num{206} \\
			
			&  &  &  &  &  & \\
			
			\textit{Covariates included} & \checkmark & \checkmark & \checkmark & \checkmark & \checkmark & \checkmark\\
			\textit{Fixed effects:}  &  &  &  &  &  & \\
			cNAICS*cSOC & \checkmark & \checkmark & \checkmark & \checkmark & \checkmark & \checkmark \\
			cNAICS*year (3-digit) & & \checkmark & & \checkmark & & \checkmark \\
			
			\midrule
			Observations & \num{202695} & \num{202695} & \num{202695} & \num{202695} & \num{202695} & \num{202695}\\
			Unique cSOC (6-digit) & \num{702} & \num{702} & \num{702} & \num{702} & \num{702} & \num{702}\\
			Unique cNAICS (4-digit) & \num{220} & \num{220} & \num{220} & \num{220} & \num{220} & \num{220}\\
			
			\midrule
			Mean outcome & \num{11.0} & \num{11.0} & \num{11.0} & \num{11.0} & \num{11.0} & \num{11.0}\\
			SD outcome & \num{2.3} & \num{2.3} & \num{2.3} & \num{2.3} & \num{2.3} & \num{2.3}\\
			\bottomrule
		\end{tabular}
	}
	\caption*{\footnotesize{Note: The table presents the outputs for the regressions (\ref{eq: impact ai on wages and employment}) when the dependent variable is the employment size in log. Panel A shows the output when OLS estimators are used, whereas Panel B is when 2SLS estimators is applied. Augmentation AI exposure and automation AI exposure are standardized to have a mean of 0 and a standard deviation equal to 1. Covariates included are: mean hourly wages (log), trade per capita and demographic characteristics (age, gender, education, ethnicity, and race) for industry (4-digit). The estimations are weighted by employment size. Standard errors are reported in brackets and are clustered at the occupation-industry cell. $\star\star\star$ Significant at the 1\% level; $\star\star$ significant at the 5\% level; $\star$ significant at the 10\% level.}}
\end{table}

\begin{table}[H]
	\centering
	\caption{Effect of AI exposure on hourly wages (log)}
	\label{tbl: app current weights wages}
	\resizebox{\textwidth}{!}{%
		\begin{tabular}[t]{lcccccc}
			\toprule
			& (1) & (2) & (3) & (4) & (5) & (6)\\
			\midrule
			&  &  &  &  &  & \\
			& \multicolumn{6}{c}{\textit{Panel A: OLS estimators}} \\
			\cline{2-7}
			&  &  &  &  &  & \\
			Automation AI   & \num{-0.091}*** & \num{-0.087}*** &                &                & \num{-0.091}*** & \num{-0.087}*** \\
			& (\num{0.006})   & (\num{0.006})   &                &                & (\num{0.006})   & (\num{0.006})   \\
			Augmentation AI &                  &                  & \num{-0.001}  & \num{0.003}   & \num{-0.001}    & \num{0.002}     \\
			&                  &                  & (\num{0.003}) & (\num{0.003}) & (\num{0.003})   & (\num{0.003})   \\
			$R^2 \ adj.$ & \num{0.994}     & \num{0.994}     & \num{0.994}   & \num{0.994}   & \num{0.994}     & \num{0.994}     \\
			$R^2 \ within \ adj.$ & \num{0.156}     & \num{0.086}     & \num{0.089}   & \num{0.022}   & \num{0.156}     & \num{0.086}     \\
			
			&  &  &  &  &  & \\
			
			& \multicolumn{6}{c}{\textit{Panel B: 2SLS estimators}} \\
			\cline{2-7}
			&  &  &  &  &  & \\
			Automation AI   & \num{-0.094}*** & \num{-0.088}*** &                 &                & \num{-0.094}*** & \num{-0.087}*** \\
			& (\num{0.006})   & (\num{0.006})   &                 &                & (\num{0.006})   & (\num{0.006})   \\
			Augmentation AI &                  &                  & \num{-0.008}** & \num{0.000}   & \num{-0.004}    & \num{0.002}     \\
			&                  &                  & (\num{0.003})  & (\num{0.003}) & (\num{0.003})   & (\num{0.003})   \\
			\addlinespace
			F-Stat (auto) & \num{12560} & \num{1657} &  &  & \num{13070} & \num{1541} \\
			F-Stat (augm) &  &  & \num{1462} & \num{208} & \num{1638} & \num{204} \\
			
			&  &  &  &  &  & \\
			
			\textit{Covariates included} & \checkmark & \checkmark & \checkmark & \checkmark & \checkmark & \checkmark\\
			\textit{Fixed effects:}  &  &  &  &  &  & \\
			cNAICS*cSOC & \checkmark & \checkmark & \checkmark & \checkmark & \checkmark & \checkmark \\
			cNAICS*year (3-digit) & & \checkmark & & \checkmark & & \checkmark \\
			
			\midrule
			Observations & \num{153664} & \num{153664} & \num{153664} & \num{153664} & \num{153664} & \num{153664}\\
			Unique cSOC (6-digit) & \num{653} & \num{653} & \num{653} & \num{653} & \num{653} & \num{653}\\
			Unique cNAICS (4-digit) & \num{220} & \num{220} & \num{220} & \num{220} & \num{220} & \num{220}\\
			
			\midrule
			Mean outcome & \num{3.2} & \num{3.2} & \num{3.2} & \num{3.2} & \num{3.2} & \num{3.2}\\
			SD outcome & \num{0.5} & \num{0.5} & \num{0.5} & \num{0.5} & \num{0.5} & \num{0.5}\\
			\bottomrule
		\end{tabular}
	}
	\caption*{\footnotesize{Note: The table presents the outputs for the regressions (\ref{eq: impact ai on wages and employment}) when the dependent variable is the employment size in log. Panel A shows the output when OLS estimators are used, whereas Panel B is when 2SLS estimators is applied. Augmentation AI exposure and automation AI exposure are standardized to have a mean of 0 and a standard deviation equal to 1. Covariates included are: employment size (log), trade per capita and demographic characteristics (age, gender, education, ethnicity, and race) for industry (4-digit). The estimations are weighted by employment size. Standard errors are reported in brackets and are clustered at the occupation-industry cell. $\star\star\star$ Significant at the 1\% level; $\star\star$ significant at the 5\% level; $\star$ significant at the 10\% level.}}
\end{table}

\clearpage

\printbibliography[heading=subbibliography,title={References for the Appendix}]

\end{refsection}

\end{document}